\documentclass[12pt,preprint]{aastex}

\shorttitle{Molecular Hydrogen in NGC 5194}
\shortauthors{Brunner et al.}

\begin{document}

\title{Warm Molecular Gas in M51: Mapping the Excitation Temperature
 and Mass of H$_2$ with the Spitzer Infrared Spectrograph}

\author{Gregory Brunner\altaffilmark{1,2}}
\affil{Department of Physics and Astronomy, Rice University, \\
    Houston, TX 77005}
\email{gbrunner@rice.edu}

\author{Kartik Sheth\altaffilmark{3}, Lee Armus\altaffilmark{3}} 
\affil{Spitzer Science Center, Caltech, Pasadena, CA 91125}

\author{Mark Wolfire\altaffilmark{4}, Stuart Vogel\altaffilmark{4}}
\affil{Department of Astronomy, University of Maryland, College Park, MD 20741}

\author{Eva Schinnerer\altaffilmark{5}}
\affil{Max-Planck-Institut f\"ur Astronomie, K\"onigstuhl 17, 69117 Heidelberg, Germany}

\author{George Helou\altaffilmark{3}}
\affil{Spitzer Science Center, Caltech, Pasadena, CA 91125}

\author{Reginald Dufour\altaffilmark{1}}
\affil{Department of Physics and Astronomy, Rice University, Houston, TX 77005}

\author{John-David Smith\altaffilmark{6}}
\affil{University of Arizona, Steward Observatory, Tucson, AZ 85721}

\and

\author{Daniel A. Dale\altaffilmark{7}}
\affil{Department of Physics and Astronomy, University of Wyoming, Laramie, WY 82071}

\altaffiltext{1}{Department of Physics and Astronomy, Rice University, Houston, TX 77005}
\altaffiltext{2}{Visiting Graduate Student Fellow, Spitzer Science Center, Caltech, Pasadena, CA 91125}
\altaffiltext{3}{Spitzer Science Center, Caltech, Pasadena, CA 91125}
\altaffiltext{4}{Department of Astronomy, University of Maryland, College Park, MD 20741}
\altaffiltext{5}{Max-Planck-Institut f\"ur Astronomie, K\"onigstuhl 17, 69117 Heidelberg, Germany}
\altaffiltext{6}{University of Arizona, Steward Observatory, Tucson, AZ 85721}
\altaffiltext{7}{Department of Physics and Astronomy, University of Wyoming, Laramie, WY 82071}

\begin{abstract}

We have mapped the warm molecular gas traced by the 
H$_2$ S(0) $-$ H$_2$ S(5) pure rotational
mid-infrared emission lines over a radial strip across the nucleus and disk of
M51 (NGC 5194) using the Infrared Spectrograph (IRS) on
the $Spitzer$ $Space$ $Telescope$.  The six H$_2$ lines 
have markedly different emission distributions.  We
obtained the H$_2$ temperature and surface 
density distributions by assuming a two temperature model: a warm (T = 100 $-$ 300 K)
phase traced by the low $J$ (S(0) -- S(2)) lines and a hot phase (T = 400 $-$ 1000 K)
traced by the high $J$ (S(2) -- S(5)) lines.  The lowest molecular gas temperatures 
are found within the spiral arms (T $\sim$ 155 K), while the highest 
temperatures are found in the inter-arm regions (T $>$ 700 K).
The warm gas surface density reaches a maximum of 11
$\mathrm{M_\sun}$ $\mathrm{pc^{-2}}$ in the north-western spiral 
arm, whereas the hot gas surface density peaks at
0.24 $\mathrm{M_\sun}$ $\mathrm{pc^{-2}}$ at the nucleus.
The spatial offset between the peaks in the warm and hot 
phases and the differences in the distributions of the H$_2$ 
line emission suggest that the warm phase is mostly produced 
by UV photons in star forming regions while the hot phase is 
mostly produced by shocks or X-rays associated with nuclear activity.  
The warm H$_2$ is found in the  dust lanes 
of M51, spatially offset from the brightest 
H$\alpha$ regions.  The warm H$_2$ is generally 
spatially coincident with the cold molecular gas 
traced by CO (J = 1 -- 0) emission, consistent with excitation of 
the warm phase in dense photodissociation regions (PDRs).
In contrast, the hot H$_2$ is most prominent in the nuclear 
region.  Here, over a 0.5 kpc radius around the nucleus of M51, 
the hot H$_2$ coincides with [O IV](25.89 $\micron$) 
and X-ray emission indicating that shocks and/or X-rays 
are responsible for exciting this phase.

\end{abstract}


\section{Introduction}

Star formation and galactic evolution are connected via the molecular
gas in a galaxy.  In the Milky Way, star formation occurs
in molecular clouds, although not all clouds
are actively forming stars.  On a global, galactic scale, star
formation may be triggered whenever the molecular gas surface density
is enhanced, for example, by a spiral density wave \citep{vog88}, by
increased pressure or gas density in galactic nuclei \citep{you91, sak99,
she05}, by hydrodynamic shocks along the leading edge of bars
\citep{she00, she02}, and in the transition region at the ends of bars
\citep{kl91,she02}.  How does this star formation affect the
surrounding molecular gas?  How is it heated and what is the
distribution of the gas temperatures?  How does the mass of the warm
and hot gas vary from region to region?  We address these questions
using spectral line maps from a radial strip across the 
grand-design spiral galaxy, M51.

M51 (the Whirlpool galaxy, NGC 5194) is a nearby,
face-on spiral galaxy that is rich in molecular gas.  Its proximity
(assumed to be 8.2 Mpc \citep{tul88}), face-on orientation, and
grand-design spiral morphology make it the ideal target for
studies of the interstellar medium (ISM) across distinct dynamical,
chemical, and physical environments in a galaxy.  Studies of the
molecular gas within M51 have revealed giant molecular associations
(GMAs) along the spiral arms \citep{vog88,ran90,aal99}, a reservoir 
of molecular gas in the nuclear region that is massive enough 
to fuel the active galactic nucleus (AGN) \citep{sco98}, and spiral density wave 
triggered star-formation in molecular clouds \citep{vog88}.
In addition to being well-studied at millimeter and radio wavelengths,
M51 has also been studied at X-ray, UV, optical, near-infrared,
infrared, and submillimeter wavelengths \citep{pal85, ter98, sco01, cal05, mat04, mei05}.  
  
In this paper we present maps of the  
H$_2$ S(0) $-$ H$_2$ S(5) pure rotational 
mid-infrared lines over a strip across M51 
created from $Spitzer$ $Space$ $Telescope$ 
Infrared Spectrograph (IRS) spectral mapping mode observations.
The mid-infrared H$_2$ lines trace the warm (T = 100 -- 1000 K) phase 
of H$_2$ and we use these lines to model the H$_2$ 
 excitation-temperature, mass \citep{rig02, hig06}, and ortho-to-para ratio 
 \citep{neu98, neu06} across the M51 strips.\footnote{While we 
 are always exploring the warm phase of H$_2$, in this paper 
 we refer to a warm and a hot phase corresponding to 
 temperatures of T = 100 -- 300 K and T = 400 -- 1000 K, 
 respectively.}  We use the inferred distributions to  place constraints 
 on the energy injection mechanisms (i.e. radiative heating, 
 shocks, turbulence) that heat the warm molecular gas 
 phase of the ISM.

\section{Observations and Data Reduction}

\subsection{Spectral Data}

\begin{figure}
\epsscale{.75}
\figurenum{1}
\plotone{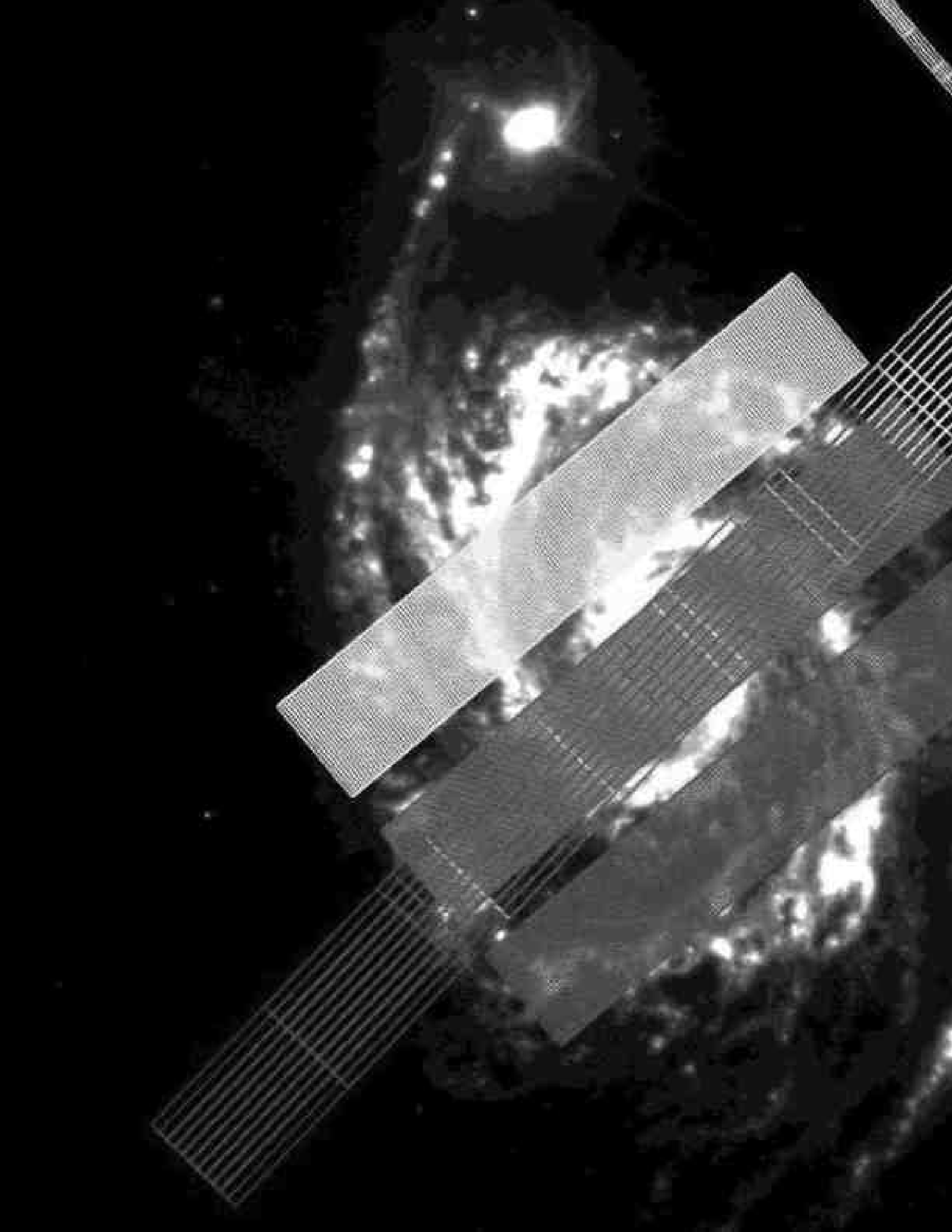}
\caption{Overlay of the IRS footprints on the $Spitzer$ IRAC 8 $\micron$ image of M51.  
The three parallel strips mark the SL (5 -- 14.5 $\micron$) observations with the white 
strip denoting the SL2 (5 -- 7.5 $\micron$) coverage and the grey strip 
denoting the SL1 (7.5 -- 14.5 $\micron$) coverage.  The central strip is where the SL1
and SL2 overlap along with the LL (14 -- 38 $\micron$) radial strip giving us complete coverage of the 
mid-infrared spectrum across the nucleus and disk of M51.  
The small off-galaxy strip is the SL background observation.}
\label{figure-1}
\end{figure}

We mapped a radial strip across \objectname{M51} using the short-low
(SL; 5 -- 14.5 $\micron$) and long-low (LL; 14 - 38 $\micron$) 
modules of the $Spitzer$ IRS in spectral
mapping mode \citep{hou04}.  The radial strips were 324$\arcsec$ $\times$
57$\arcsec$ and 295$\arcsec$ $\times$ 51$\arcsec$ in the SL and LL,
respectively.  Each slit position was mapped twice with half-slit spacings.  In
total, 1,412 spectra were taken in the SL and 100 were taken in the LL.  
Integration times for individual 
spectra were 14.6 s in both the SL and LL.  Dedicated off-source 
background observations were taken for the SL observations.  Backgrounds for the
LL observations were taken from outrigger data collected while the
spacecraft was mapping in the adjacent module.  Figure \ref{figure-1} 
presents the astronomical observation requests (AORs)
overlaid on the $Spitzer$ Infrared Array Camera (IRAC) 8 $\micron$ image of M51.

The spectra were assembled from the basic calibration data (BCD) into
spectral data cubes for each module using CUBISM \citep{ken03, smi04, smi07a}.
Background subtraction and bad pixel removal were done within CUBISM.
The individual BCDs were processed using the S14.0 version of
the Spitzer Science Center (SSC) pipeline.  In CUBISM, the SL and LL
data cubes have 1$\farcs$85 and 5$\farcs$08 pixels,
respectively.  The pixel size is half the full width at half 
maximum (FWHM) of the point spread function (PSF)
at the red end of a given module.  In principle, the PSF should 
vary with wavelength but since the PSF is
undersampled at the blue end of the module, it is approximately
constant across a given module.  So the approximate resolution of the
SL and LL modules and maps of spectral features observed 
in the SL and LL modules is 3$\farcs$7 and 10$\farcs$1, respectively.

We created continuum-subtracted line flux maps of the H$_2$
S(0) $-$ H$_2$ S(5) lines using a combination of PAHFIT
\citep{smi07b} and our own code.  PAHFIT is a spectral fitting routine
that decomposes IRS low resolution spectra into broad 
PAH features, unresolved line emission, and grain continuum with the main advantage
being that it allows one to recover the full line flux of any blended
features.  Several H$_2$ lines are blended with PAH features and atomic 
lines in IRS low resolution spectra:
H$_2$ S(1) with the 17.0 $\micron$ PAH complex,
H$_2$ S(2) with the 12.0 and 12.6 $\micron$ PAH complexes, 
H$_2$ S(4) with the 7.8 and 8.6 $\micron$ PAH complexes, and 
H$_2$ S(5) with the [Ar II](6.9 $\micron$) line.  
PAHFIT also solves for the foreground dust emission and 
dereddens the emitted line intensities.  Our code
concatenates SL1 and SL2, and LL1 and LL2 data cubes into two cubes,
one for SL and one for LL and smoothes each map in the cubes by a 3
$\times$ 3 pixel box, conserving the flux, to increase the
signal-to-noise ratio of the spectra.  Then, for each pixel,
the spectrum is extracted and PAHFIT is run to decompose it.  Our code saves the
location of the pixel on the sky along with the PAHFIT output 
(i.e. integrated line flux, line FWHM, line equivalent width, the 
uncertainty in the line flux, the fit to the entire spectrum and 
the fit to the continuum) for each spectrum and uses this 
information to construct line flux maps for all of the mid-infrared 
features.  In addition to creating line maps, our code creates maps 
of line FWHM, line equivalent width, and uncertainty in the 
flux and data cubes of the fit to the entire spectrum, the fit to 
the continuum, a continuum-subtracted data cube, and a 
residual data cube.  These parameters (the uncertainty in 
the line flux, line FWHM, and line equivalent width) are all determined 
within PAHFIT.  For a more thorough discussion of how 
PAHFIT determines quantities such as the uncertainty 
in the integrated line flux, line FWHM, and line 
equivalent width, see Smith et al. (2007b). 

\begin{figure}
\epsscale{1.1}
\figurenum{2}
\plottwo{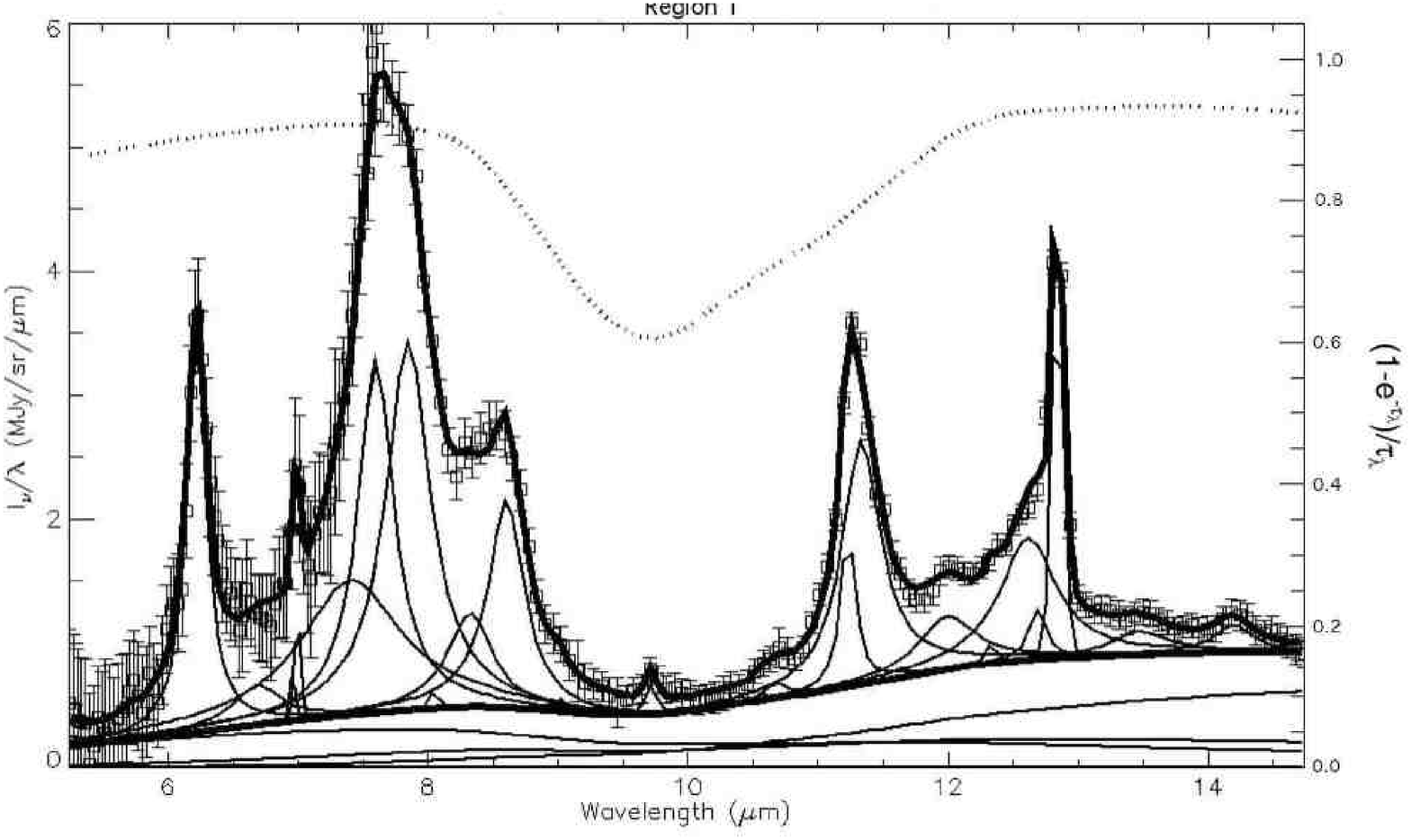}{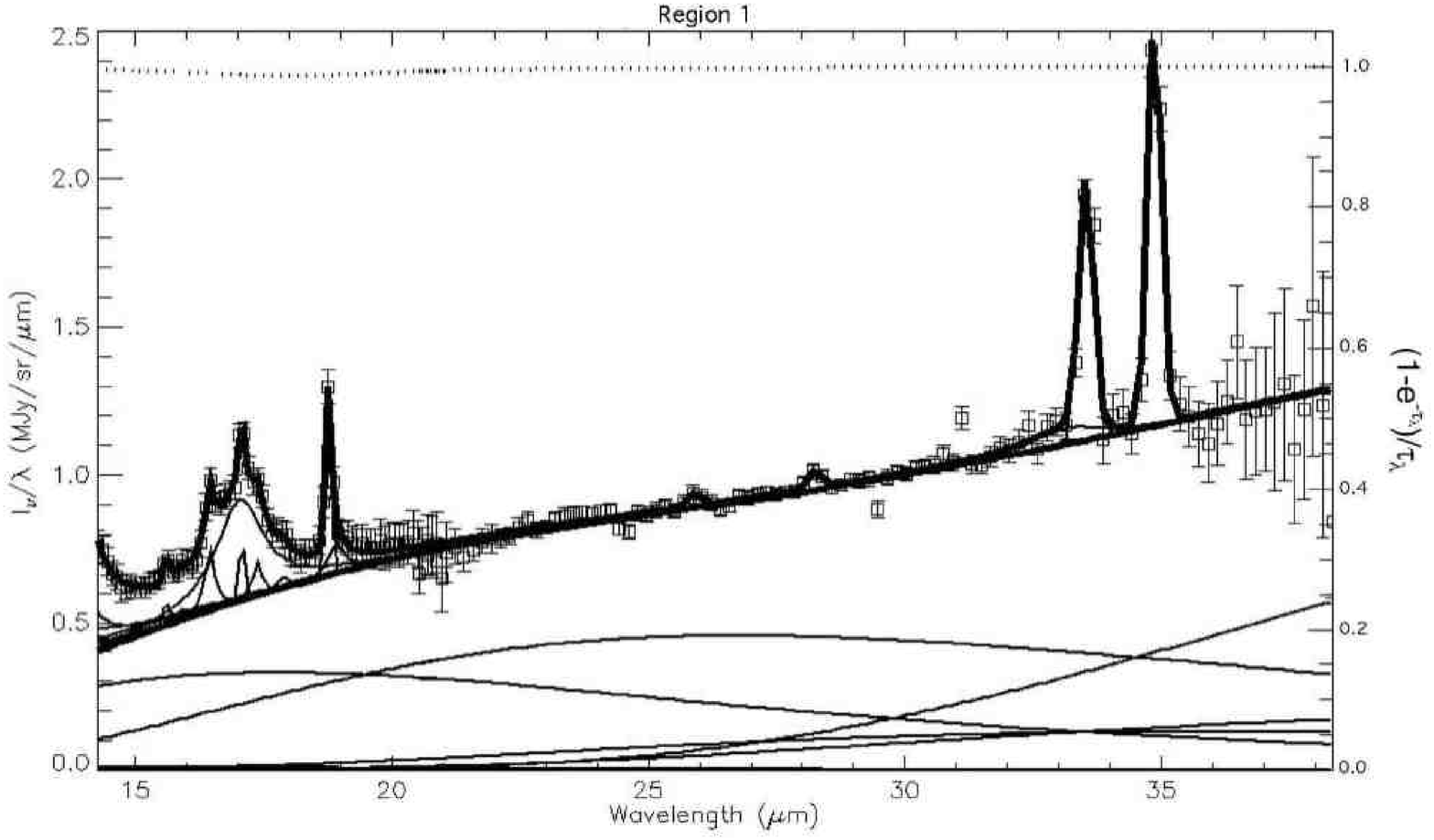}
\plottwo{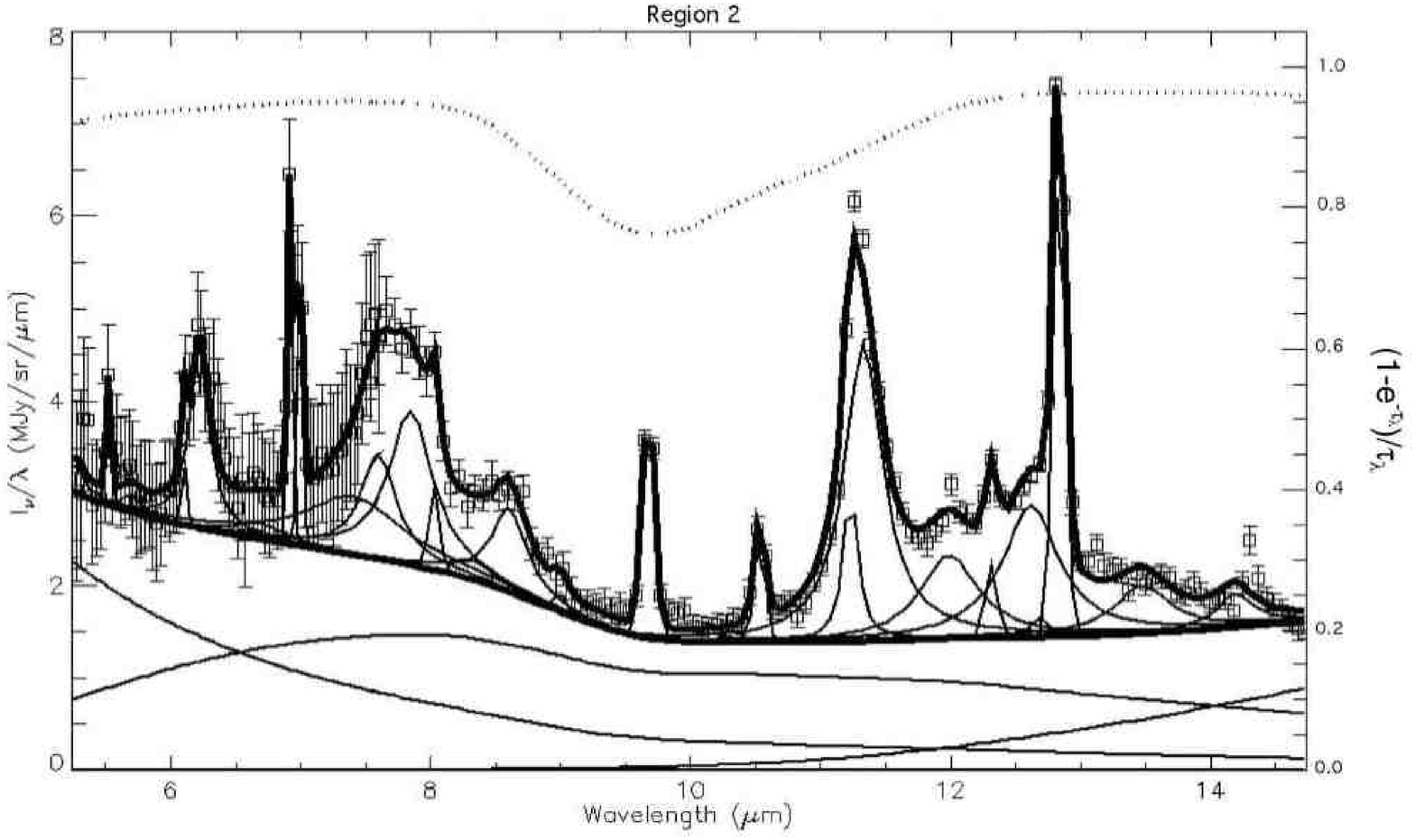}{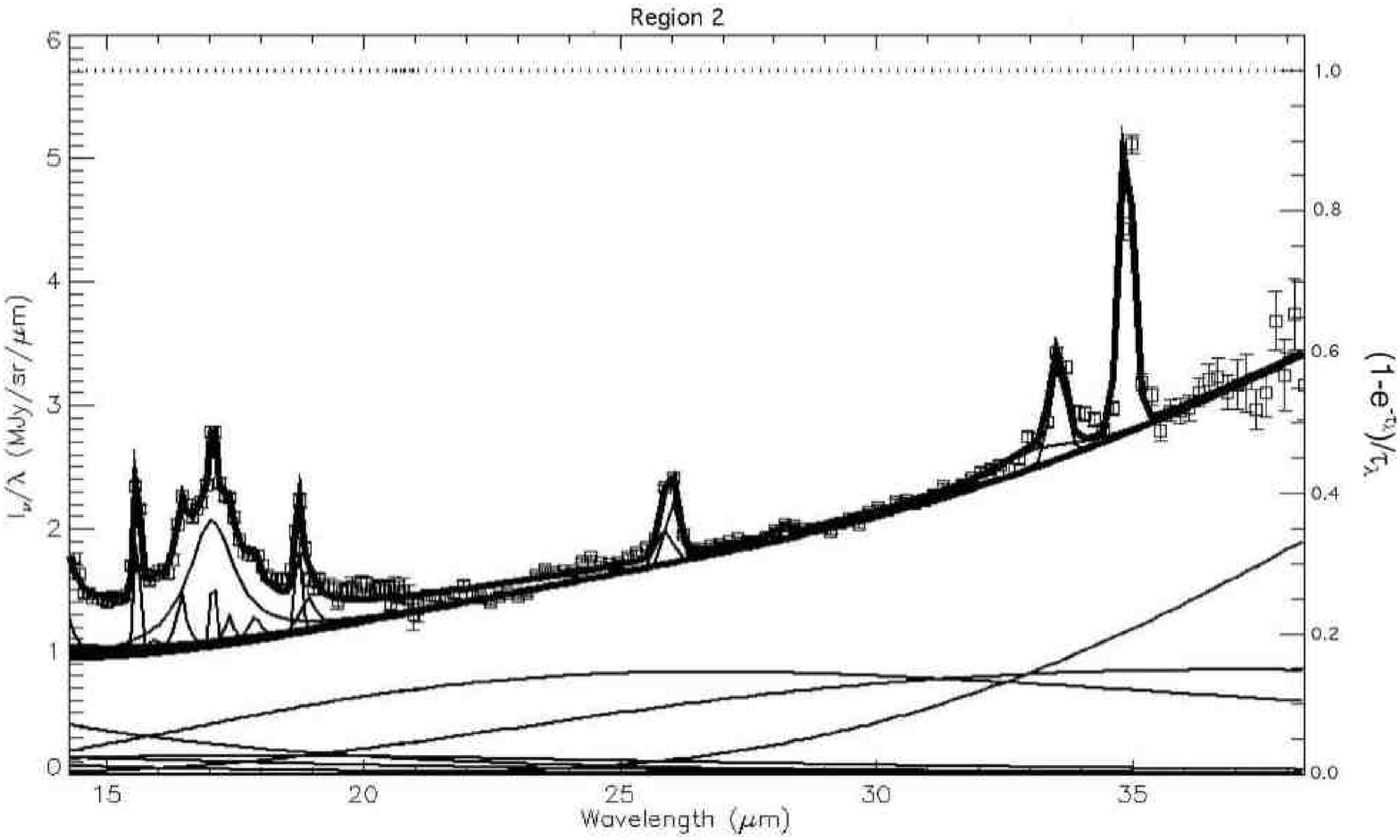}
\epsscale{1.1}
\plottwo{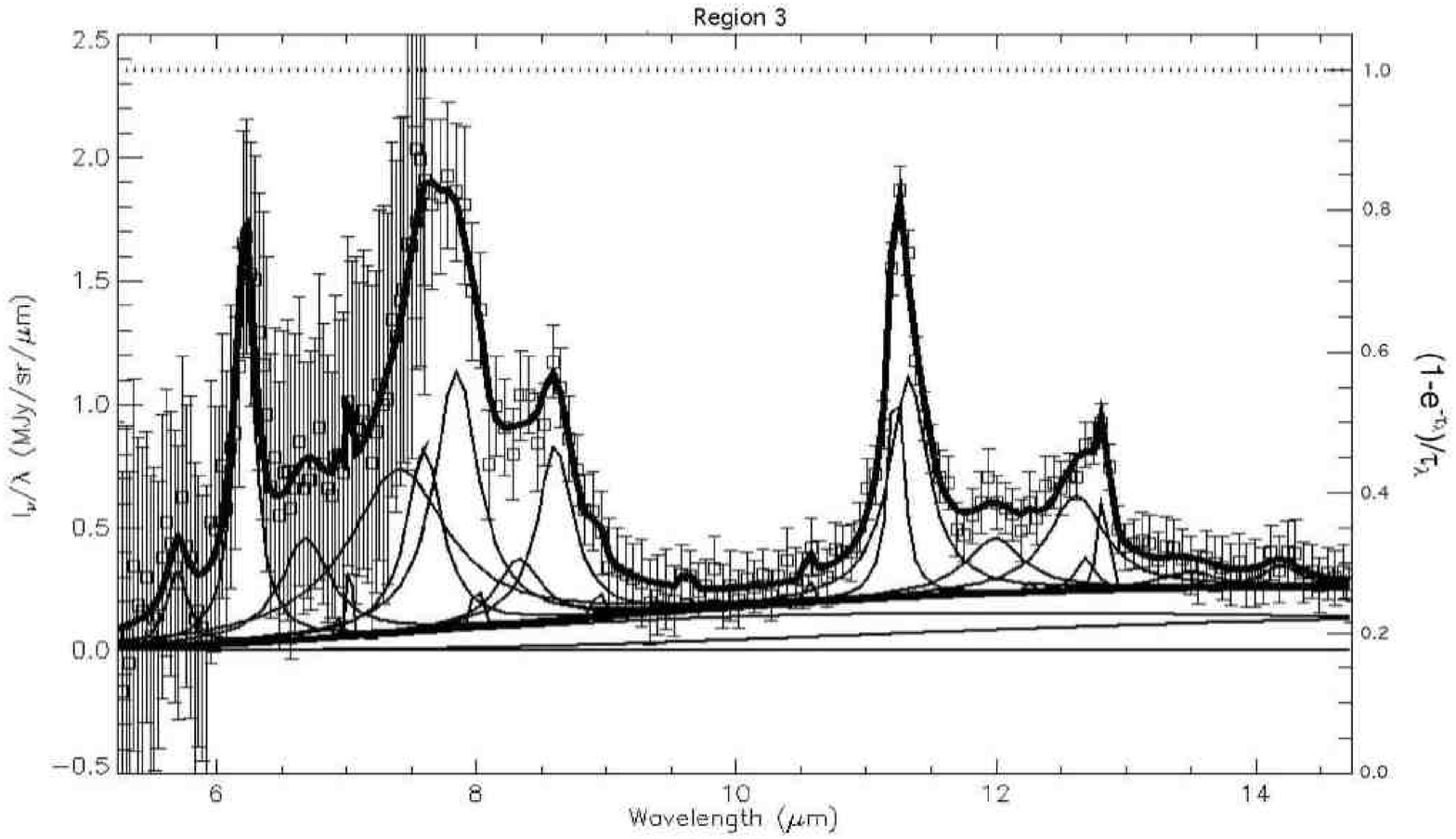}{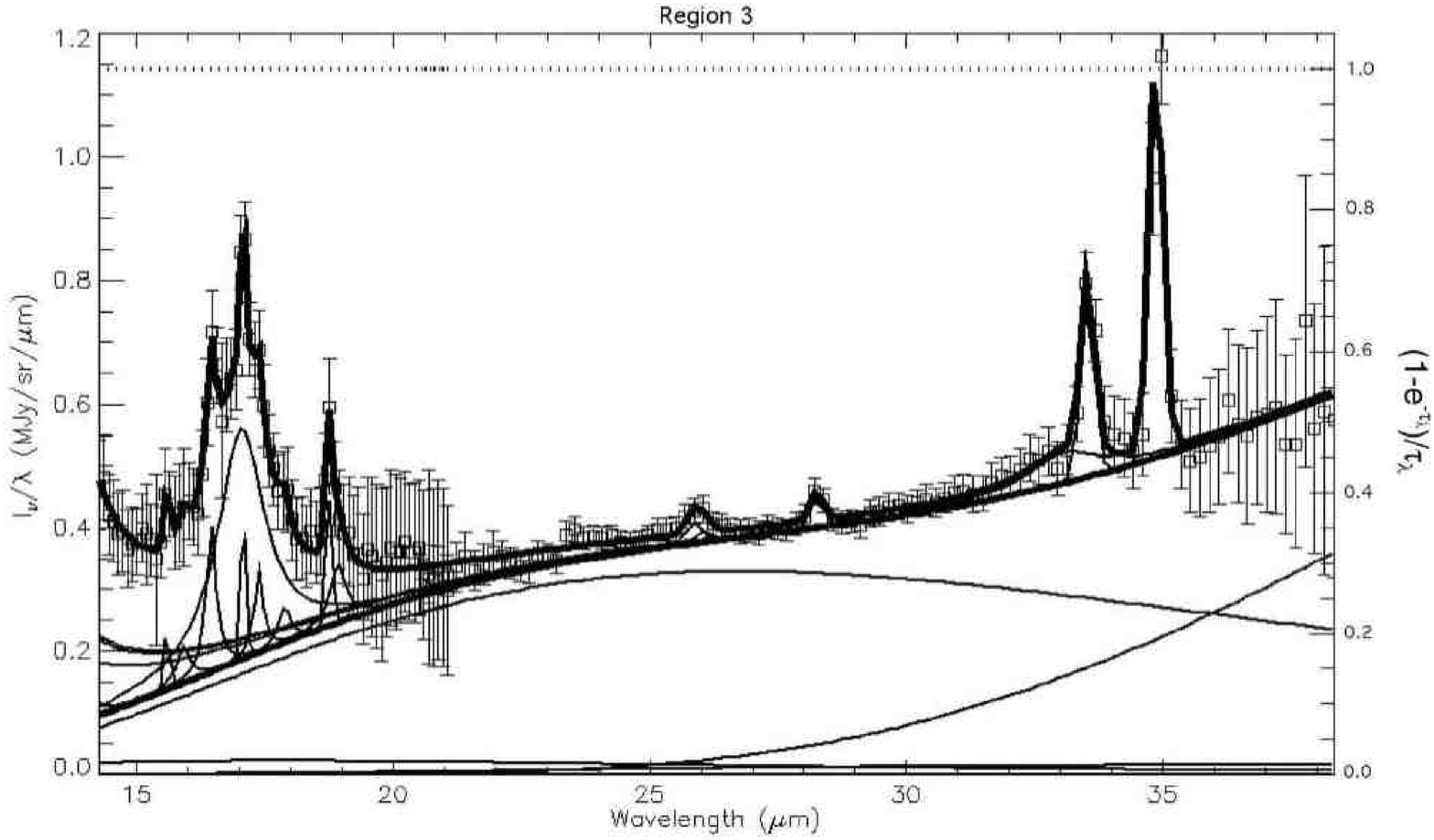}
\caption{Shown are sample $Spitzer$ IRS low resolution 
spectra and the PAHFIT spectral decomposition for 3 different 
single pixel regions across the M51 strip.  The SL and LL spectra 
are plotted separately because we decomposed the SL and LL 
data cubes separately.  The locations of the regions are 
marked on the maps of the H$_2$ surface density in Figure 
\ref{figure-5}.}
\label{figure-2}
\end{figure}

Figure \ref{figure-2} presents sample $Spitzer$ IRS low resolution 
spectra and the PAHFIT spectral decomposition for 3 different 
single pixel regions across the M51 strip.  The SL and LL spectra 
are plotted separately because we decomposed the SL and LL 
data cubes separately.  The locations of the regions are 
marked on the maps of the H$_2$ surface density in Figure \ref{figure-5}.  
While we have decomposed the entire spectrum and created maps of all of 
PAH features and spectral lines, in this paper we focus primarily 
on the H$_2$ line maps.

\subsection{Ancillary Data: CO (J = 1 -- 0), Optical $V$ Band, H$\alpha$, and X-ray Observations}

In this section we briefly discuss the ancillary data that we 
have used in order to understand H$_2$ excitation in M51.  
We take note of the image resolutions for comparison to the
$Spitzer$ IRS beam.  The Berkely Illinois Maryland Array (BIMA) CO (J = 1 -- 0) map was
acquired as part of the BIMA Survey of Nearby Galaxies (SONG)
\citep{reg01, hel03}.  At the distance of M51, the SONG beam
(5$\farcs$8 $\times$ 5$\farcs$1) subtends 220 pc $\times$ 190 pc.  
The optical $V$ band and continuum-subtracted H$\alpha$ + [N II] images of M51 were obtained from the $Spitzer$ $Infrared$ $Nearby$ $Galaxies$ $Survey$ (SINGS) archive.\footnote{http://data.spitzer.caltech.edu/popular/sings/} 
The native pixel scale for both images is 0$\farcs$3 
and the angular resolutions are $\sim$ 1$\arcsec$.  X-ray 
emission from M51 was observed by the Advanced CCD 
Imaging Spectrometer (ACIS) on the $Chandra$ $X$-$Ray$ 
$Observatory$ on 20 June 2000.  The resolution of 
the image is $\sim$ 1$\arcsec$.  The X-ray image that 
we use has been presented and discussed in 
Terashima and Wilson (2001).

\section{Results}

\subsection{The Distribution of H$_2$ Emission}

We have detected and mapped H$_2$ emission from the six
lowest pure rotational H$_2$ lines (Figure \ref{figure-3}).  The
maps reveal remarkable differences in the distribution of
H$_2$ emission in M51.  H$_2$ S(0) emission is
strongest in the northwest inner spiral arm peaking at a flux of 3.7
$\times$ $\mathrm{10^{-18}}$ W $\mathrm{m^{-2}}$ and decreases by a
factor of 2 in the nuclear region.  In contrast, the H$_2$
S(1) emission peaks in the nucleus of the galaxy at
1.0 $\times$ $\mathrm{10^{-17}}$ W $\mathrm{m^{-2}}$ and has an
extension of equal brightness towards the northwest inner spiral arm.  In the
spiral arm itself, the H$_2$ S(0) peak is offset from the
H$_2$ S(1) emission by 10$\arcsec$ ($\sim$ 380 pc).  
We detect emission from the H$_2$ S(0) and 
H$_2$ S(1) lines to the outer limit of our radial 
strip,  $\sim$ 6 kpc from the nucleus of the galaxy.  In the outer spiral arms,
the H$_2$ S(0) flux is a factor of 2 times lower than in
the inner northwest spiral arm and the H$_2$ S(1) flux
is a factor of 5 times lower than in the nucleus.
 
The $\mathrm{ H_2}$ S(2) $-$ H$_2$ S(5) maps show different
molecular gas distributions within \objectname{M51} through each
H$_2$ line.  The brightest H$_2$ S(2) emission is
from the nucleus at 2.2 $\times$ $\mathrm{10^{-18}}$ W
$\mathrm{m^{-2}}$.  We also see bright H$_2$ S(2) emission
from the inner northwest spiral arm at half the flux of the
nuclear peak.  The H$_2$ S(3) peak at the nucleus is 1.4
$\times$ $\mathrm{10^{-17}}$ W $\mathrm{m^{-2}}$, a factor of $\sim$ 6
greater than the H$_2$ S(2) nuclear peak.  There is also a
linear bar-like structure in H$_2$ S(3) emission across the
nucleus of the galaxy at a PA$\sim$-10$^0$ and length of $\sim$ 1.5 kpc. 
The emission peaks in the
H$_2$ S(2) and H$_2$ S(3) maps are not spatially
coincident.  For instance, in the northwest inner spiral arm the brightest
H$_2$ S(3) emission is further down the spiral arm than 
the brightest H$_2$ S(2) emission.  Offsets like
these suggest variations in the excitation temperature from 
region to region within a galaxy, and even within a spiral arm.

The H$_2$ S(4) and H$_2$ S(5) lines are brightest at
the nucleus with fluxes of 3.1 and 8.0 $\times$
$\mathrm{10^{-18}}$ W $\mathrm{m^{-2}}$, respectively.  The H$_2$
S(4) line shows emission in the nucleus and in the spiral arm to the
west.  In the spiral arm to the west of the nucleus, the
H$_2$ S(4) flux is 2.1 $\times$ $\mathrm{10^{-18}}$ W
$\mathrm{m^{-2}}$.  This is notable because the spiral arm to the
southwest of the nucleus is very bright in CO and and studies have
revealed very high molecular gas column densities in the southwest
inner spiral arm \citep{lor90, aal99}.  H$_2$ S(5) emission is
asymmetric in the nucleus and mimics the morphology of the
H$_2$ S(3) line with extended emission to the north of the
nucleus.  The trail of intensity peaks along the long northeast edge
of the H$_2$ S(4) and H$_2$ S(5) maps is not real.  It is likely due to 
difficulty smoothing the maps near the edges.  These regions do 
not effect the determination of the H$_2$ temperature and surface 
density due to the offset of the SL strip relative to the LL strip.

Based on the uncertainty maps generated while creating the H$_2$ line maps, 
errors in the H$_2$ S(0) line flux that range from 15\% to 75\% with the 
largest uncertainties being found in the inter-arm regions and the outer 
northwest and southeast spiral arms.  Errors in the H$_2$ S(1) line 
fluxes range from 10\% to 70\% with the largest uncertainties being 
found at the distant northwest and southeast portions of the M51 strip.  
Errors in the H$_2$ S(2) and H$_2$ S(3) line fluxes range from 10\% 
to 85\% with the largest uncertainties being found toward the inter-arm 
regions.  Errors in the H$_2$ S(4) and H$_2$ S(5) line fluxes are 
$\sim$ 30\% and 20\% within a 1 kpc radius of the nucleus and become higher 
moving radially away. 

\subsection{Mapping H$_2$ Excitation Temperature and Surface Density across M51}

\subsubsection{Modeling H$_2$ Excitation Temperature and Surface Density}

The pure rotational lines of molecular hydrogen provide a powerful
probe of the conditions of the ISM by placing constraints on the
energy injection that excites H$_2$.  For example, 
Neufeld et al. (2006) discuss shock excitation of H$_2$ and 
Kaufman et al. (2006) discuss H$_2$ excitation in photodissociation regions (PDRs).  
Using the maps of H$_2$ emission, we modeled the H$_2$ 
temperature and mass distribution over the radial strip in M51 
following the methods described in Rigopoulou et al. (2002) 
and Higdon et al. (2006).

First, we smoothed the H$_2$ S(1) $-$ H$_2$ S(5)
maps to the resolution of the H$_2$ S(0) map, 10$\farcs$1.
The maps were then interpolated to the same spatial grid.  Excitation
diagrams across the strip were derived from the Boltzmann equation
\begin{equation}
N_i/N = (g(i)/Z(\mathrm{T_{ex}}))exp(-T_i/\mathrm{T_{ex}})
\end{equation}
where $g(i)$ is the statistical weight of state $i$,
Z($\mathrm{T_{ex}}$) is the partition function, $\mathrm{T_i}$ is the
energy level of a given state, and $\mathrm{T_{ex}}$ is the excitation
temperature.  N and $\mathrm{N_i}$ are the total column density and
the column density of a given state $i$ and $\mathrm{N_i}$ is
determined directly from the measured extinction-corrected flux by
\begin{equation}
N_i = 4 \pi \times flux(i)/(\Omega A(i)h\nu (i))
\end{equation}
where A($i$) is the Einstein $A$-coefficient, $\nu$($i$) is the
frequency of state $i$, $\Omega$ is the solid angle of the beam, and
$h$ is Planck's constant.  Table \ref{tbl1} lists the values for the
wavelength, rotational state, Einstein $A$-coefficient, energy, and
statistical weight of the pure rotational levels of H$_2$.

In order to derive temperature and surface density 
distributions we assume a two temperature model for the 
H$_2$.  To determine the hot (T = 400 -- 1000 K) 
phase temperature, we do a least squares fit 
to the H$_2$ S(2) -- H$_2$ S(5) column densities in the 
excitation diagram at every pixel in our maps.   
We then subtract the contribution of the hot phase from the lower $J$ 
lines and do a least squares fit to the column densities of 
H$_2$ S(0) -- H$_2$ S(2) lines at every pixel in our 
maps to determine the temperature distribution 
of the warm (T = 100 -- 300 K) phase.  The warm and hot 
phase surface density distributions are derived from the 
column densities of the H$_2$ S(0) and H$_2$ S(3) lines, 
respectively.  The column densities are easily converted 
to mass surface density by determining the mass of the 
warm and hot H$_2$ within every pixel. 

\begin{figure}
\figurenum{4}
\plottwo{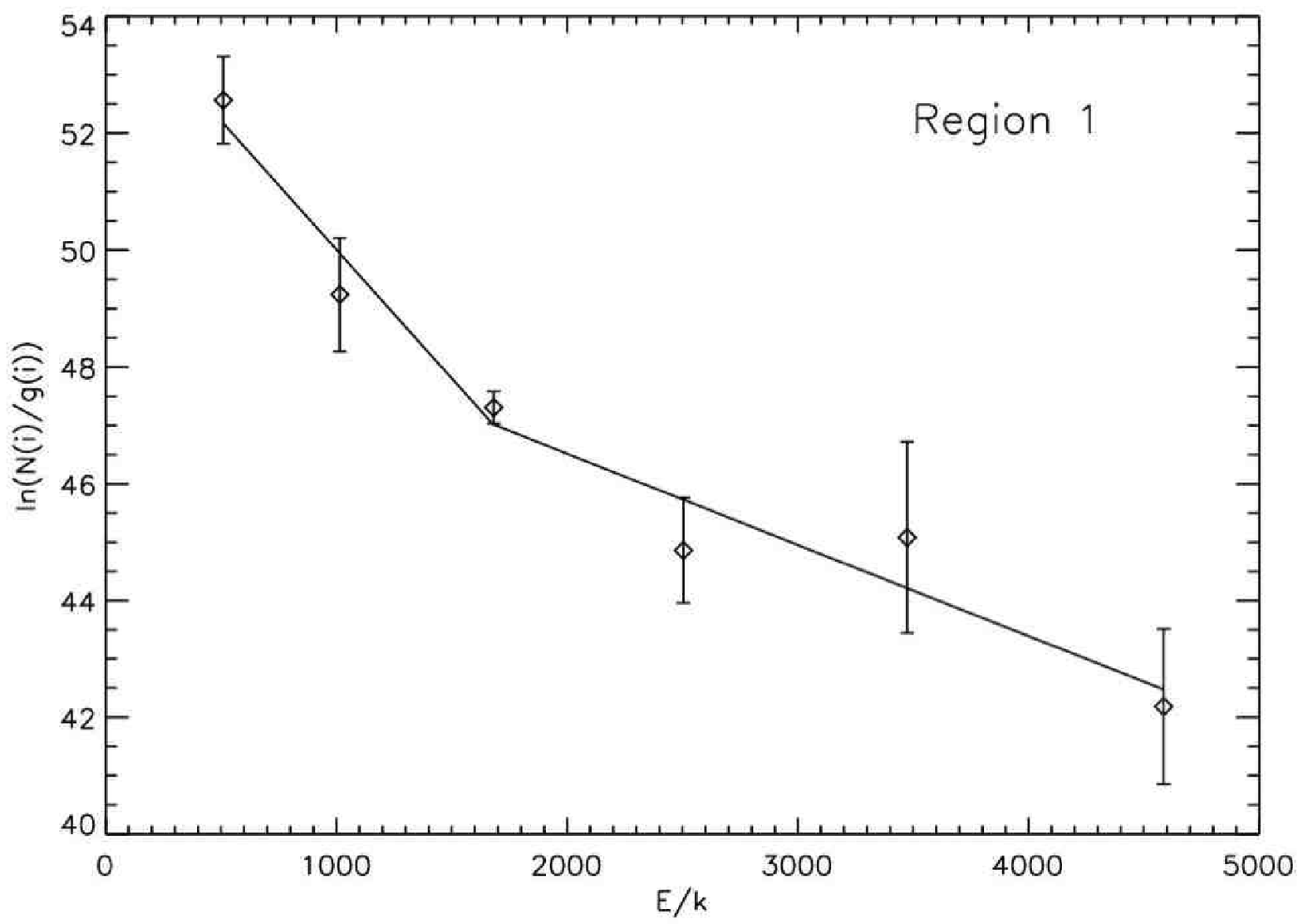}{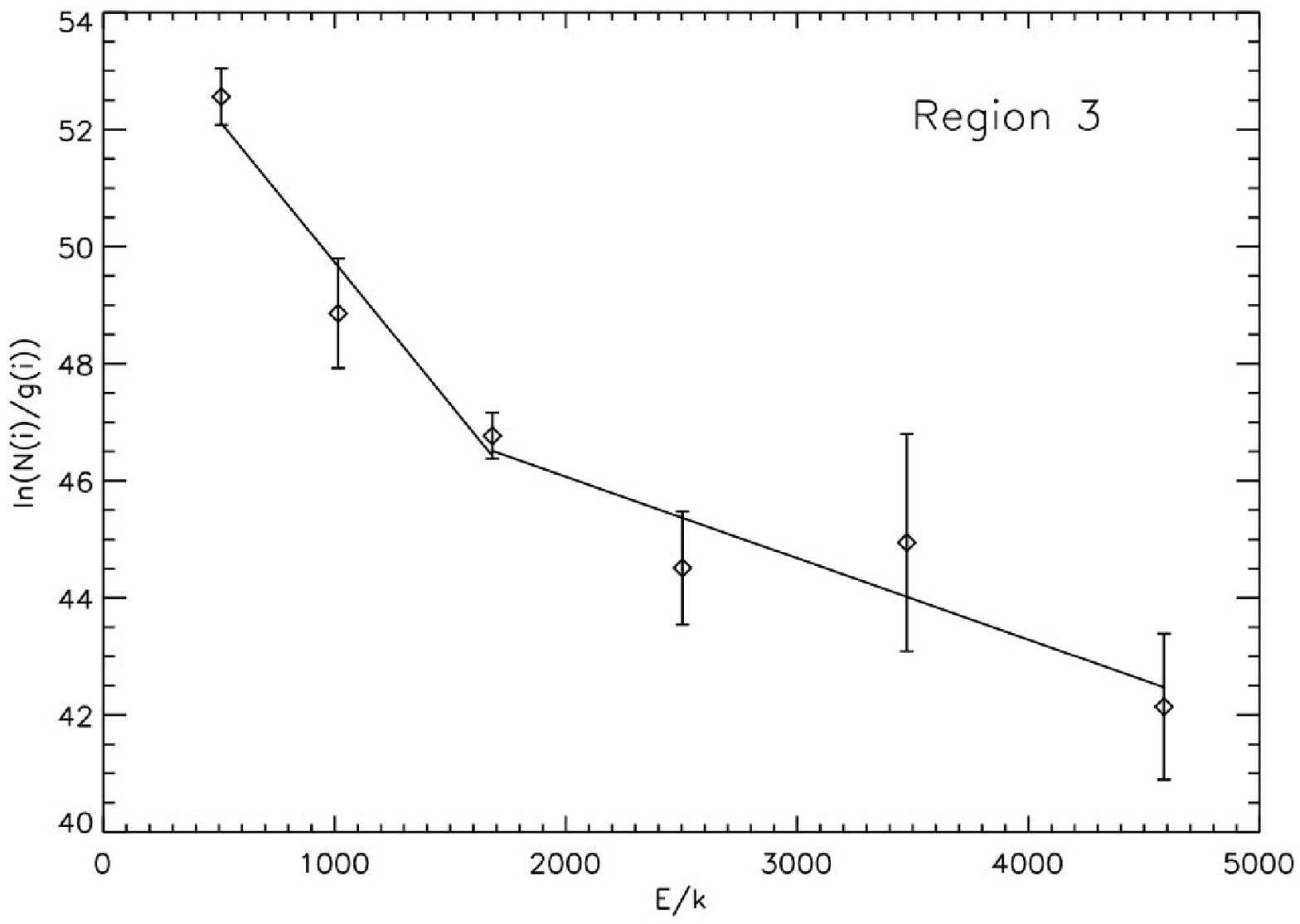}
\epsscale{.5}
\plotone{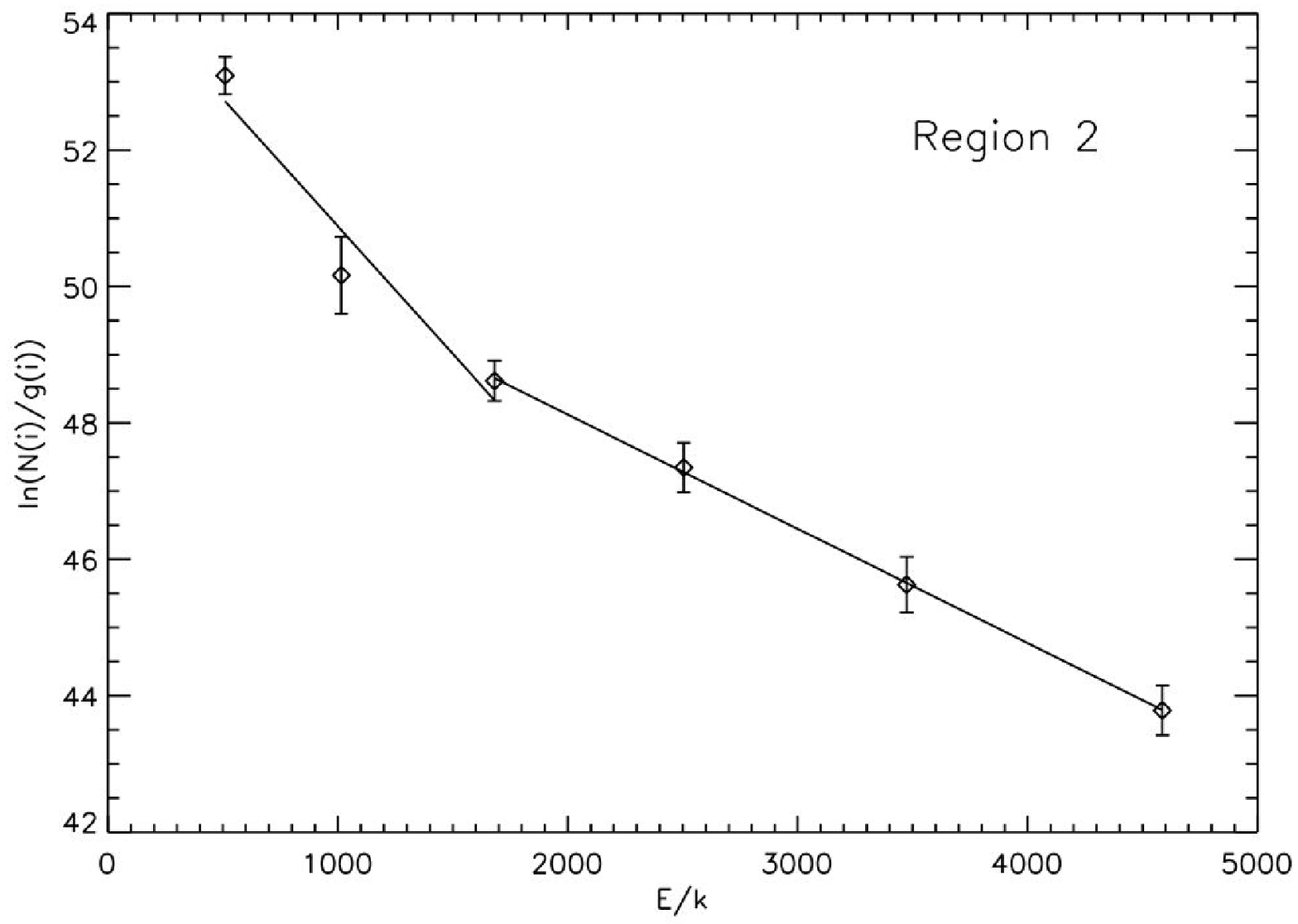}
\caption{Shown are excitation diagrams and the fits to the warm
and hot H$_2$ phases taken from three different single pixel regions along the M51 strip. 
The different regions are marked on the H$_2$ surface density maps 
in Figure \ref{figure-5}.}
\label{figure-4}
\end{figure}

\begin{figure}
\epsscale{1.1}
\figurenum{5}
\plottwo{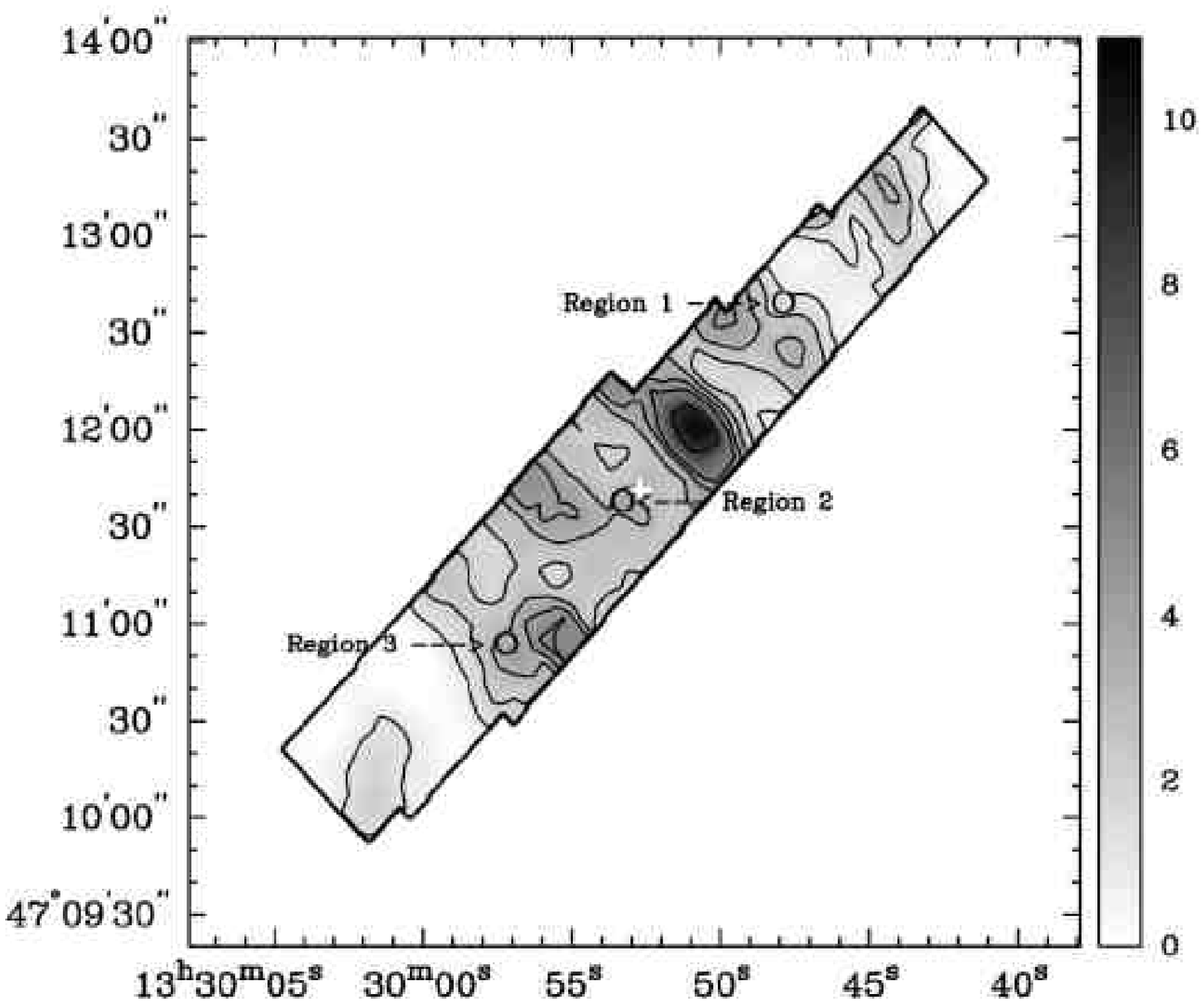}{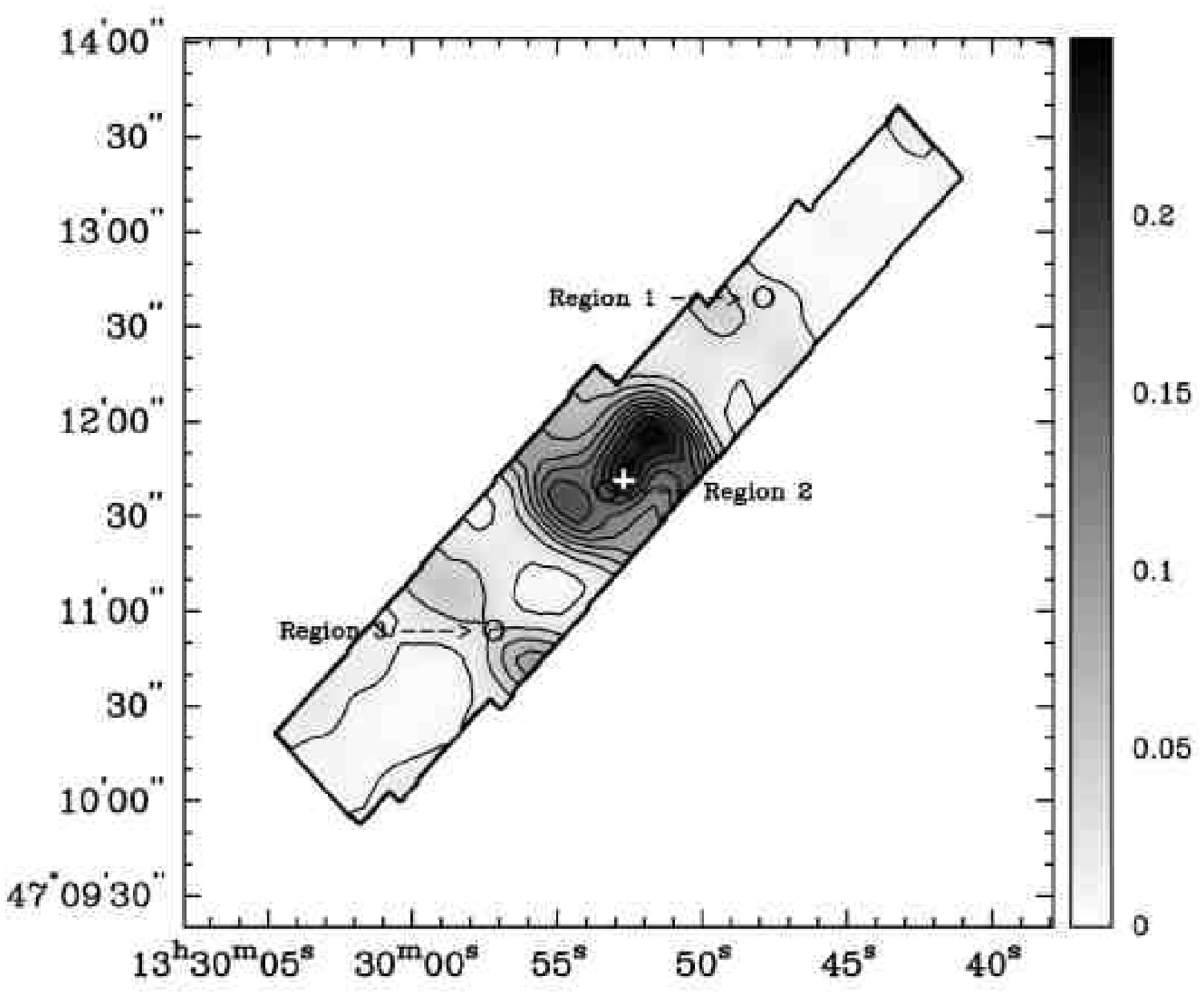}
\caption{Shown are the warm (T = 100 $-$ 300 K) 
H$_2$ ($left$) and hot (T = 400 $-$ 1000 K) 
H$_2$ ($right$) surface density distributions.  The surface density
 distributions are in units of $\mathrm{M_\sun}$ 
$\mathrm{pc^{-2}}$.   Contours are overplotted for clarity.  
The warm H$_2$ surface density contour levels are at 
1.10, 2.21, 3.32, 4.43, 5.55, and 8.85 $\mathrm{M_\sun}$ $\mathrm{pc^{-2}}$.  
The hot H$_2$ contour 
levels are at intervals of 10 \% of 0.25 $\mathrm{M_\sun}$ $\mathrm{pc^{-2}}$.  
The hot H$_2$ surface density distribution is derived 
from the fit to the H$_2$ S(2) $-$ 
H$_2$ S(5) lines and the warm H$_2$ 
surface density distribution is derived from the fit to the 
H$_2$ S(0) $-$ H$_2$ S(2) lines, 
corrected for the contribution of the hot H$_2$ phase.
The three circles denote the regions of the spectra and excitation diagrams in Figures 
\ref{figure-2} and \ref{figure-4}, respectively.}
\label{figure-5}
\end{figure}

Figure \ref{figure-4} shows excitation diagrams and the 
fits to the warm and hot H$_2$ phases for three
different regions across the M51 strip.  The three 
regions are marked on the H$_2$ surface density 
maps in Figure \ref{figure-5}.  In the nuclear region 
(Region 2) the ortho and para levels appear to lie along the 
same curve indicating an ortho-to-para ratio (OPR) of 3; 
the excitation diagrams do not exhibit the ``zigzag" characteristic of a 
non-equlibrium H$_2$ OPR \citep{neu98, fue99}.  The 
slope of the curve appears to decrease as the rotational 
state increases indicating that we are sampling a continuous 
range of H$_2$ temperatures with the higher rotational 
states (S(2) -- S(5)) probing a hotter phase of H$_2$ 
than the lower rotational states (S(0) -- S(2)).  In the 
southeast and northwest spiral arms (Regions 1 and 
3, respectively) the lower $J$ (H$_2$ S(0) - 
H$_2$ S(3)) levels exhibit an OPR of $\sim$ 3.  The 
H$_2$ S(4) measurement shows significant 
scatter in the excitation diagrams outside of the nuclear 
region of M51.  This would indicate that the OPR is 
less than 3, however, due to the low signal-to-noise 
ratio of the H$_2$ S(4) map, we do not 
believe that the OPR determined from the H$_2$ 
S(4) flux reflects the OPR of the warm H$_2$. 
We assume an OPR of 3 when deriving the H$_2$ 
temperature and mass distributions, which is consistent 
with Roussel et al. (2007) who found an OPR of 3  for the 
nuclear region and four of seven HII regions studied in M51.

\subsubsection{Warm and Hot H$_2$ Excitation Temperature and Surface Density Distributions}

Figure \ref{figure-5} presents the warm ($left$) and hot ($right$)
H$_2$ surface density distributions across the M51 
strip.\footnote{Note that the non-rectangular shape of the strip 
is due to the offset between the SL and LL strips.}  The highest 
gas surface density for the warm H$_2$ phase is in 
the inner northwest spiral arm at 11 $\mathrm{M_\sun}$ 
$\mathrm{pc^{-2}}$. The gas surface density in the outer 
northwest and southeast spiral arms is maximum at the 
center of the spiral arms at 3.5 $\mathrm{M_\sun}$ 
$\mathrm{pc^{-2}}$ and 1.0 $\mathrm{M_\sun}$ $\mathrm{pc^{-2}}$ 
respectively.  The hot phase surface density is highest in 
the nucleus and interior to the inner spiral arm at 0.24 
$\mathrm{M_\sun}$ $\mathrm{pc^{-2}}$.  The gas surface 
density of the hot phase in the spiral arms is 3 -- 5 times 
lower than that of the nuclear region.

\begin{figure}
\figurenum{6}
\epsscale{.55}
\plotone{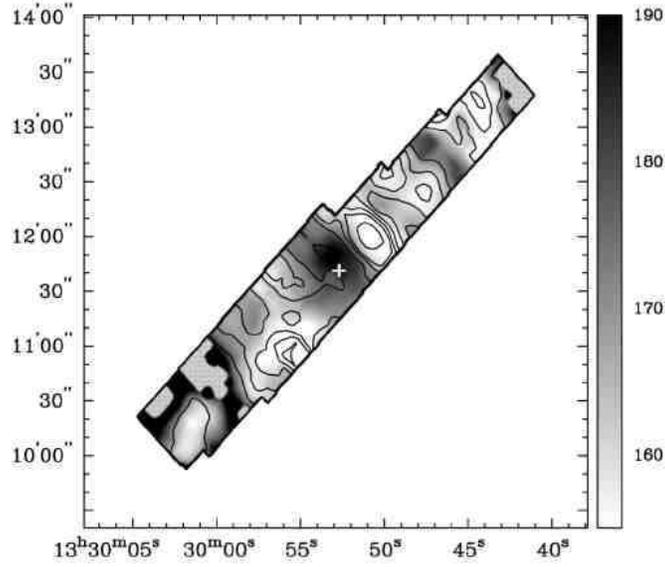}
\caption{The warm (T = 100 $-$ 300 K) H$_2$ surface 
density distribution (in contours) compared 
to the warm H$_2$ temperature distribution 
(in grayscale, in units Kelvin).  The warm H$_2$ 
temperature and surface density distributions are derived from the fit to the excitation 
diagrams across the strip for the H$_2$ S(0) $-$ H$_2$ S(2) lines, 
corrected for the contribution of the hot (T = 400 $-$ 1000 K) H$_2$ phase.  
Surface density contour levels are at 
1.10, 2.21, 3.32, 4.43, 5.55, and 8.85 $\mathrm{M_\sun}$ $\mathrm{pc^{-2}}$ 
(same as in Figure \ref{figure-5}). The non-rectangular shape of the 
map is due to the slight offset of the $Spitzer$ IRS SL 
strip relative to the LL strip.  In both the northwest and southeast corners 
of the maps, the temperature and surface density distributions could not 
be determined due to no detection of the H$_2$ S(0) line.}
\label{figure-6}
\end{figure}

\begin{figure}
\figurenum{7}
\epsscale{.55}
\plotone{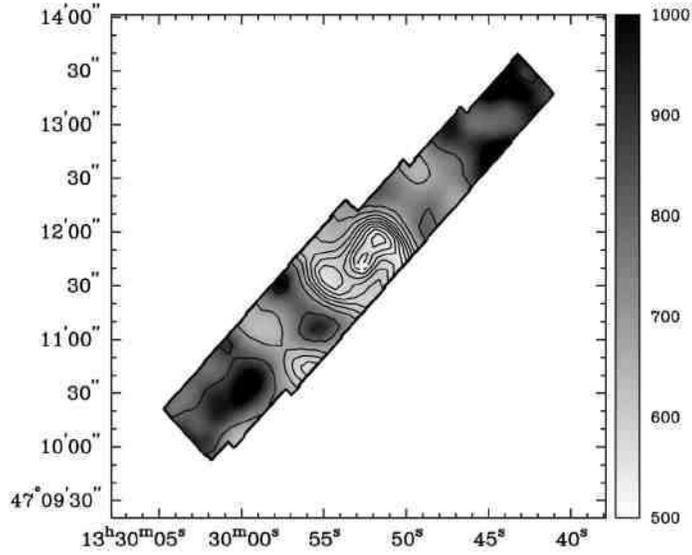}
\caption{The hot (T = 400 $-$ 1000 K) H$_2$ surface density distribution (in contours) 
compared to the hot H$_2$ temperature distribution (in grayscale, in units Kelvin).  
The hot H$_2$ temperature and surface density distributions are derived from the 
fit to the excitation diagrams across the strip for the H$_2$ S(2) $-$ H$_2$ S(5) lines.  Surface density 
contour levels are at intervals of 10\% of 0.25 $\mathrm{M_\sun}$ $\mathrm{pc^{-2}}$ (same as in Figure \ref{figure-6}).  The high temperature and surface density of the hot phase towards the northwest and southeast ends of the strips is due to the lack of detection of extended emission from the H$_2$ S(4) and H$_2$ S(5) lines.  In these regions (beyond the northwest and southeast spiral arms) we are unable to accurately determine the hot phase temperature and surface density.}
\label{figure-7}
\end{figure}

Figures \ref{figure-6} and \ref{figure-7} show the distribution 
of the warm and hot H$_2$ temperatures (in grayscale) with 
the contours of the warm and hot H$_2$ surface densities overlaid, respectively.  In 
both cases we see that the temperature and surface density 
are inversely correlated with the hottest temperatures 
corresponding to regions of lowest surface density.  
For both the warm and hot phases, we see the temperature 
is higher in the inter-arm regions than in the spiral arms. 
This is real and does not result from lower signal-to-noise 
in the inter-arm regions.   We tested this by extracting spectra 
over 0.76 kpc$^2$ inter-arm regions between the northwest inner arm and northwest 
arm and between the southeast inner arm and southeast arm and found warm H$_2$ 
temperatures of 177 and 175 K and hot H$_2$ temperatures 
of 691 and 690K, respectively.  These temperatures are higher 
than the 150 - 165 K warm H$_2$ temperatures and the 550 - 650 
K hot H$_2$ temperatures found in the spiral arms.

\begin{figure}
\epsscale{.55}
\figurenum{8}
\plotone{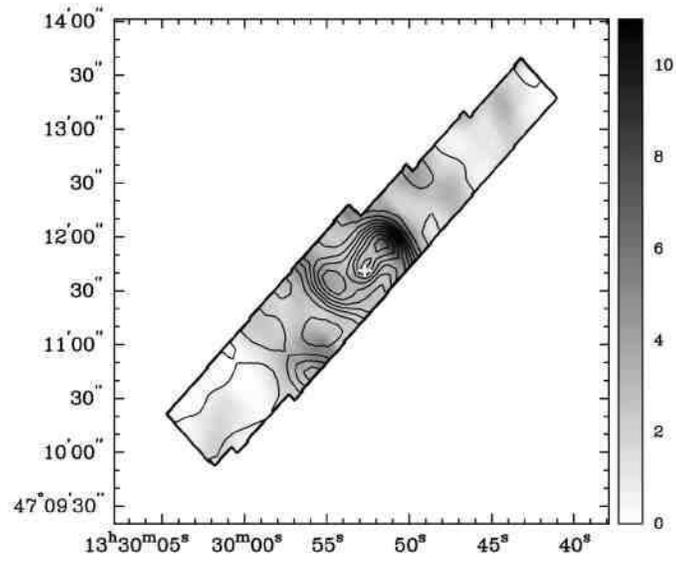}
\caption{The warm (T = 100 $-$ 300 K) H$_2$ surface density (in grayscale) compared to
 the hot (T = 400 $-$ 1000 K) H$_2$ surface density (in contours).  Contours levels for the 
 hot H$_2$ surface density distribution are at intervals of 10\% of the maximum surface density 
 (0.25 $\mathrm{M_\sun}$ $\mathrm{pc^{-2}}$).  The grayscale is in units of 
 $\mathrm{M_\sun}$ $\mathrm{pc^{-2}}$.}
\label{figure-8}
\end{figure}

Figure \ref{figure-8} compares the warm (in grayscale) 
and hot (in contours) H$_2$ surface density distributions.   
The warm H$_2$ mass distribution peaks
in the northwest inner spiral arm and the hot H$_2$ 
mass distribution peaks in the nucleus and in the 
region interior to the northwest spiral arm.  The warm-to-hot 
H$_2$ mass ratio is not constant across 
the galaxy but is lowest ($\sim$ 12) in the nucleus of the 
galaxy and increases to 170 and 136 in the southeast 
and northwest spiral arms, respectively.

\section{Discussion}

\subsection{Effects of Beam Averaging on Studies of Extragalactic H$_2$}

Previous studies have used aperture-averages over entire galactic
nuclei to derive the physical conditions of the molecular gas
\citep{rig02, hig06, rou07}.  In M51, Roussel et al. (2007) find that within
the central 330 $\mathrm{arcsec^2}$ (4.61 $\times$ 10$^5$ pc$^2$),
the warm H$_2$ phase has a temperature of 180 K and 
a surface density of 3.2 $\mathrm{M_\sun}$ $\mathrm{pc^{-2}}$ 
(a total mass of $\mathrm{M_{warm}}$ = 1.5 $\times$ $\mathrm{10^6}$
$\mathrm{M_\sun}$).  Roussel et al. (2007) have also measured the temperature of
the hot phase (though they do not measure the mass in the hot phase)
and find a hot H$_2$ temperature of 521 K.

Having spatially resolved spectra over a strip across M51, we can 
investigate the behavior of the warm and hot H$_2$ phases 
on smaller scales.  In the nuclear region of M51 we see that the 
warm phase temperature peaks at 192 K and decreases 
radially towards the inner spiral arms.  The warm H$_2$ surface 
density at the nucleus is 4.4 $\mathrm{M_\sun}$ $\mathrm{pc^{-2}}$ 
and decreases over a 0.5 kpc radius surrounding the 
nucleus.  To check the consistency of 
our results against those of Roussel et al. (2007), we averaged the 
warm phase temperature over a similar 412 
$\mathrm{arcsec^2}$ (5.76 $\times$ 10$^2$ pc$^2$) aperture and
found that the warm phase temperature and surface density are
186 K and 2.8 $\mathrm{M_\sun}$ $\mathrm{pc^{-2}}$, 
respectively.  Our results are consistent with previous studies of 
warm H$_2$ done at lower resolution and suggest that 
previous studies of the warm H$_2$ temperature and 
mass have yielded average values rather than a maximum (or minimum) values.

\subsection{Distinguishing the H$_2$ Excitation Mechanisms}

The warm-to-hot H$_2$ surface density ratio varies across 
M51 suggesting that the H$_2$ excitation mechanisms 
have different effects on the warm and hot phases. 
The largest warm H$_2$ surface densities are found in 
the spiral arms suggesting that the warm phase is 
associated with star formation activity.  The largest 
hot H$_2$ surface densities are found in the nuclear 
region of M51.  This suggests that where nuclear 
activity is the dominant excitation mechanism, a 
larger fraction of the H$_2$ can be maintained in 
the hot phase and that H$_2$ is generally more 
efficiently heated by nuclear activity.  This is further supported 
by the fact that the hot H$_2$ surface density is $\sim$ 660 K 
in the nuclear region and decreases to $\sim$ 530 K and 
550 K in the inner northwest and southeast spiral 
arms, respectively.  Roussel et al. (2007) 
find that H$_2$ is generally heated by massive stars in 
PDRs, however, Seyferts and LINERs show evidence for 
the dominance of other excitation mechanisms such as 
X-rays or shocks.  By comparing the spatial distribution 
of the warm and hot H$_2$ to optical $V$ band imagery, H$\alpha$, and CO 
(J = 1 -- 0) emission, we can investigate the relationship 
between star formation and H$_2$.  Additionally, by 
comparing the warm and hot H$_2$ distributions to X-ray 
observations and the spatial distribution of [O IV](25.89 $\micron$) 
emission, we can investigate X-ray and shock heating 
of H$_2$ within M51.

\begin{figure}
\epsscale{1.1}
\figurenum{9}
\plottwo{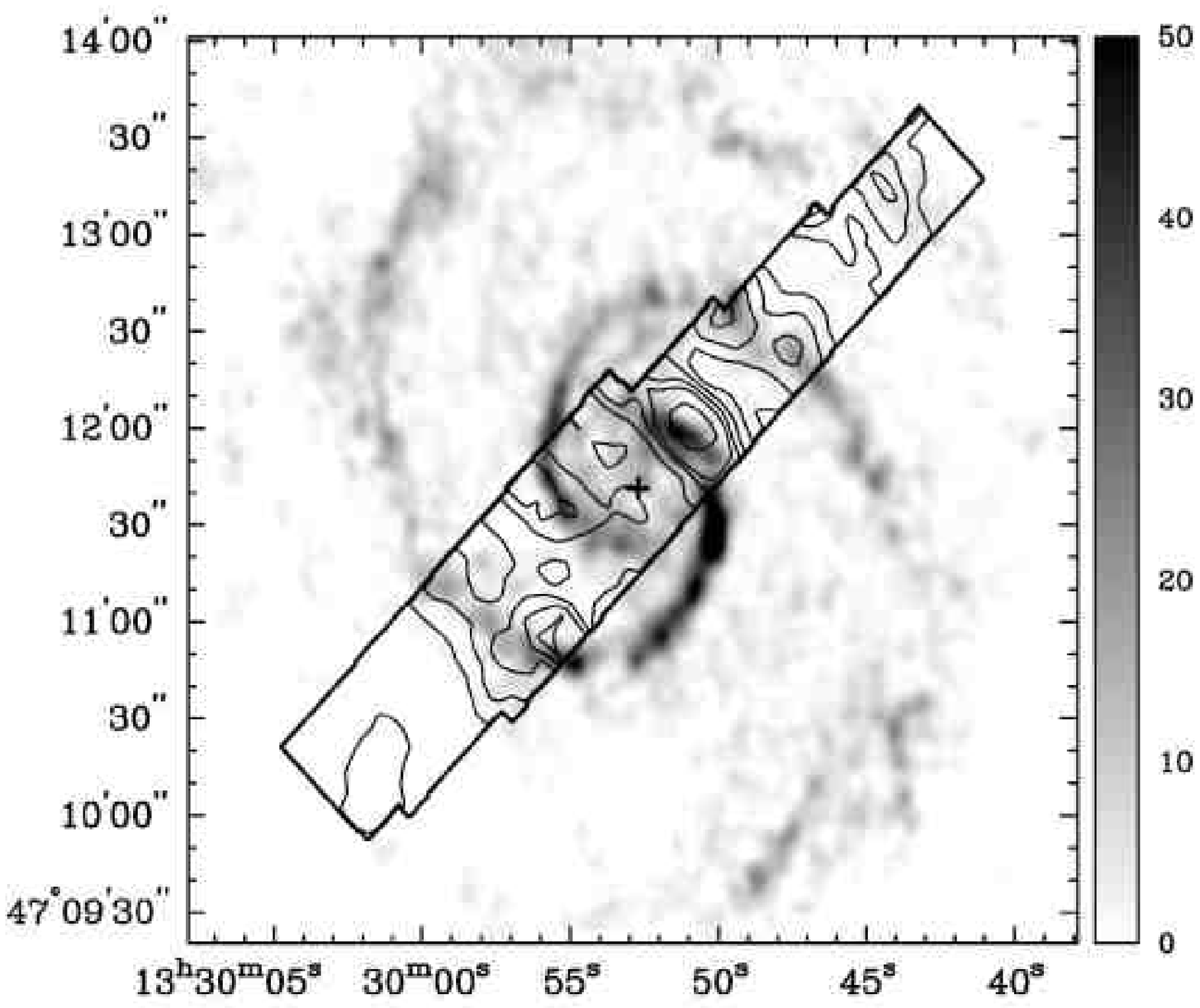}{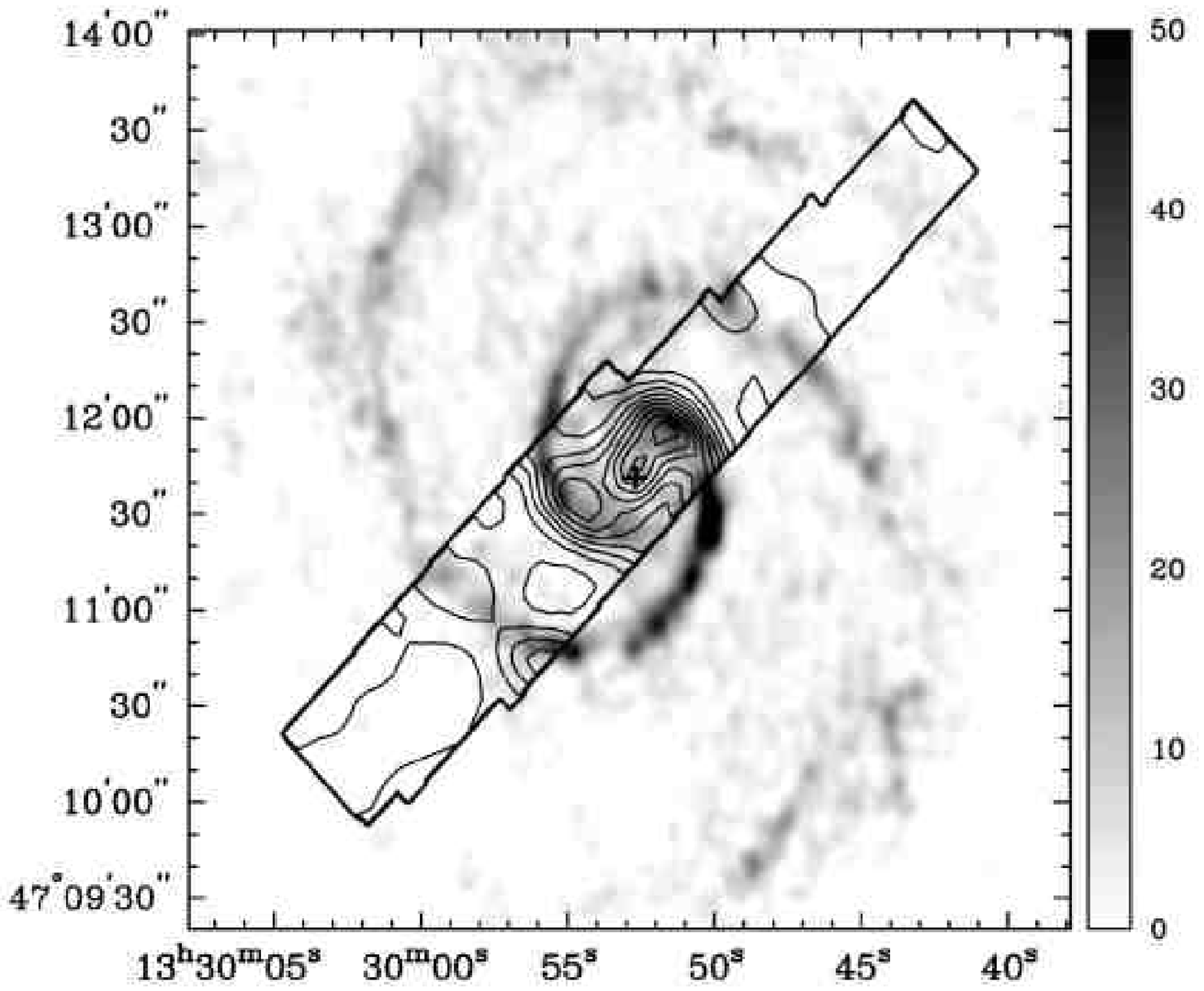}
\caption{$Left$: Comparison of  CO intensity (in grayscale) to the warm (T = 100 $-$ 300 K) 
H$_2$ surface density (in contours).  The CO intensity is in units of Jy km $\mathrm{s^{-1}}$. 
The warm H$_2$ surface density contours are the same as in Figures \ref{figure-5} and \ref{figure-6}.  
$Right$: Comparison of CO intensity (in grayscale) to the hot (T = 400 $-$ 1000 K) 
H$_2$ surface density (in contours).  The CO intensity is in units of Jy km 
$\mathrm{s^{-1}}$. The hot H$_2$ surface density contours are the same as in 
Figures \ref{figure-5} and \ref{figure-7}.}
\label{figure-9}
\end{figure}

Kaufman et al. (2006) show that within galaxies, where the
telescope beam size is generally kiloparsecs across, H$_2$
emission could serve to probe the average physical 
conditions in the surfaces of molecular clouds.  In Figure 
\ref{figure-9}, we compare the warm ($left$) and hot
H$_2$ ($right$) mass distributions to the cold molecular 
gas traced by CO (J = 1 $-$ 0) emission.  The warm and hot 
H$_2$ phases appear to trace the bright CO emission
in the northwest and southeast spiral arms. Comparison 
of CO to the individual H$_2$  S(0) $-$ H$_2$ S(3) line intensity 
maps in Figure \ref{figure-10} also shows that the H$_2$ in 
the spiral arms traces the bright CO emission.  
In the spiral arms, both the warm and hot H$_2$ 
phases are found in PDRs and are associated 
with star formation activity.  The most 
striking result is that in the inner spiral arms, we see that the
CO is offset toward the nucleus from the 
warm H$_2$ mass (in Figure \ref{figure-9}).
The offset between the peaks in CO and warm H$_2$ mass is
7$\arcsec$ in the northwest inner spiral arm and 5$\arcsec$ in the
southeast inner spiral arm.  We believe that these offsets are real with
one possible explanation being that the H$_2$ is tracing the
regions of active star-formation within the giant molecular
associations.  

\begin{figure}
\epsscale{1.1}
\figurenum{11}
\plottwo{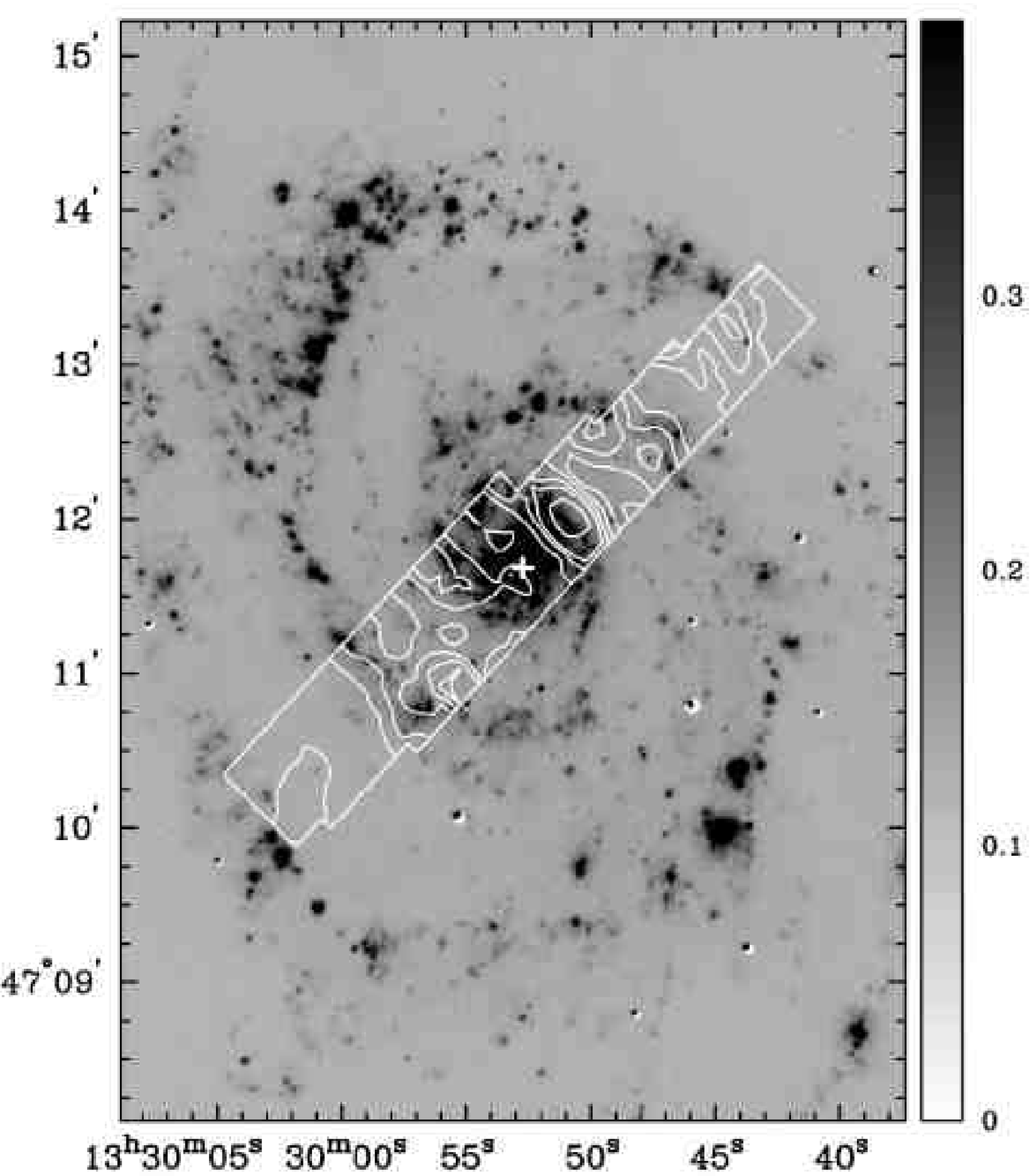}{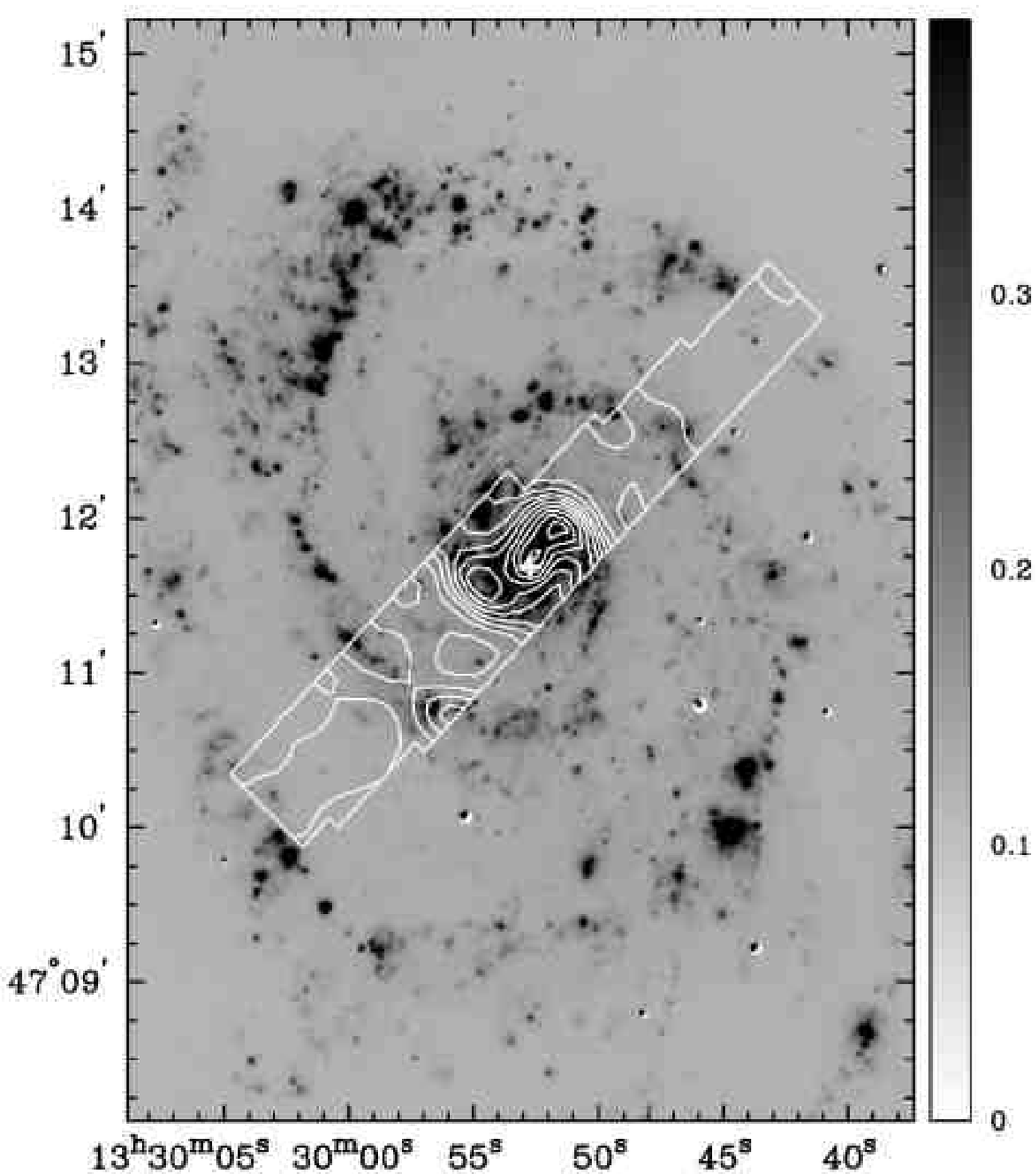}
\caption{$Left$: Comparison of  H$\alpha$ (in grayscale) to the warm (T = 100 $-$ 300 K) 
H$_2$ surface density (in contours).  The H$\alpha$ image is in units of counts 
$\mathrm{s^{-1}}$. The warm H$_2$ surface density contours are the same as in 
Figures \ref{figure-5} and \ref{figure-6}.  $Right$: Comparison of H$\alpha$ (in grayscale) 
to the hot (T = 400 $-$ 1000 K) H$_2$ surface density (in contours).  The H$\alpha$ 
image is in units of counts $\mathrm{s^{-1}}$. The hot H$_2$ surface density 
contours are the same as in Figures \ref{figure-5} and \ref{figure-7}.}
\label{figure-11}
\end{figure}

\begin{figure}
\epsscale{1.1}
\figurenum{12}
\plottwo{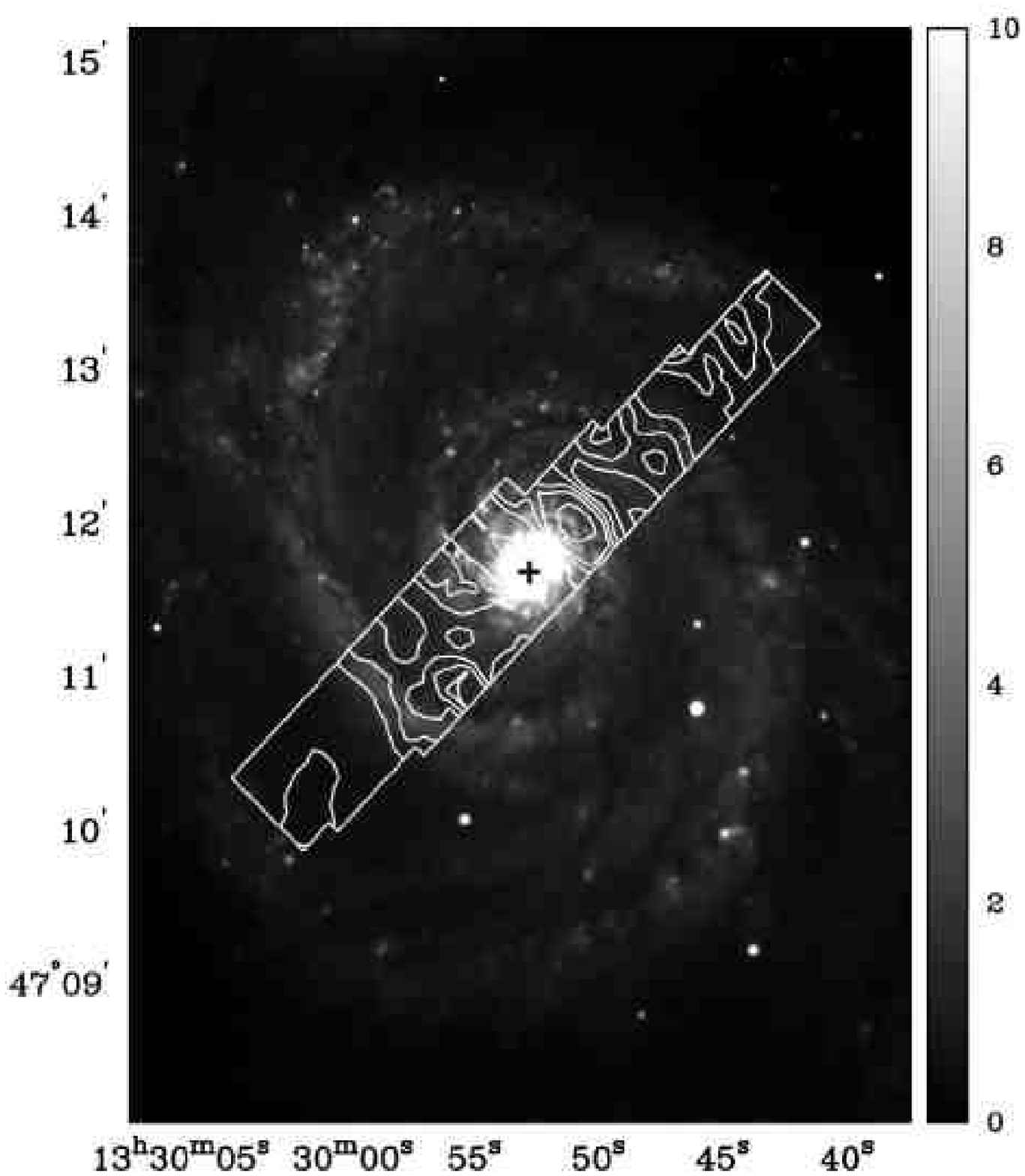}{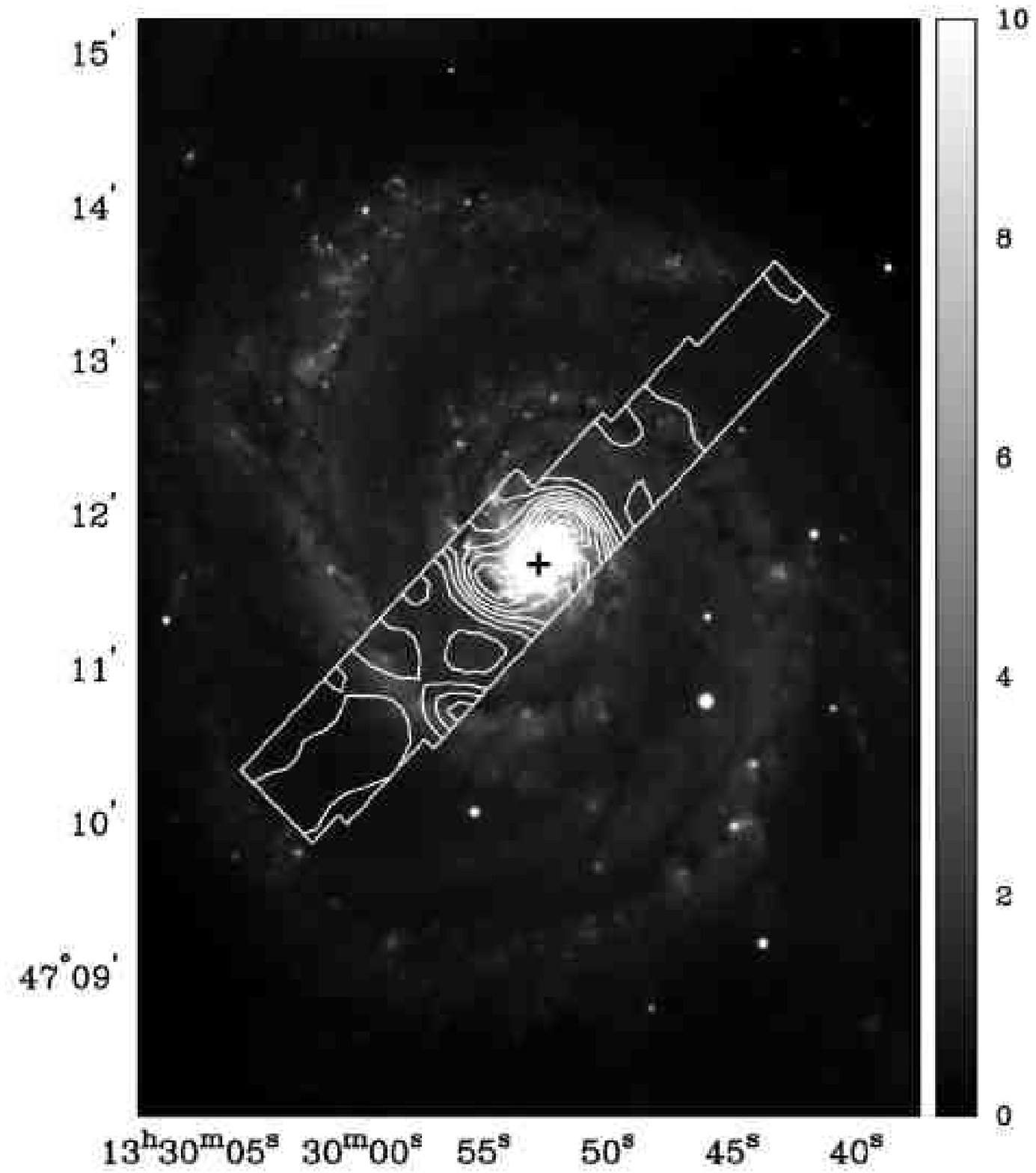}
\caption{$Left$: Comparison of  the $V$ band image (in grayscale) to the warm (T = 100 $-$ 300 K) 
H$_2$ surface density (in contours).  The $V$ band shows the dust lanes within M51.
The $V$ band image is in units of counts $\mathrm{s^{-1}}$. 
The warm H$_2$ surface density contours are the same as in 
Figures \ref{figure-5} and \ref{figure-6}.  $Right$: Comparison of $V$ band image (in grayscale) 
to the hot (T = 400 $-$ 1000 K) H$_2$ surface density (in contours).  The $V$ band 
image is in units of counts $\mathrm{s^{-1}}$. The hot H$_2$ surface density 
contours are the same as in Figures \ref{figure-5} and \ref{figure-7}.}
\label{figure-12}
\end{figure}

In Figures \ref{figure-11} and \ref{figure-12}, we compare the warm ($left$) and hot ($right$)
H$_2$ mass distributions to H$\alpha$ emission and an optical $V$ 
band image, respectively .  The H$\alpha$ image reveals the 
brightest HII regions and the $V$ band image shows 
the dust lanes in M51.  In general,
the warm and hot H$_2$ concentrations are not cospatial with
the brightest H$\alpha$ emission regions in the spiral arms with the one
exception being that the warm H$_2$ mass in 
the northwest and southeast inner spiral
arms appears to trace the H$\alpha$ emission.  The warm H$_2$
mass contours show that local peaks in H$_2$ mass are found
within the dust lanes, with the H$\alpha$ emission spatially 
offset towards the leading edge of the spiral arms.  
An example of this is in the northwest spiral
arms where we see the H$_2$ mass offset from the H$\alpha$
spiral arms with local peaks being found in the dust lanes.
In Figures \ref{figure-13} and \ref{figure-14}, we compare the H$_2$ S(0) $-$
H$_2$ S(3) line intensity maps to H$\alpha$ emission and $V$ band imagery, respectively.
Comparison of the H$_2$ S(0) map to H$\alpha$ reveals that
the strongest H$_2$ emission in the northwest and southeast
inner spiral arms is coincident with H$\alpha$ emission; however, the
other H$_2$ S(0) spiral arms show the strongest emission in
the dust lanes, offset from the H$\alpha$ spiral arms.  The largest
offsets are seen in the southeast spiral arm where the H$_2$
S(0) emission is offset from the H$\alpha$ spiral arm by $\sim$
15$\arcsec$ (560 pc).  H$_2$ S(1) emission appears to follow
the dust lanes and the H$_2$ S(1) intensity subsides into the
H$\alpha$ spiral arms.  H$_2$ S(2) and H$_2$ S(3)
emission is also found in the dust lanes; however, there are instances
(such as in the southeast spiral arm) where the H$_2$
emission appears to be found straddling the dust lane and H$\alpha$
spiral arm.

\begin{figure}
\epsscale{1.1}
\figurenum{15}
\plottwo{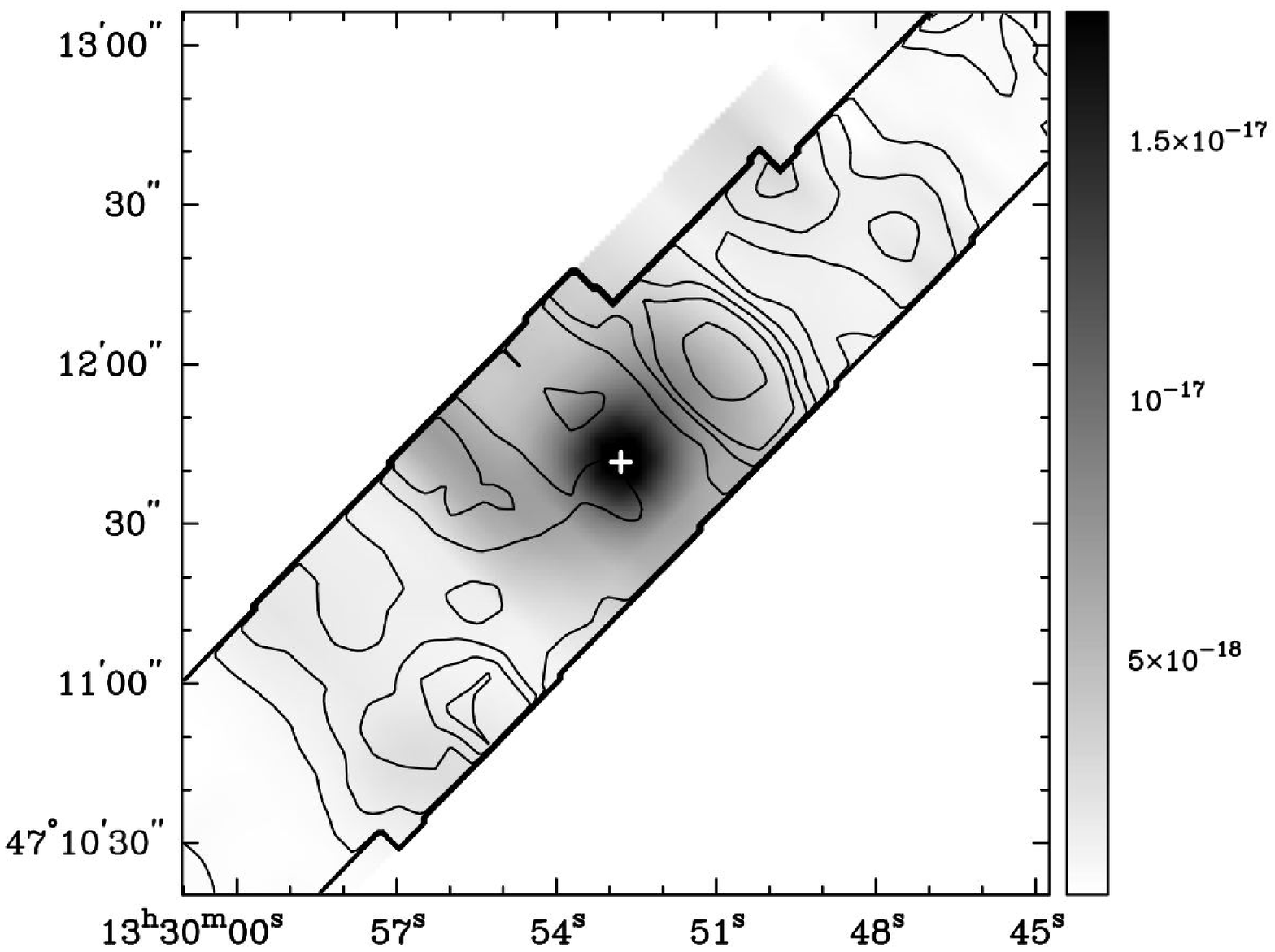}{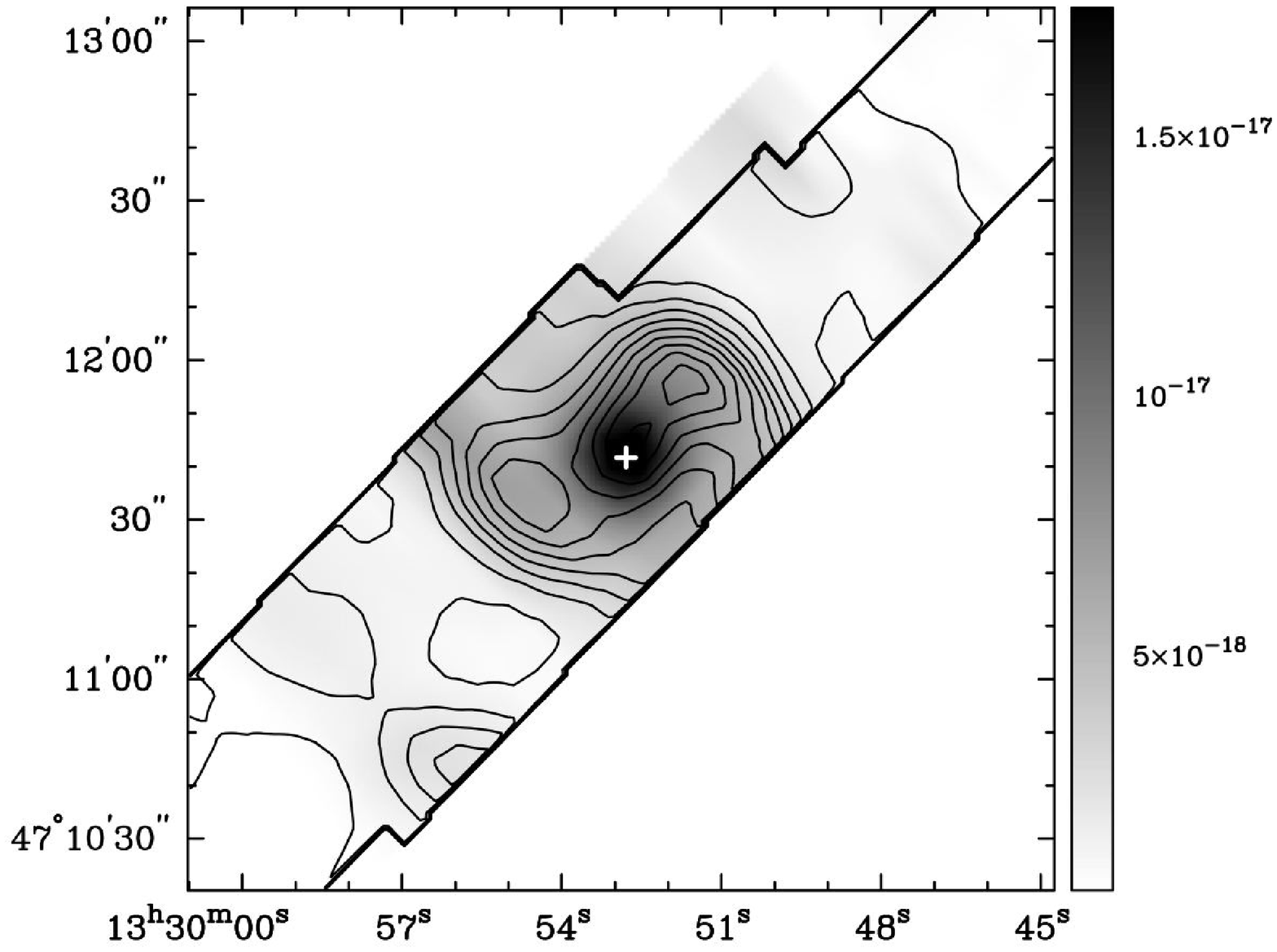}
\caption{$Left$:  Comparison of the [O IV](25.89 \micron) emission (in grayscale) to 
the warm (T = 100 K - 300 K) H$_2$ surface density distribution (in contours).  
Hot H$_2$ surface density contours are the same as in Figures \ref{figure-5} and \ref{figure-6}.  
$Right$: Comparison of the [O IV](25.89 \micron) emission (in grayscale) to the hot 
(T = 400 - 1000 K) H$_2$ surface density distribution (in contours).  Hot H$_2$ 
surface density contours are the same as in Figures \ref{figure-5} and \ref{figure-7}.  The [O IV](25.89 \micron)
 emission is in units of W $\mathrm{m^{-2}}$.}
\label{figure-15}
\end{figure}

The [O IV](25.89 \micron) line can be excited in fast shocks 
(v$_s$ $>$ 100 km s$^{-1}$) \citep{lutz98},
by photoionization in Wolf-Rayet stars \citep{ss99}, or by an
AGN \citep{smi04}.  We have mapped the blended 
[O IV](25.89 \micron) + [Fe II](25.99 \micron) line across M51.  
While the map contains the contribution of both lines, we have examined the SINGS 
high resolution spectra and find that averaged over a 
365 arcsec$^2$ region around the nucleus of M51, the [O IV] line flux is 2.7 times 
higher than the [Fe II] flux.\footnote{SINGS data cubes and spectra for M51 can be found at 
http://irsa.ipac.caltech.edu/data/SPITZER/SINGS/galaxies/ngc5194.html} 
In the high resolution data cubes of the nuclear region, 
the [O IV] line flux dominates the [O IV] + [Fe II] blend and the [O IV] 
emission distribution derived from the low resolution spectra accurately reflects 
the distribution seen in the high resolution spectra.
In Figure \ref{figure-15}, we compare 
the [O IV] surface brightness to the warm ($left$) and hot ($right$) H$_2$ distributions.  
The [O IV] emission is brightest in the nuclear region at 1.2 $\times$
$\mathrm{10^{-17}}$ W $\mathrm{m^{-2}}$ and within a 0.5 kpc radius of the nucleus, the peak is coincident with the peak in the mass of the hot H$_2$.  [O IV]
surface brightness decreases from the nucleus to the inner spiral arm by 50 \%.
We resolve weaker [O IV] emission within the warm and hot
H$_2$ spiral arms.  The [O IV] surface brightness
is a factor of $\sim$ 6 lower in the spiral arms than the peak
intensity found in the nucleus.  The [O IV] emission within the nuclear 
region of M51 is likely due to the weak Seyfert 2 nucleus \citep{ford85} 
and is possibly associated with shocked gas from the outflows of the AGN.  
The peak of the [O IV] emission coincides with the nuclear peak in hot 
H$_2$ mass, indicating that the hot H$_2$ phase in the 
nuclear region of the galaxy is AGN or shock heated. 
In the nuclear region we observe a factor of 12 times greater
warm H$_2$ mass than the hot H$_2$ mass.  
The warm H$_2$ mass is much
greater within the spiral arms than within the nucleus and the
warm-to-hot mass ratio is lowest in the nuclear region where 
the [O IV] flux is greatest.  In general, nuclear activity and shocks 
heat the ISM more efficiently causing the higher surface 
density of hot H$_2$.

\begin{figure}
\epsscale{1.1}
\figurenum{16}
\plottwo{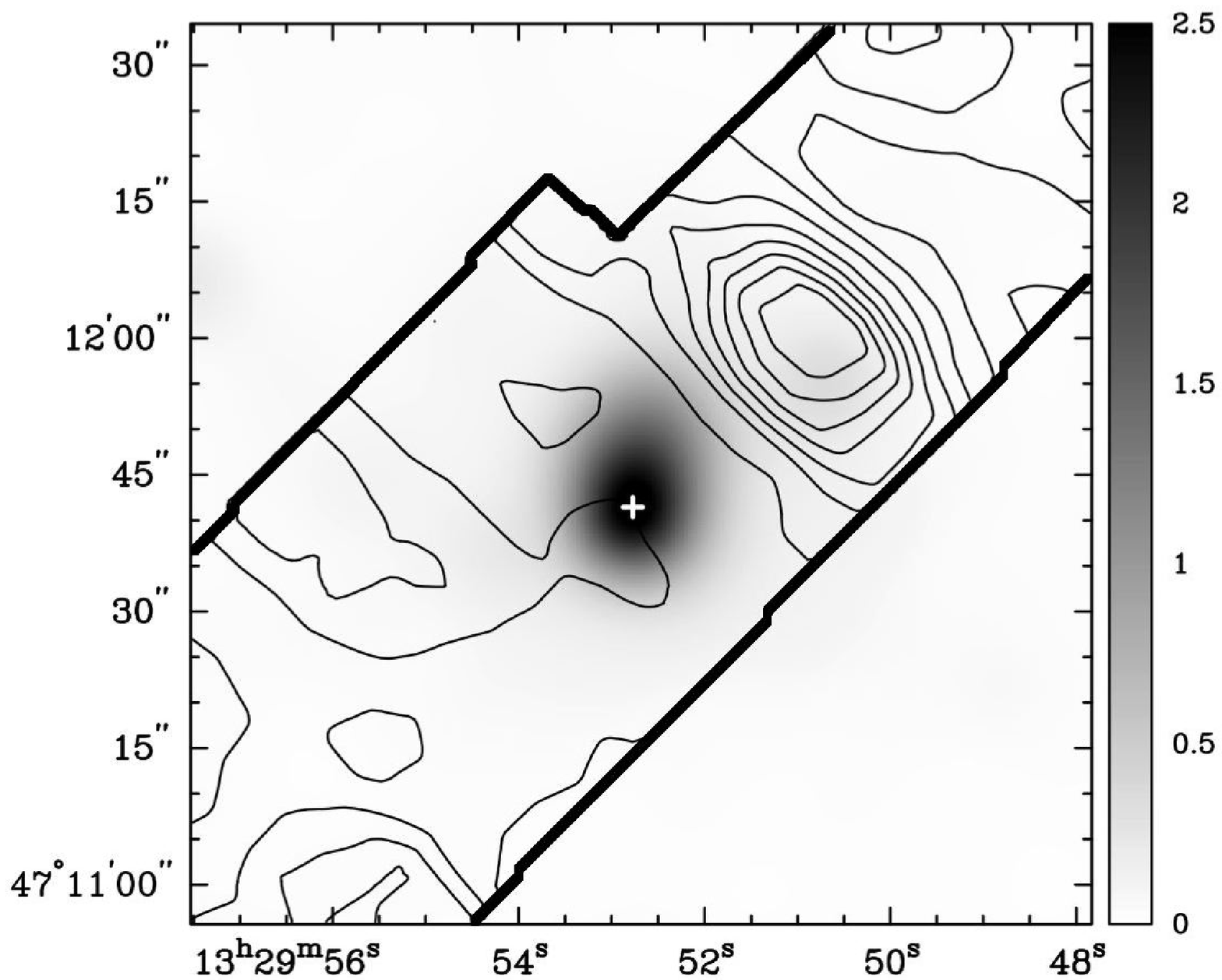}{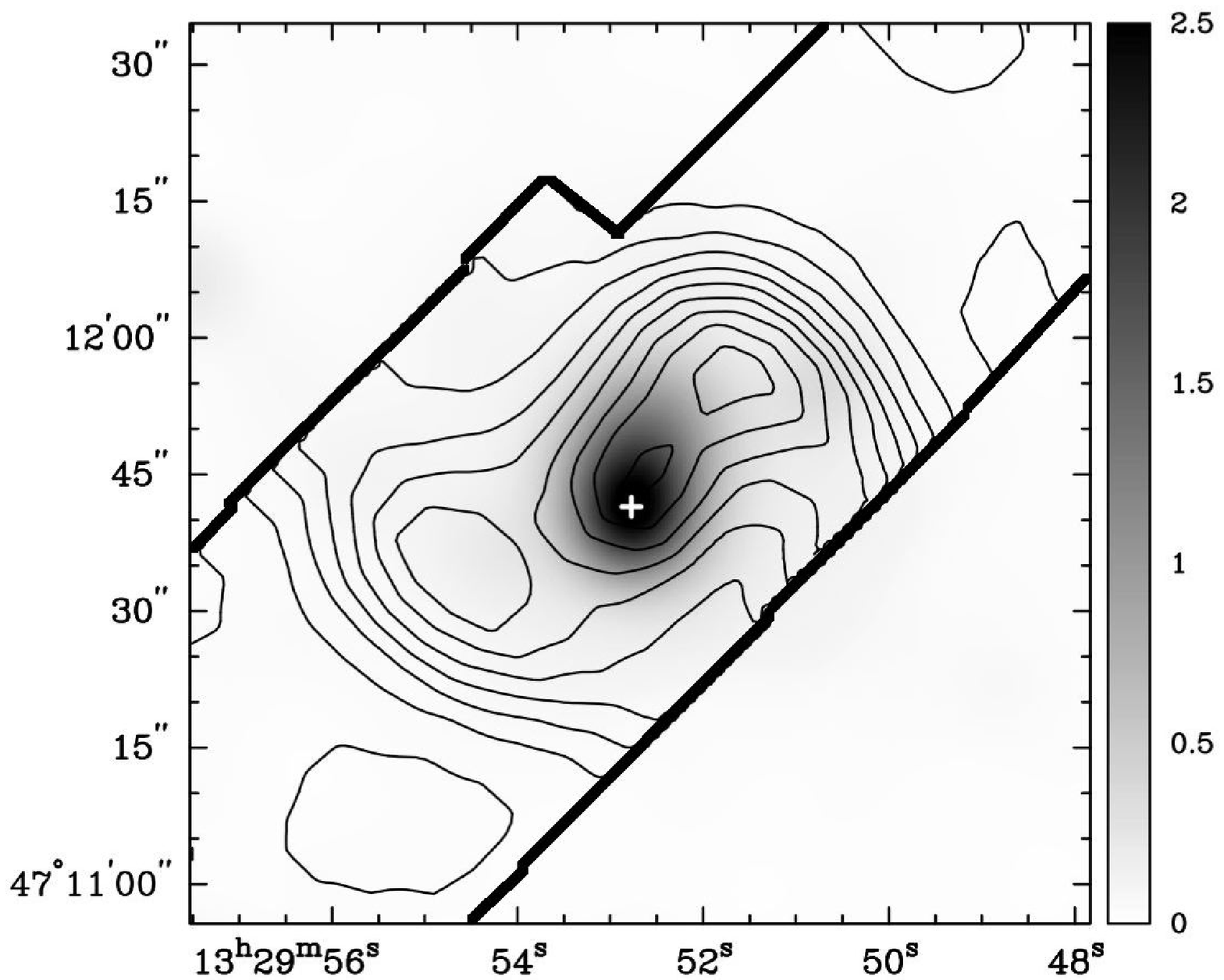}
\caption{$Left$:  Comparison of the smoothed 0.5 $-$ 10 keV X-ray emission band 
(in grayscale) to the warm (T = 100 $-$ 300 K) H$_2$ surface density distribution 
(in contours).   The X-ray image has been smoothed to the same resolution as the 
warm H$_2$ surface density map.  X-ray emission is in units of counts.  The warm H$_2$ 
surface density contours are the same as in Figures \ref{figure-5} and \ref{figure-6}.  $Right$: 
Comparison of the smoothed 0.5 $-$ 10 keV X-ray emission band (in grayscale) to the hot 
(T = 400 $-$ 1000 K) H$_2$ surface density distribution (in contours).  The hot
H$_2$ surface density distribution contours are the same as in Figures \ref{figure-5} 
and \ref{figure-7}.}
\label{figure-16}
\end{figure}

\begin{figure}
\epsscale{1.1}
\figurenum{17}
\plottwo{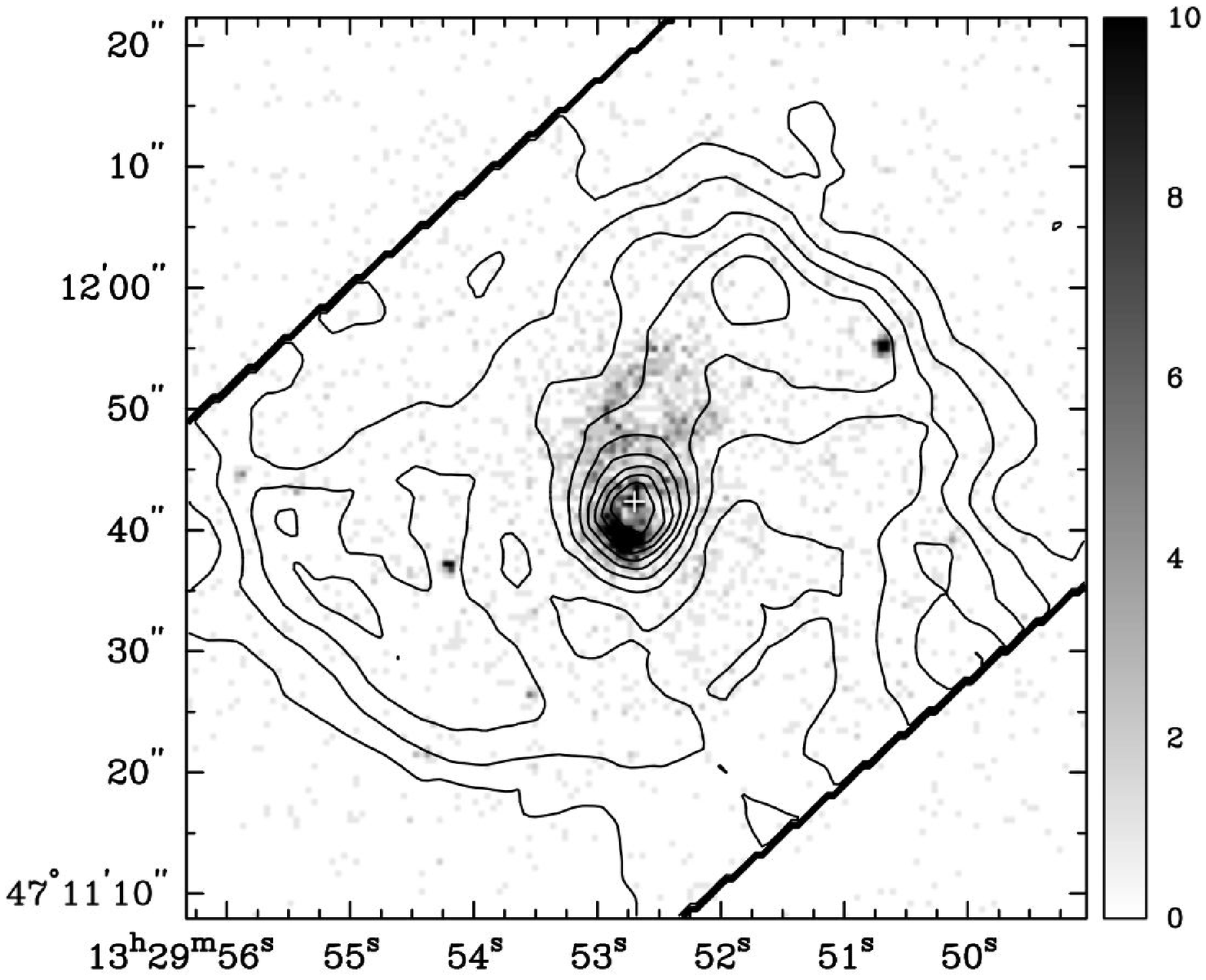}{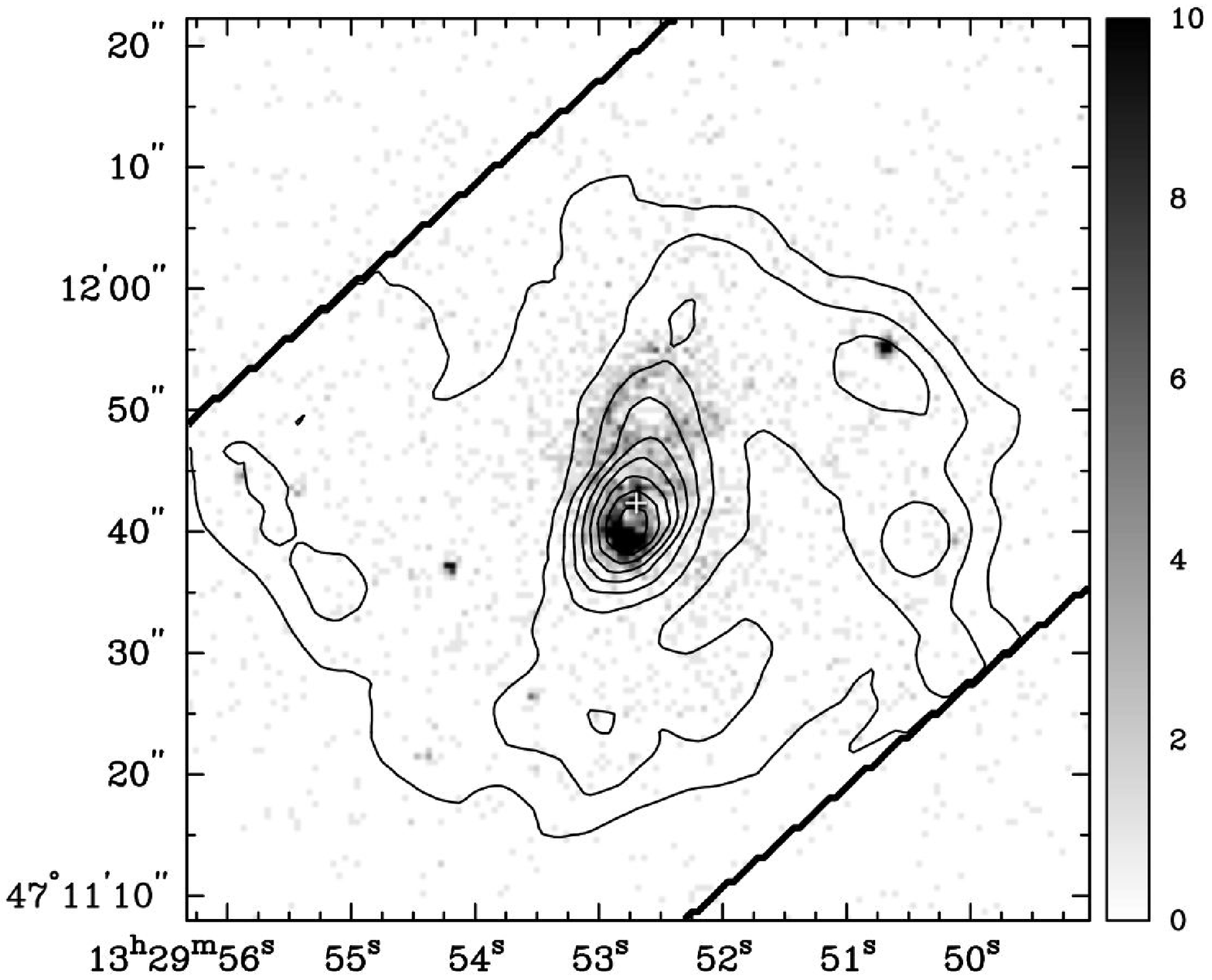}
\plottwo{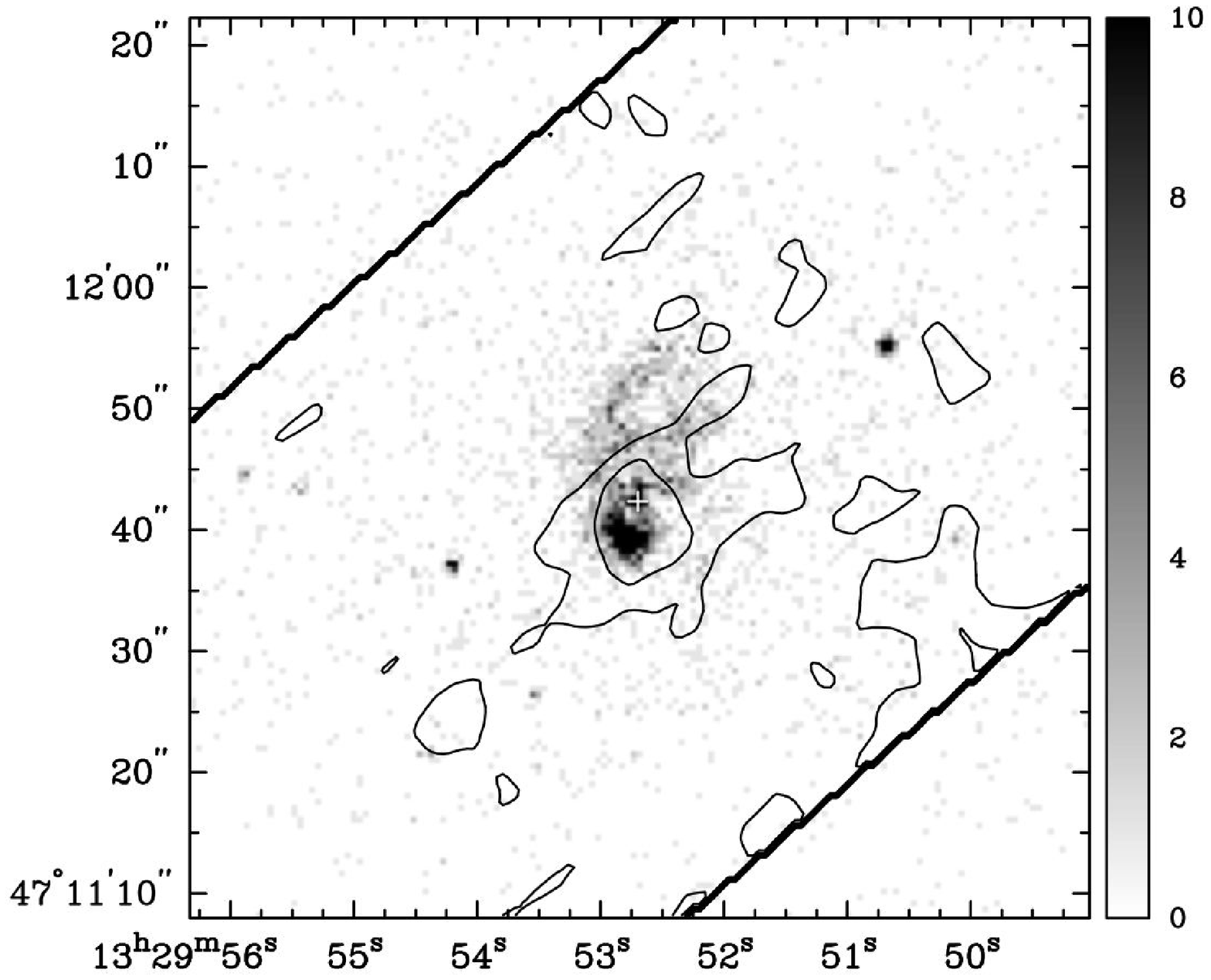}{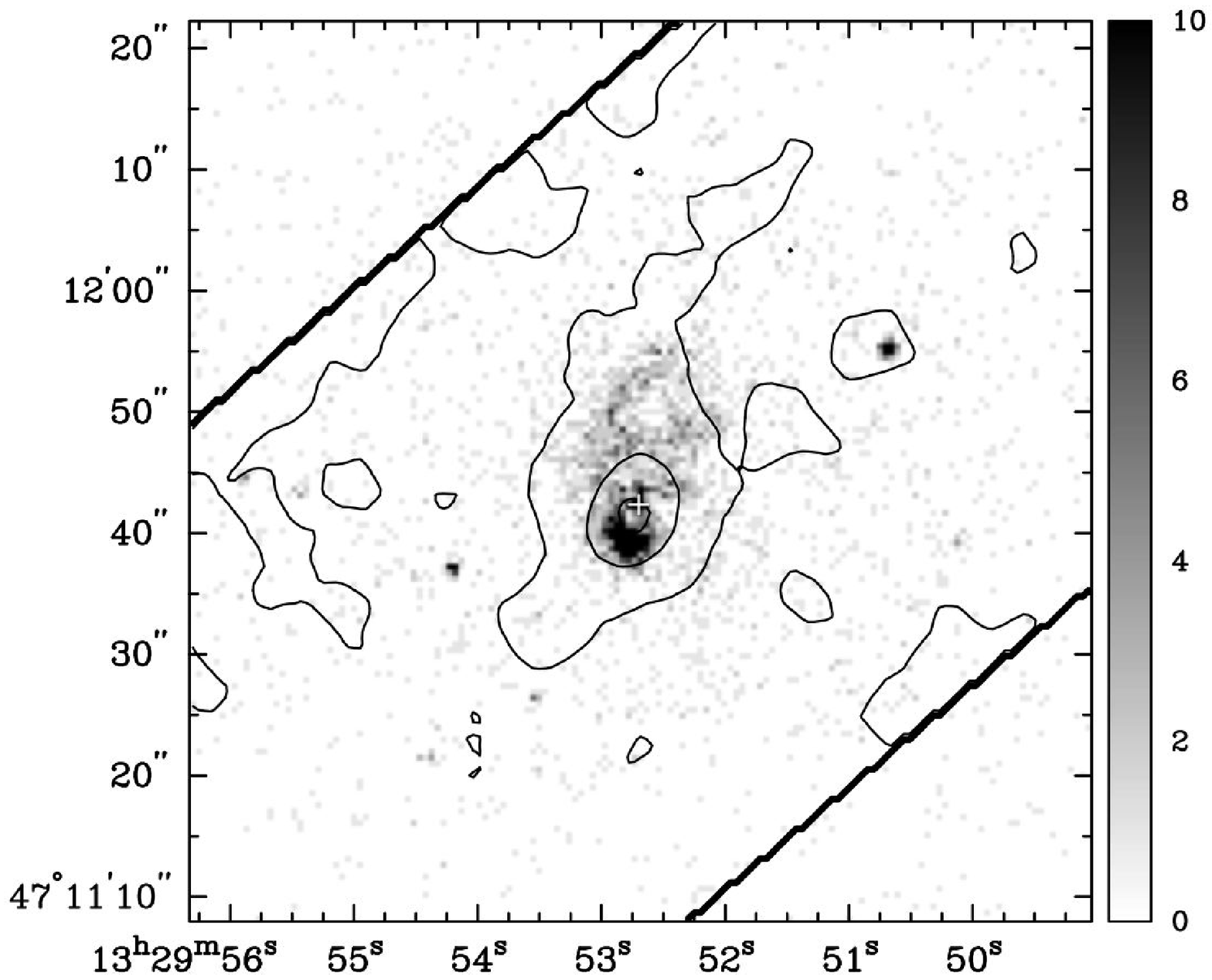}
\caption{Comparison of the 0.5 $-$ 10 keV X-ray emission band (in grayscale) to the 
H$_2$ S(2) ($top$ $left$), H$_2$ S(3) ($top$ $right$), 
H$_2$ S(4) ($bottom$ $left$), and H$_2$ S(5) ($bottom$ $right$) 
emission in the nuclear region of M51.  X-ray emission is in units of counts.  
The H$_2$ S(2) and H$_2$ S(3) emission contours are at 
10 \% of their peak values (2.20 $\times$ ${10^{-18}}$ and 1.35 $\times$ ${10^{-17}}$ 
W $\mathrm{m^{-2}}$, respectively).  The H$_2$ S(4) contours are at 1.0 
$\times$ ${10^{-18}}$ and 2.0 $\times$ ${10^{-18}}$ W $\mathrm{m^{-2}}$ and the 
H$_2$ S(5) contours are at 8.0 $\times$ ${10^{-19}}$, 4.0 $\times$ ${10^{-18}}$, 
and 7.3 $\times$ ${10^{-18}}$ W $\mathrm{m^{-2}}$.}
\label{figure-17}
\end{figure}

X-ray studies of M51 have revealed bright X-ray emission from 
the nucleus, the extranuclear cloud (XNC, to the south of the nucleus), 
and the northern loop \citep{wil01,mad07}.  A radio jet that is believed to be shock heating 
the ISM has been observed emanating from the south of the nucleus 
towards the XNC in 6 cm imagery \citep{cra92}.  In Figure \ref{figure-16}, 
we compare the smoothed 0.5 - 10 keV band X-ray
image to the warm ($left$) and hot ($right$) H$_2$ mass
distributions.  The 0.5 - 10 keV band has been smoothed to the
resolution of the warm and hot H$_2$ mass distributions and
the nucleus, XNC, and northern loop are indistinguishable in the
smoothed image.  X-ray emission is brightest in the nucleus and
decreases into the northwest spiral arm that contains the greatest
H$_2$ mass.  There appears to be very little connection
between the 0.5 - 10 keV X-ray band and the warm H$_2$ mass
distribution.  The peak in X-ray emission is coincident with the hot H$_2$
mass peak. The brightest 0.5 - 10 keV X-ray emission originates
from the nucleus and is oriented north-to-south, similar to the [O
  IV](25.89 $\micron$) emission.  The peak in X-ray emission is
located within the peak in hot H$_2$ mass suggesting that
X-rays play an important role in producing the hot H$_2$
phase.  While there is a correlation between X-ray emission and the
hot H$_2$ phase, H$_2$ excitation by X-rays cannot
be distinguished from H$_2$ excitation by shocks. In Figure 
\ref{figure-17}, we compare the X-ray surface brightness to the H$_2$ S(2) $-$ 
H$_2$ S(5) maps and find that the nuclear H$_2$ emission 
appears to be correlated with the X-ray source.

\section{Conclusions}

We have spectrally mapped a strip across M51 using the $Spitzer$ IRS
low resolution modules.  We used the spatially resolved spectra to map
H$_2$ S(0) $-$ H$_2$ S(5) lines across
the strip.  We find:\\
\\
1.  The distribution of H$_2$
emission in M51 varies with H$_2$ rotational level.
H$_2$ S(0) emission is brightest in the spiral arms of the
galaxy while the higher $J$ transitions show the strongest emission
towards the nucleus.  The H$_2$ S(1) line is brightest
in the nuclear region and is offset from the peak in H$_2$ S(0) 
intensity in the inner northwest spiral arm by 10$\arcsec$.
The H$_2$ S(2) and H$_2$ S(3) maps show
H$_2$ emission in the nucleus, spiral arms, and inter-arm
regions of M51 and bar structure aligned north-to-south is apparent in
H$_2$ S(3) emission.  H$_2$ S(4) and H$_2$
S(5) emission is resolved in the nuclear region of M51.\\
\\
2.  The different distributions of H$_2$ emission in M51
indicate significant spatial variations in H$_2$
temperature and surface density.  Using the low $J$ (S(0) -- S(2)) lines to trace the warm (T = 100
$-$ 300 K) H$_2$, we find that the warm H$_2$
temperature is highest in the nuclear region at 192 K and
the warm H$_2$ surface density peaks in the northwest inner spiral arm
at 11 $\mathrm{M_\sun}$ $\mathrm{pc^{-2}}$.  Using the
higher $J$ (S(2) -- S(5)) lines to trace the hot (T = 400 $-$ 1000 K) H$_2$,
we find that the hot H$_2$ temperature is lowest
in the inner spiral arms (500 $-$ 550 K) and increases to $\sim$ 600 K
in the nucleus where the largest hot H$_2$ surface density
is found to be 0.24 $\mathrm{M_\sun}$ $\mathrm{pc^{-2}}$.\\
\\
3.  The warm and the hot H$_2$ surface density distributions are not
cospatial and the warm-to-hot surface density ratio varies across M51.  The warm
H$_2$ surface density distribution peaks in the northwest spiral arm and is offset from
the hot mass peak by 11$\arcsec$.  The hot H$_2$ surface density distribution 
shows two peaks, one in the nucleus of M51 and one located interior to the 
northwest inner spiral arm of M51.  Variations in the
warm-to-hot H$_2$ ratio and differences in the
distributions of the H$_2$ line emission across 
M51 suggest that the warm H$_2$ is mostly produced 
by UV photons in star forming regions while the hot H$_2$ is 
mostly produced by shocks or X-rays associated with nuclear activity.\\
\\
4. The warm H$_2$ follows the cold molecular gas traced by CO 
in the spiral arms of M51 indicating that the warm phase is associated 
with the surface layers of dense molecular clouds.  The H$_2$ S(0) $-$
H$_2$ S(3) contours trace the CO; however, within the spiral
arms, the peaks in H$_2$ can be offset from the peaks in CO
intensity.\\
\\
5.  Comparing the distributions of H$_2$ to 
$V$ band imagery and H$\alpha$ emission reveals
that the warm and hot H$_2$ in the spiral arms is found in the dust lanes
rather than spatially coincident with the H$\alpha$ emission.\\
\\
6.  [O IV](25.89 $\micron$) emission and X-ray
intensity peak in the nuclear region of M51 and their 
peaks are spatially coincident with the peak in hot H$_2$ 
surface density.  This implies that the hot H$_2$ 
is more efficiently heated by the AGN, shocks (possibly associated
with the AGN), or X-rays associated with the AGN.  The spatial
distributions of the [O IV] emission and X-ray surface brightness are
very similar preventing the characterization of the primary excitation 
mechanism (shocks or X-rays) of the hot H$_2$ phase in the nuclear region.\\ 
\\

\acknowledgments

The author graciously acknowledges the Spitzer Science Center Spitzer
Visiting Graduate Student Fellowship program and committee for
providing support for this research.  The author would like to
specifically acknowledge the program coordinators, Phil Appleton and
Alberto Noriega-Crepso. The authors would also like to thank the 
anonymous referee who helped to clarify our results and improve our discussion.
Partial support for the completion and preparation for publication of this study by the
author was provided by AURA grant GO10822.1 to Rice
University.\\ 
\\
{\it Facilities:} \facility{Spitzer Science Center
  (SSC)}, \facility{Spitzer Space Telescope (SST)},
\facility{Berkely-Illinois-Maryland Array (BIMA)}.

\clearpage

\clearpage

\clearpage

\begin{figure}
\epsscale{1.1}
\plottwo{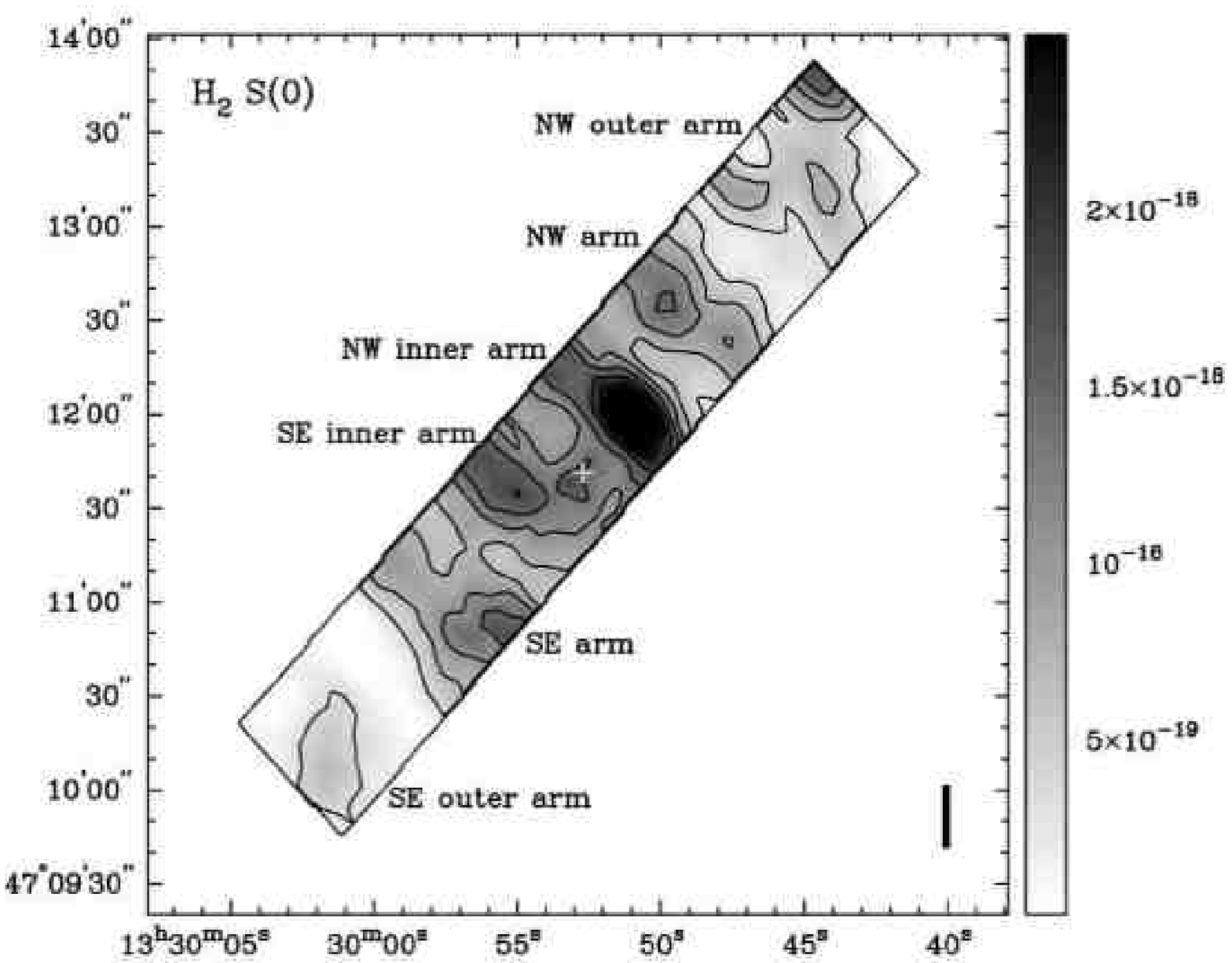}{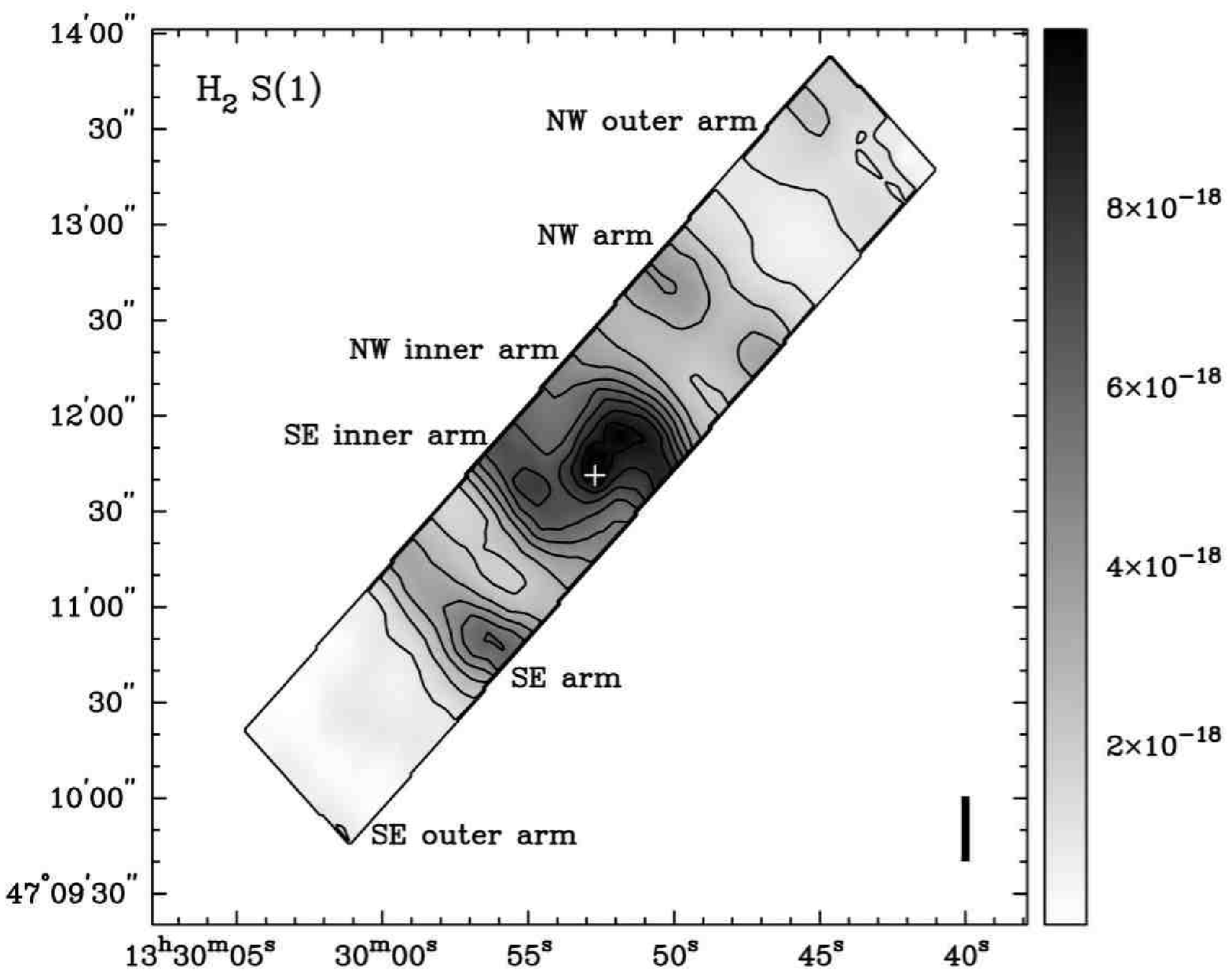}
\plottwo{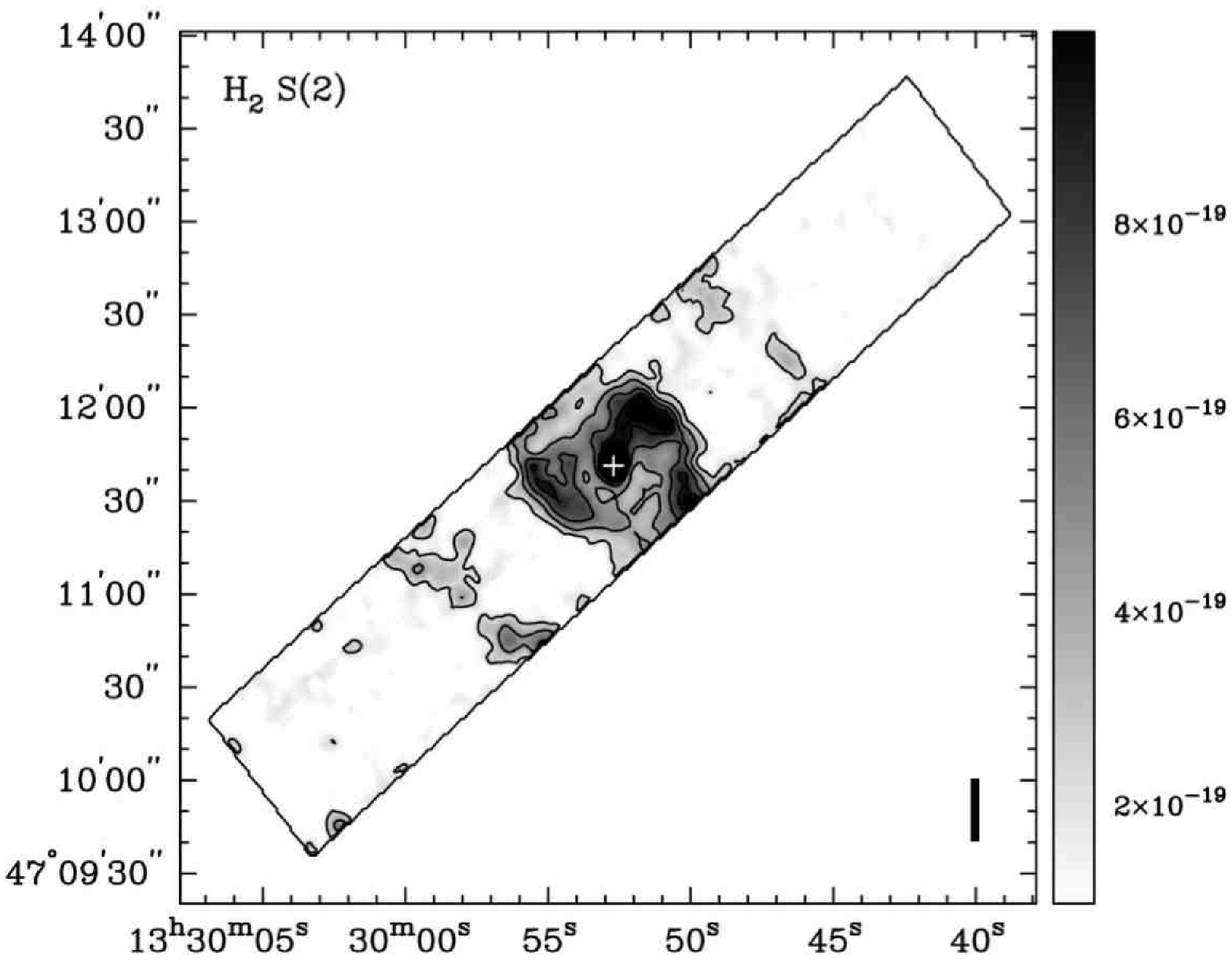}{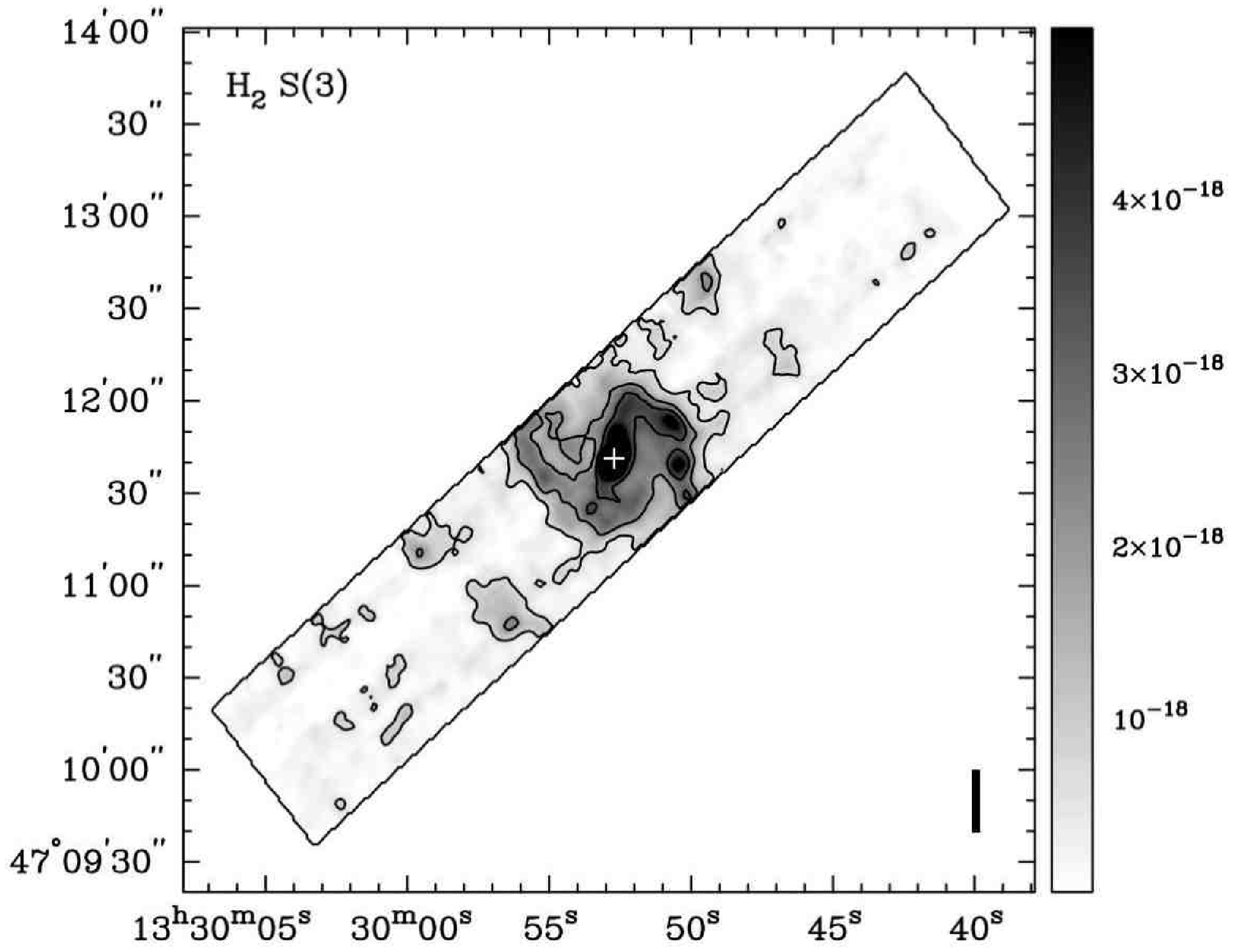}
\epsscale{1.1}
\plottwo{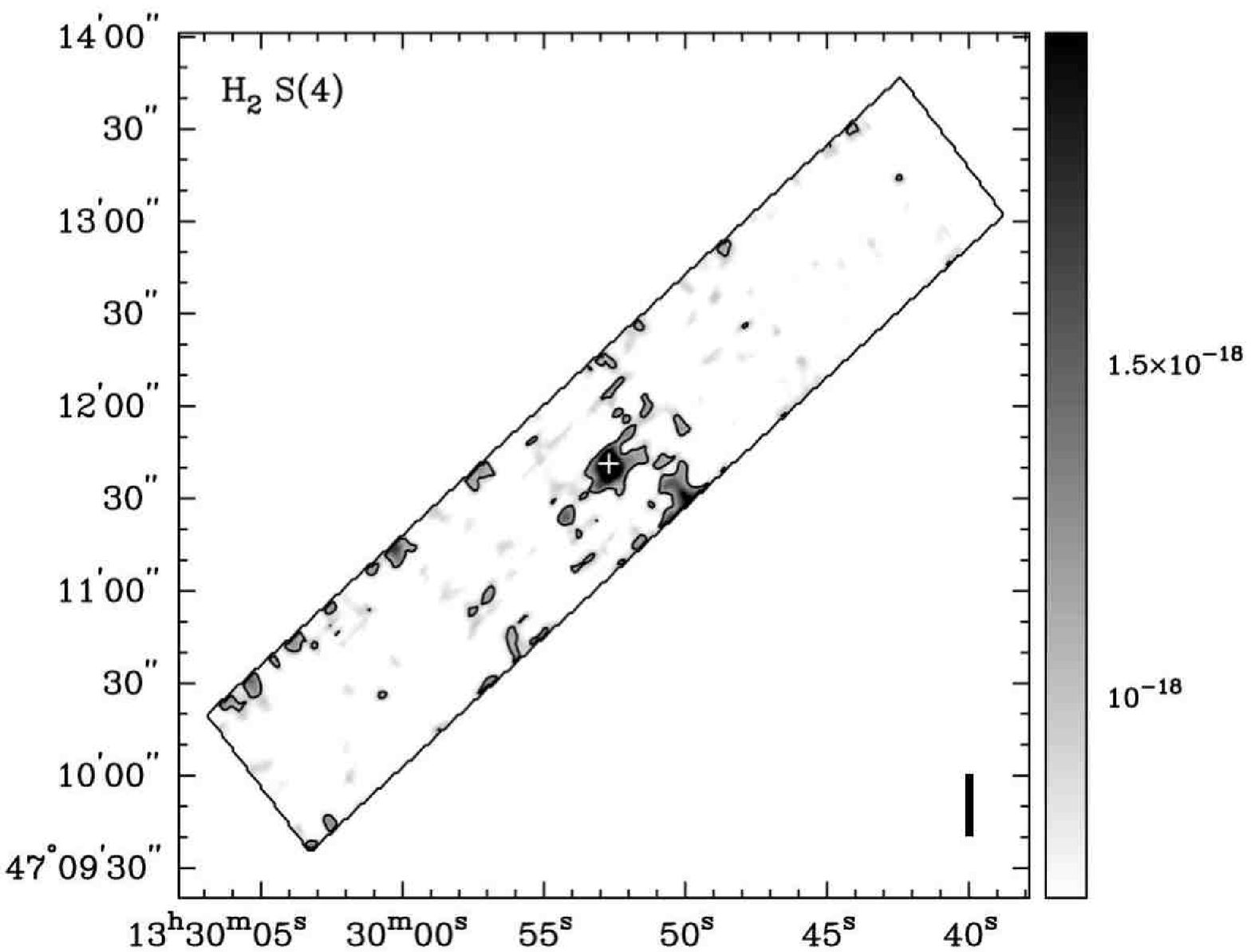}{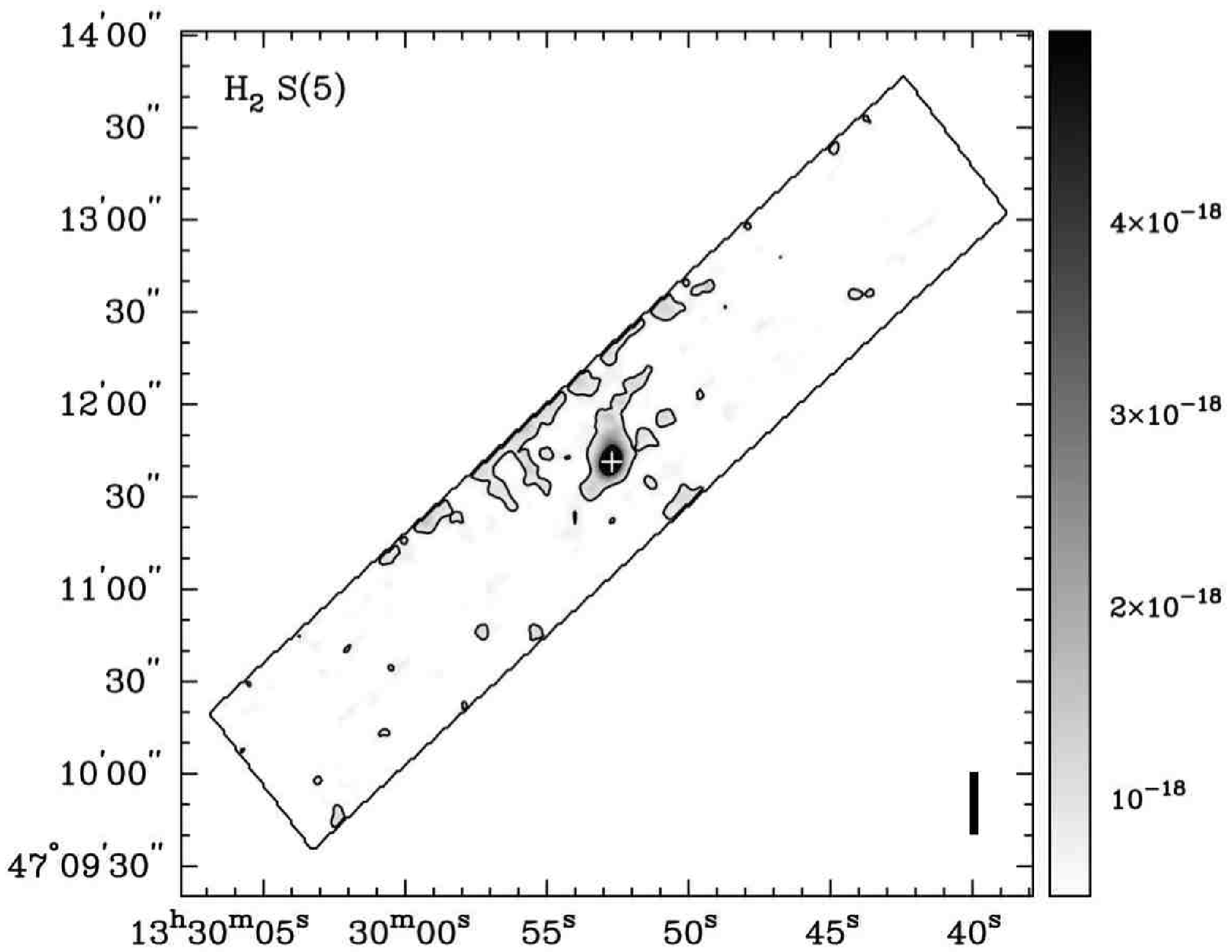}
 \end{figure}
 \clearpage
 
 \begin{figure}
 \figurenum{3}
 \caption{Maps of the H$_2$ S(0) ($top$ $left$), H$_2$ S(1)
 ($top$ $right$), H$_2$ S(2) ($middle$ $left$), H$_2$ S(3) 
 ($middle$ $right$), H$_2$ S(4) ($bottom$ $left$), and H$_2$ 
 S(5) ($bottom$ $right$) line fluxes across the SL and LL strips that we mapped 
 with the $Spitzer$ IRS.  The H$_2$ S(0) and H$_2$ S(1) 
 maps are created from the LL data cube.    
The H$_2$ S(2), H$_2$ S(3), H$_2$ S(4), and 
H$_2$ S(5) maps are created from the SL data cube.  
The grayscale is in units of W $\mathrm{m^{-2}}$.  Contour levels are at 
3.7 $\times$ ${10^{-19}}$, 7.3 $\times$ ${10^{-19}}$, 1.1 $\times$ ${10^{-18}}$, 1.5 $\times$ ${10^{-18}}$, 1.8 $\times$ ${10^{-18}}$, 2.2 $\times$ ${10^{-18}}$, and 2.9 $\times$ ${10^{-18}}$ W $\mathrm{m^{-2}}$ for H$_2$ S(0); 
1.1 $\times$ ${10^{-18}}$, 2.1 $\times$ ${10^{-18}}$, 3.2 $\times$ ${10^{-18}}$, 4.3 $\times$ ${10^{-18}}$, 5.4 $\times$ ${10^{-18}}$, 6.4 $\times$ ${10^{-18}}$, 7.5 $\times$ ${10^{-18}}$, 8.6 $\times$ ${10^{-18}}$, and 9.6 $\times$ ${10^{-18}}$ W $\mathrm{m^{-2}}$ for H$_2$ S(1); 
2.2 $\times$ ${10^{-19}}$, 4.4 $\times$ ${10^{-19}}$,  6.7 $\times$ ${10^{-19}}$, 8.9 $\times$ ${10^{-19}}$, and 1.1 $\times$ ${10^{-18}}$ W $\mathrm{m^{-2}}$ for H$_2$ S(2);
1.3 $\times$ ${10^{-18}}$, 4.0 $\times$ ${10^{-18}}$, 6.7 $\times$ ${10^{-18}}$, 9.4 $\times$ ${10^{-18}}$, and 1.2 $\times$ ${10^{-17}}$ W $\mathrm{m^{-2}}$ for H$_2$ S(3);  
1.0 $\times$ ${10^{-18}}$ and 2.0 $\times$ ${10^{-18}}$ W $\mathrm{m^{-2}}$ for H$_2$ S(4); 
8.0 $\times$ ${10^{-19}}$, 4.0 $\times$ ${10^{-18}}$, and 7.3 $\times$ ${10^{-18}}$W $\mathrm{m^{-2}}$ for H$_2$ S(5).  
The vertical axis is the right ascension and the horizontal axis is the declination.  
In all of the maps, north is up, east is to the left, and the cross 
denotes the nucleus of the galaxy.  The different spiral 
arm regions are labeled on the H$_2$ S(0) and H$_2$ S(1) 
maps in order to aid in discussing the molecular gas morphologies.  
The bar in the bottom right corner of the maps 
represents 1 kiloparsec.  The box around 
the maps represents the SL or LL strip that we mapped.}
  \label{figure-3}
  \end{figure}

\clearpage

\clearpage

\clearpage

\clearpage

\clearpage

\clearpage

\clearpage

\begin{figure}
\epsscale{1.1}
\figurenum{10}
\plottwo{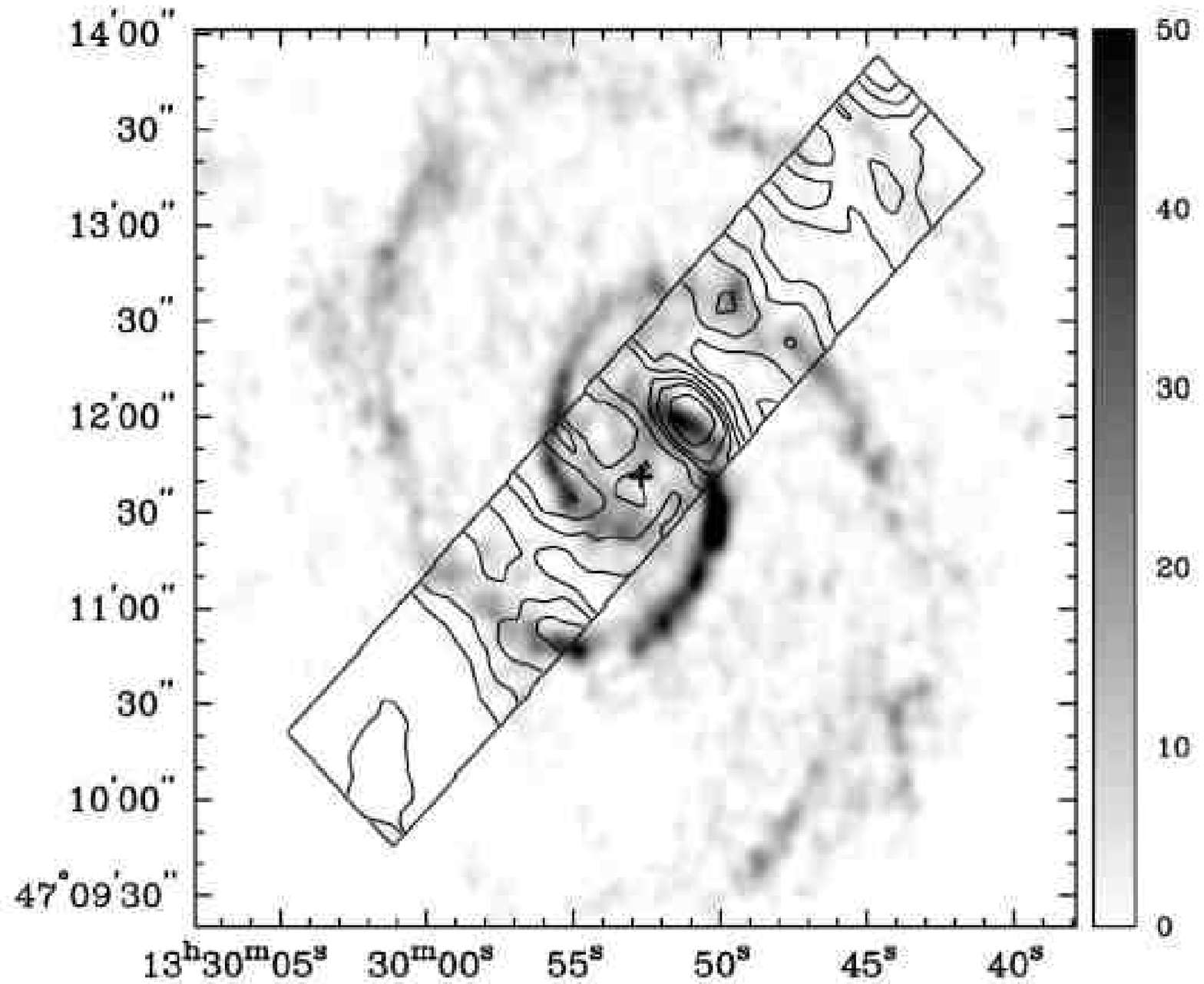}{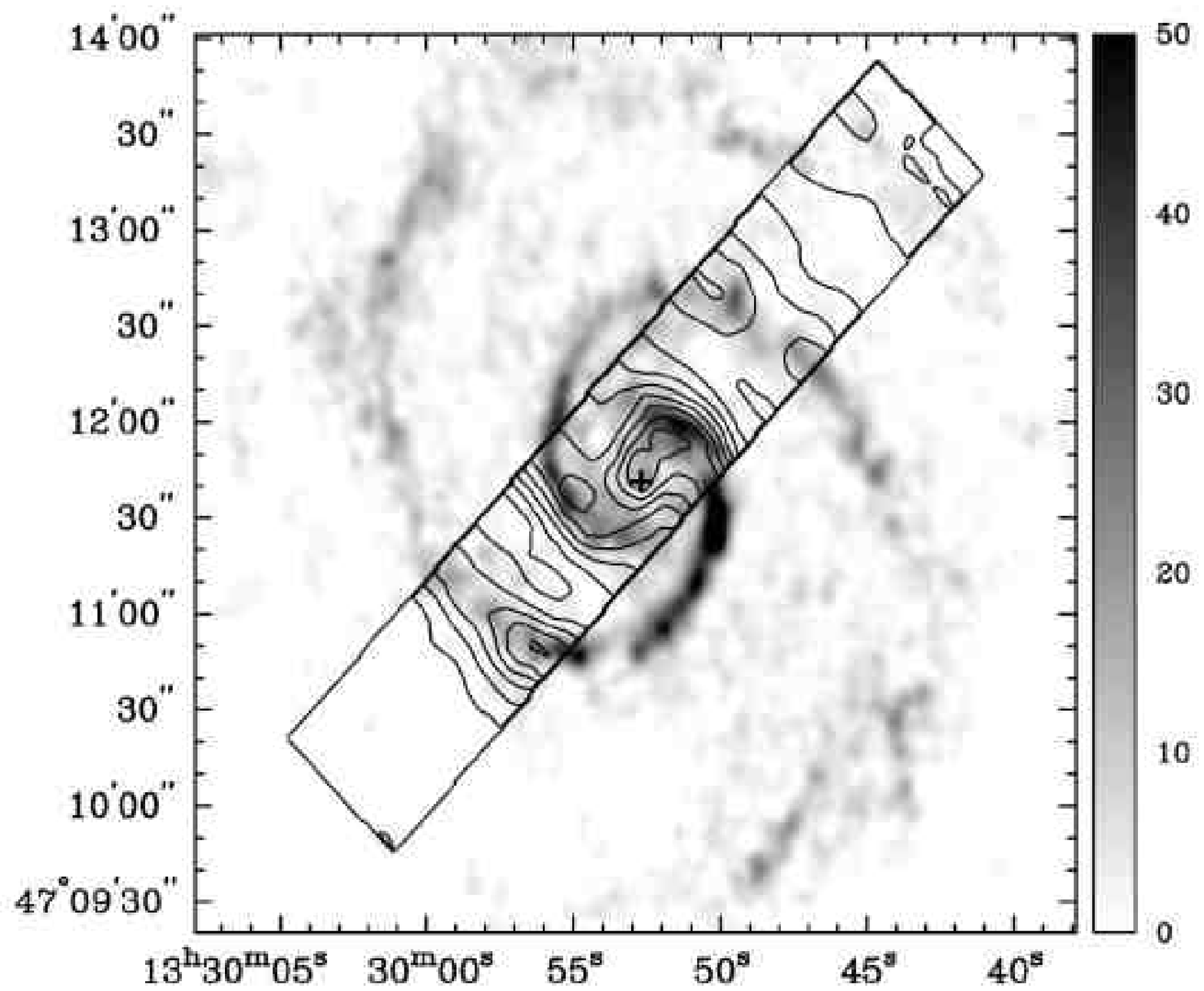}
\plottwo{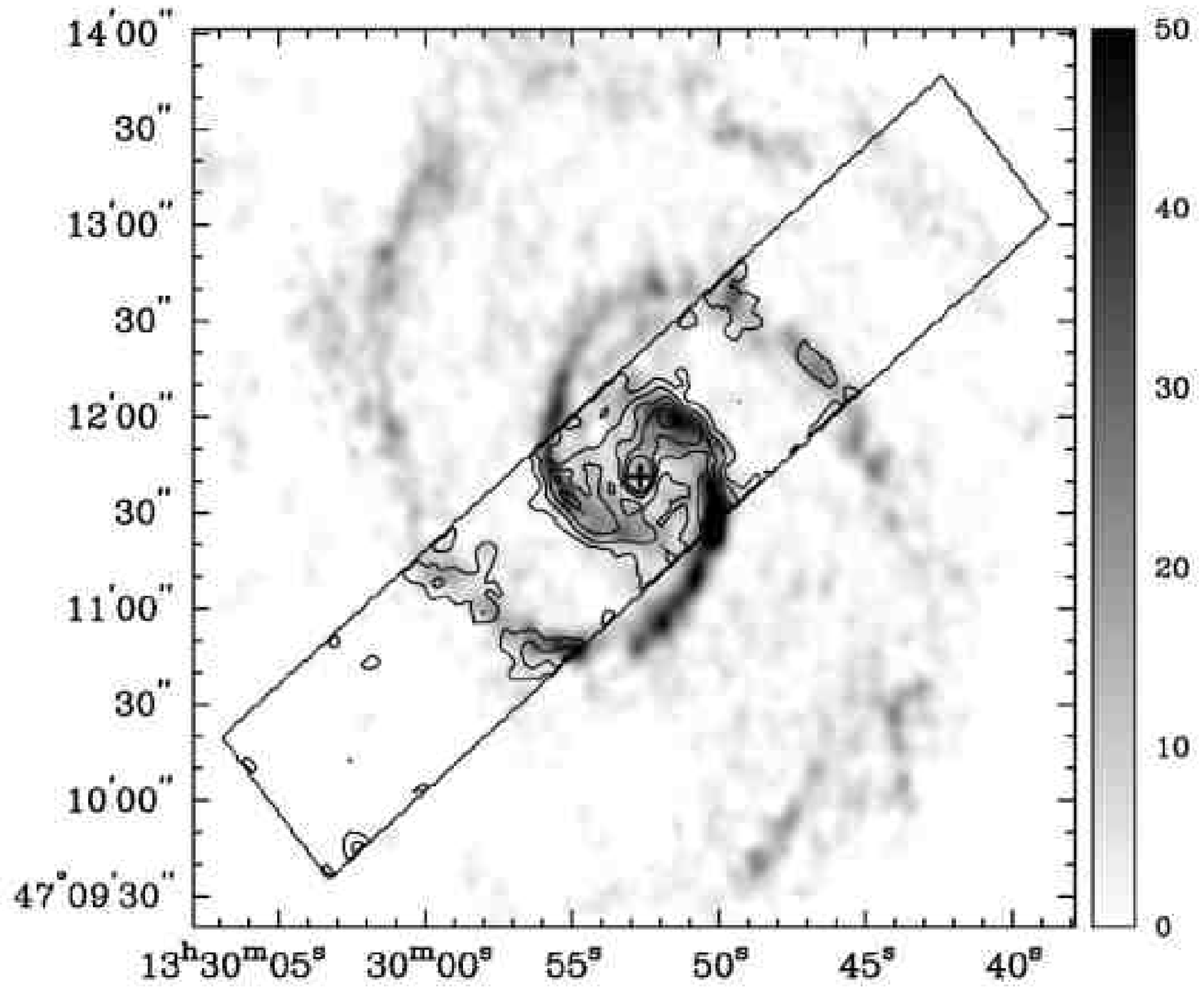}{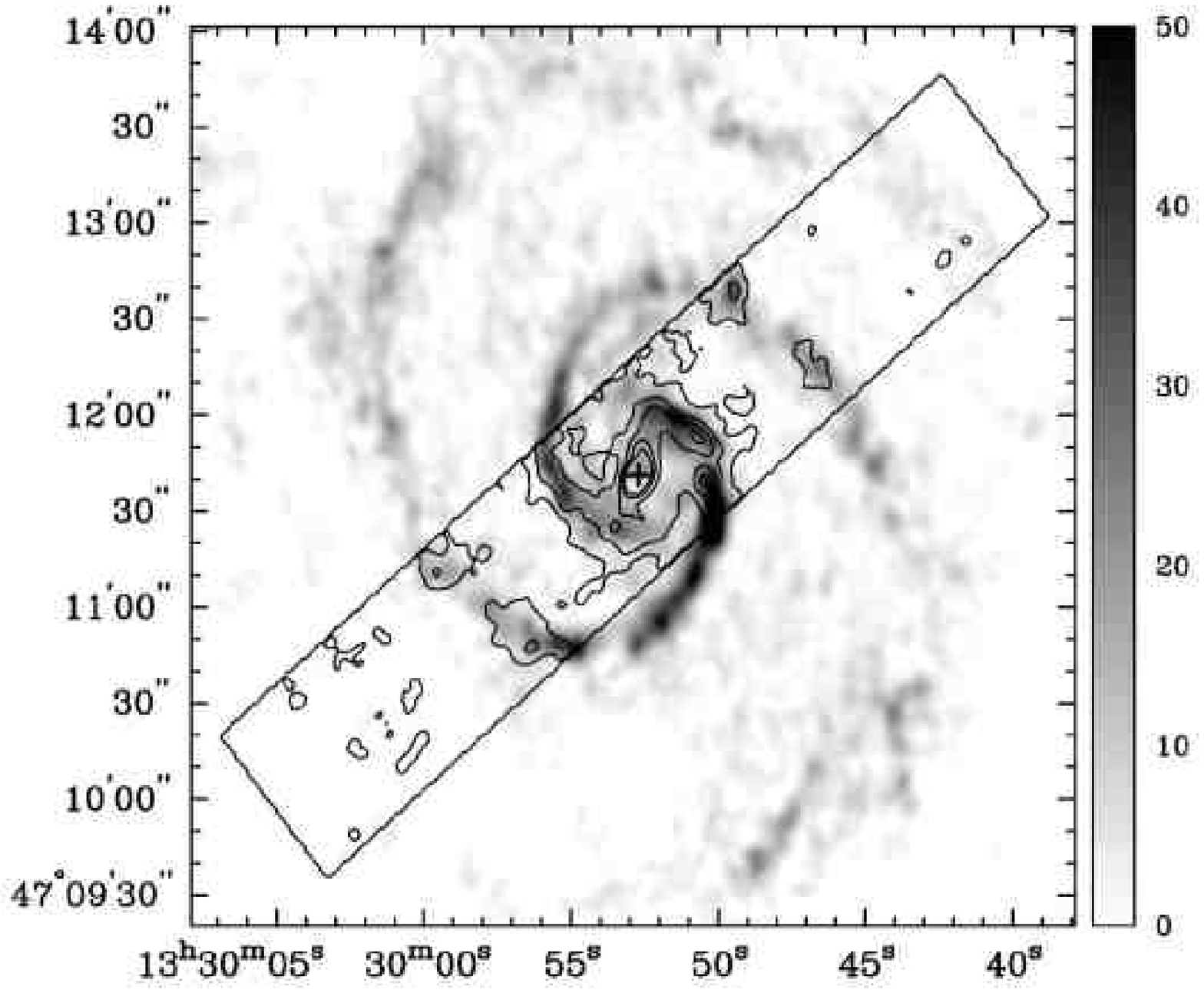}
\caption{Comparison of the CO emission to the H$_2$ S(0) ($top$ $left$), 
 H$_2$ S(1) ($top$ $right$),  H$_2$ S(2) ($bottom$ $left$),  and 
 H$_2$ S(3) ($bottom$ $right$) emission.  The CO emission maps are in 
 units of Jy km $\mathrm{s^{-1}}$.  Contour levels for H$_2$ S(0), 
 H$_2$ S(1), H$_2$ S(2), and H$_2$ S(3) are the 
 same as in Figure \ref{figure-3}.}
\label{figure-10}
\end{figure}

\clearpage

\clearpage

\clearpage

\begin{figure}
\epsscale{1.1}
\figurenum{13}
\plottwo{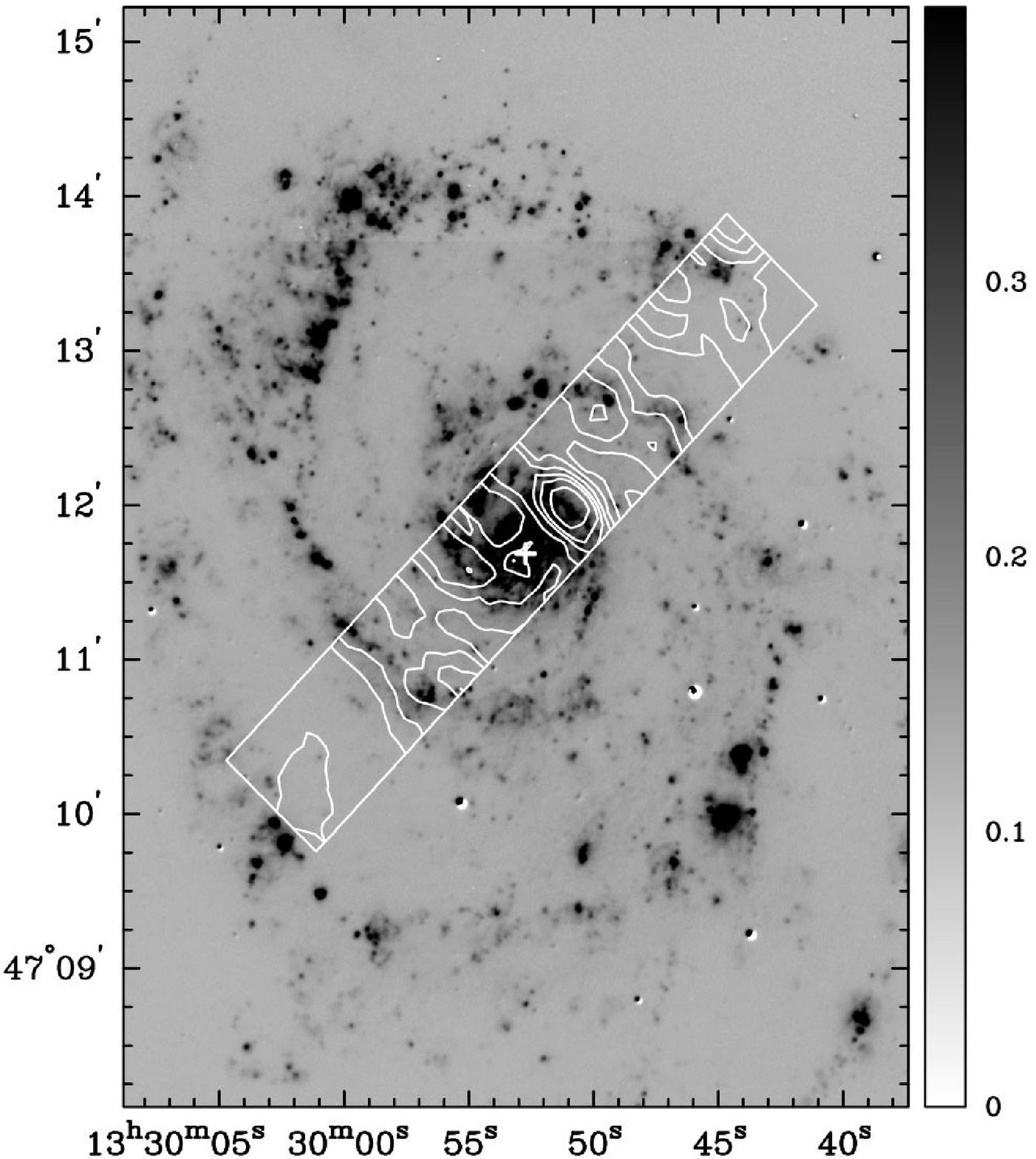}{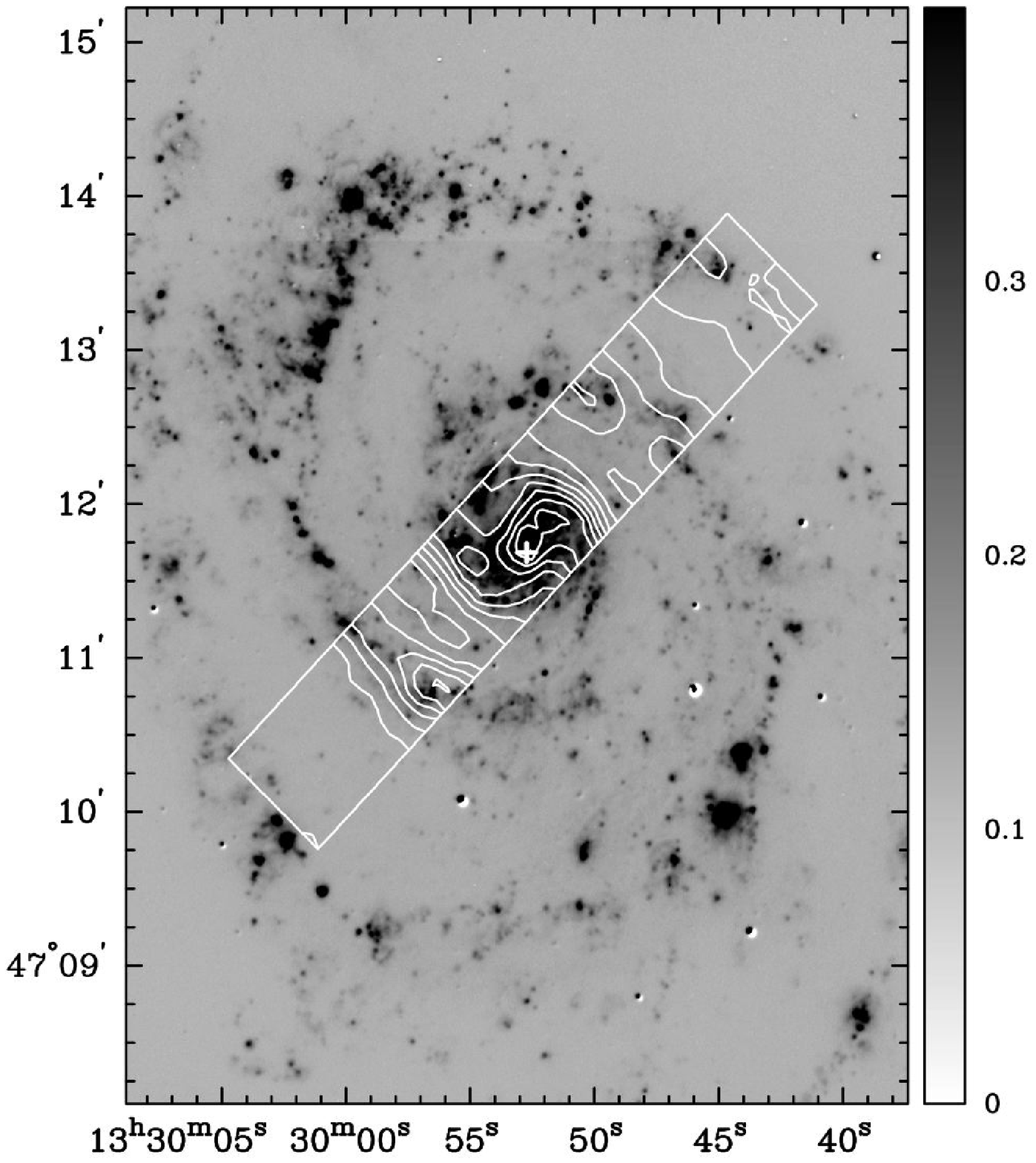}
\plottwo{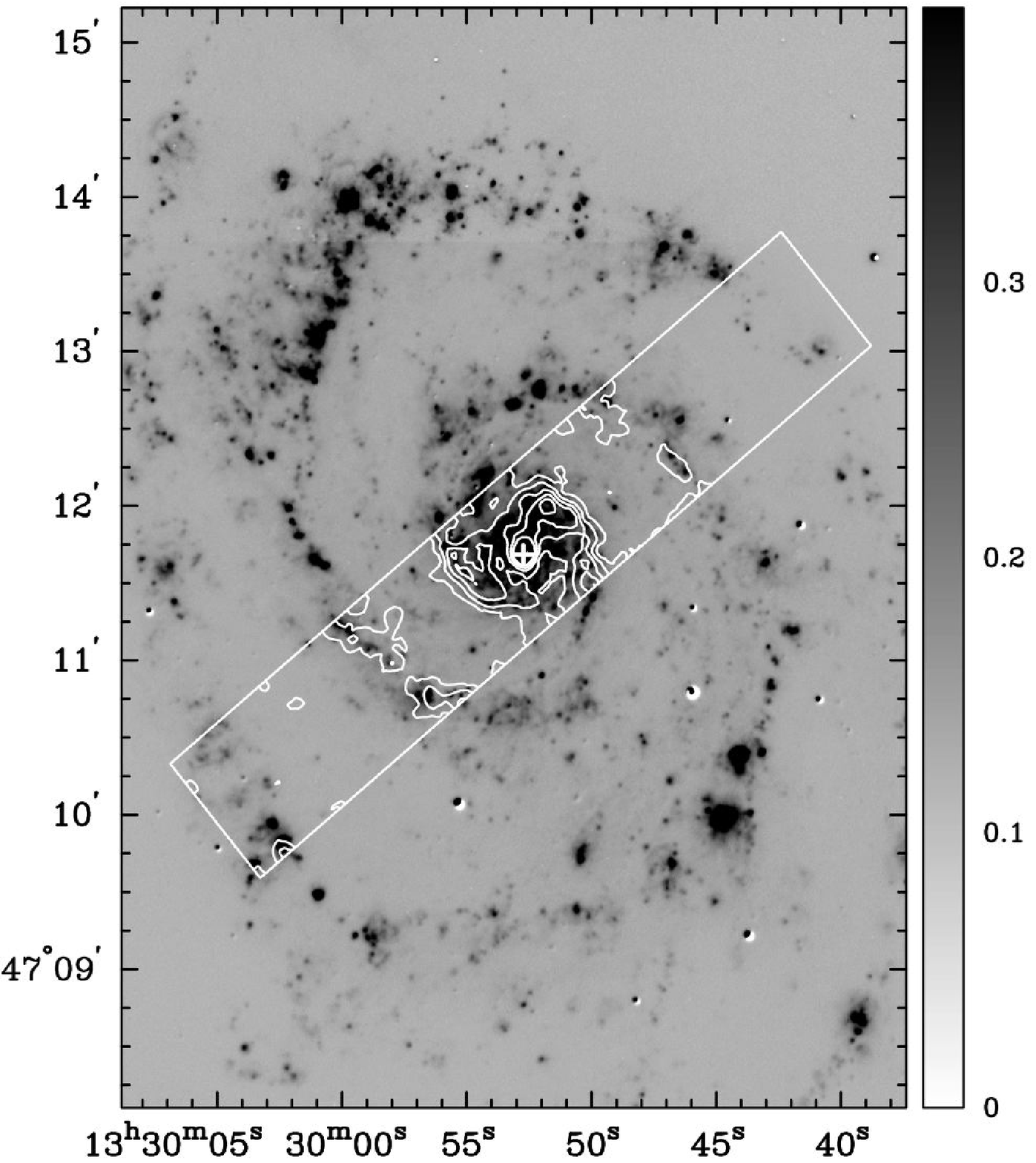}{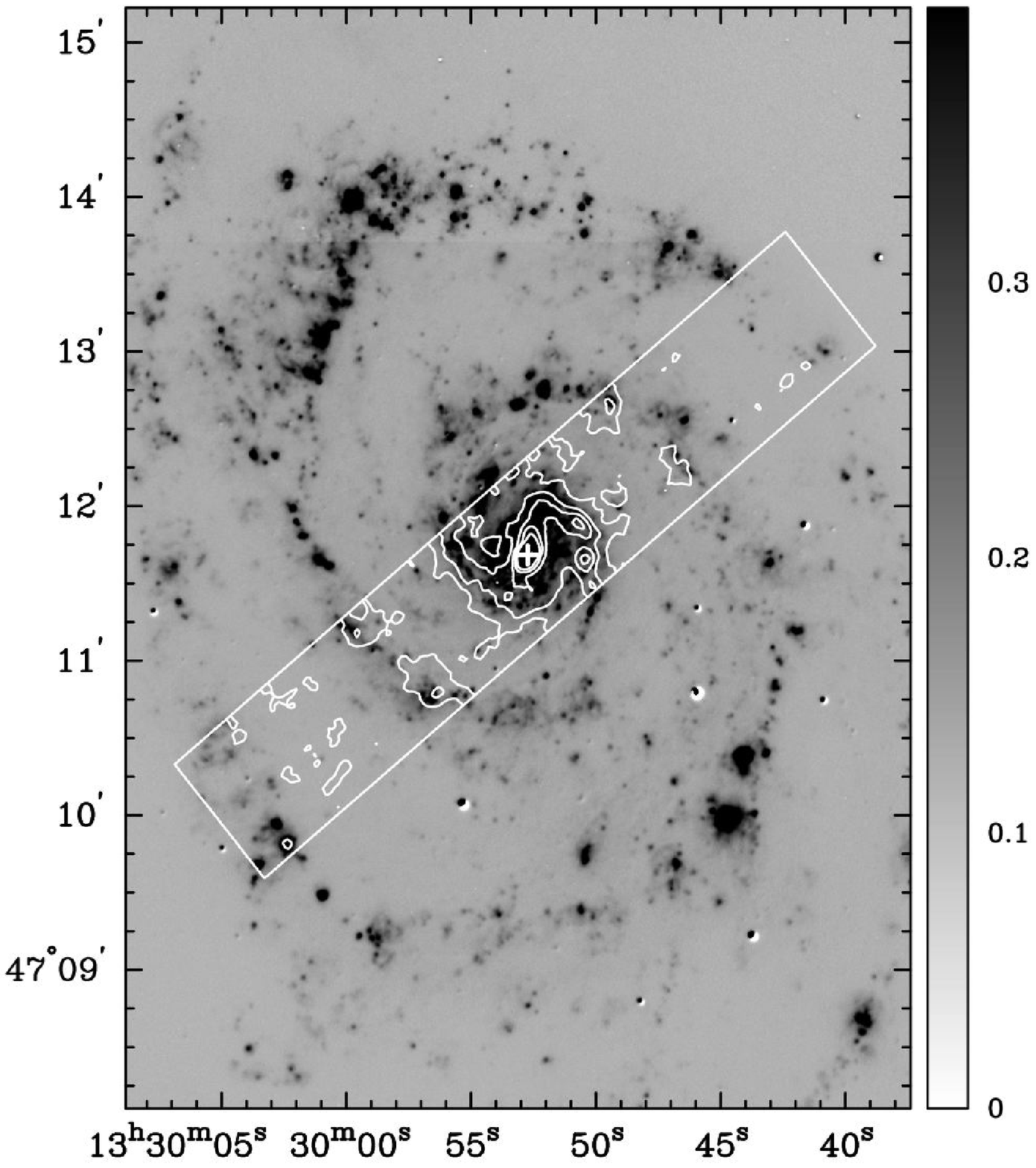}
\caption{Comparison of H$\alpha$ emission to the H$_2$ S(0) ($top$ $left$),  
H$_2$ S(1) ($top$ $right$),  H$_2$ S(2) ($bottom$ $left$),  and 
H$_2$ S(3) ($bottom$ $right$) emission.  The H$\alpha$ image is in units 
of counts s$^{-1}$.  Contour levels for H$_2$ S(0), H$_2$ S(1), 
H$_2$ S(2), and H$_2$ S(3) are the same as in Figure \ref{figure-3}.}
\label{figure-13}
\end{figure}

\clearpage

\begin{figure}
\epsscale{1.1}
\figurenum{14}
\plottwo{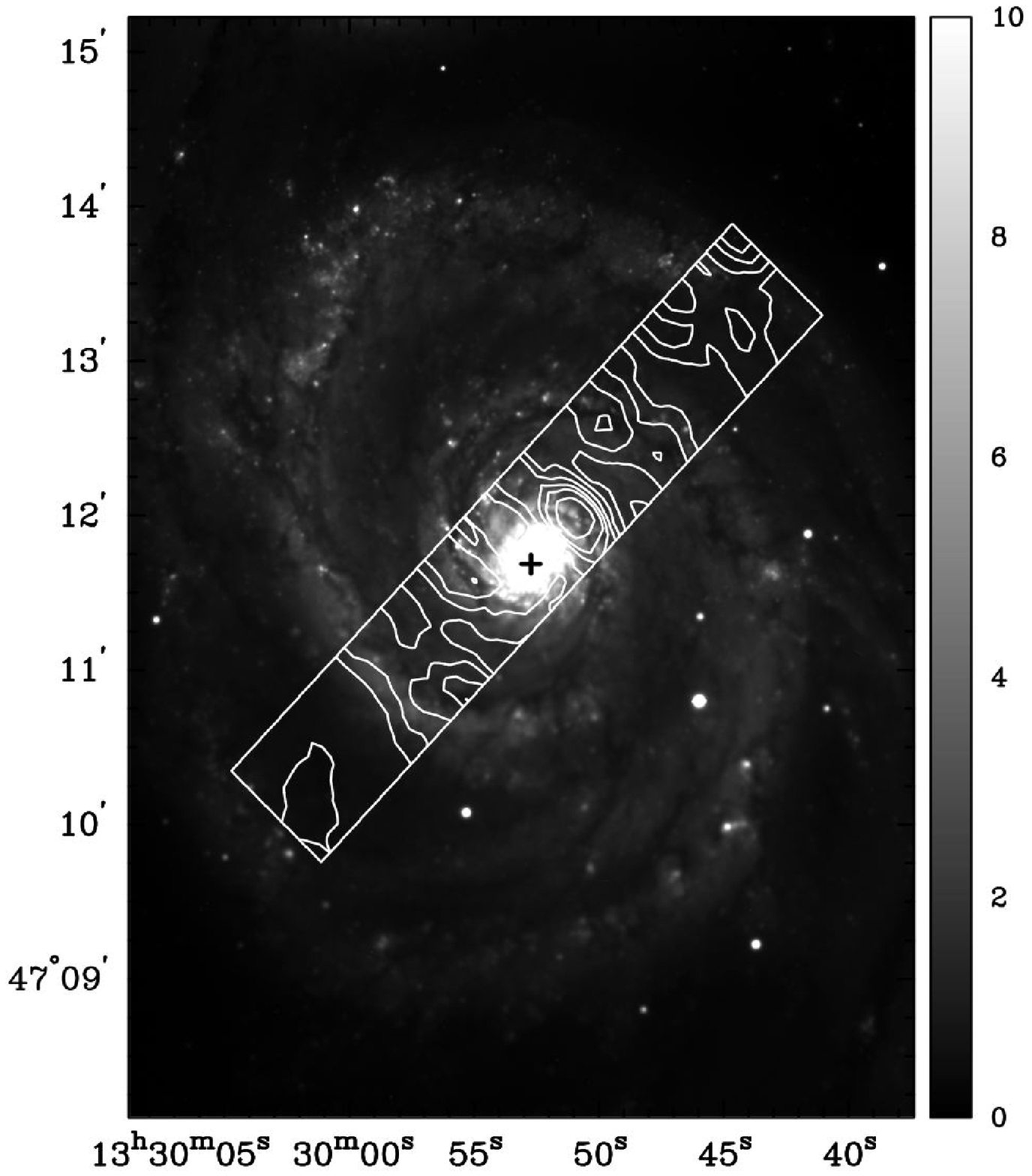}{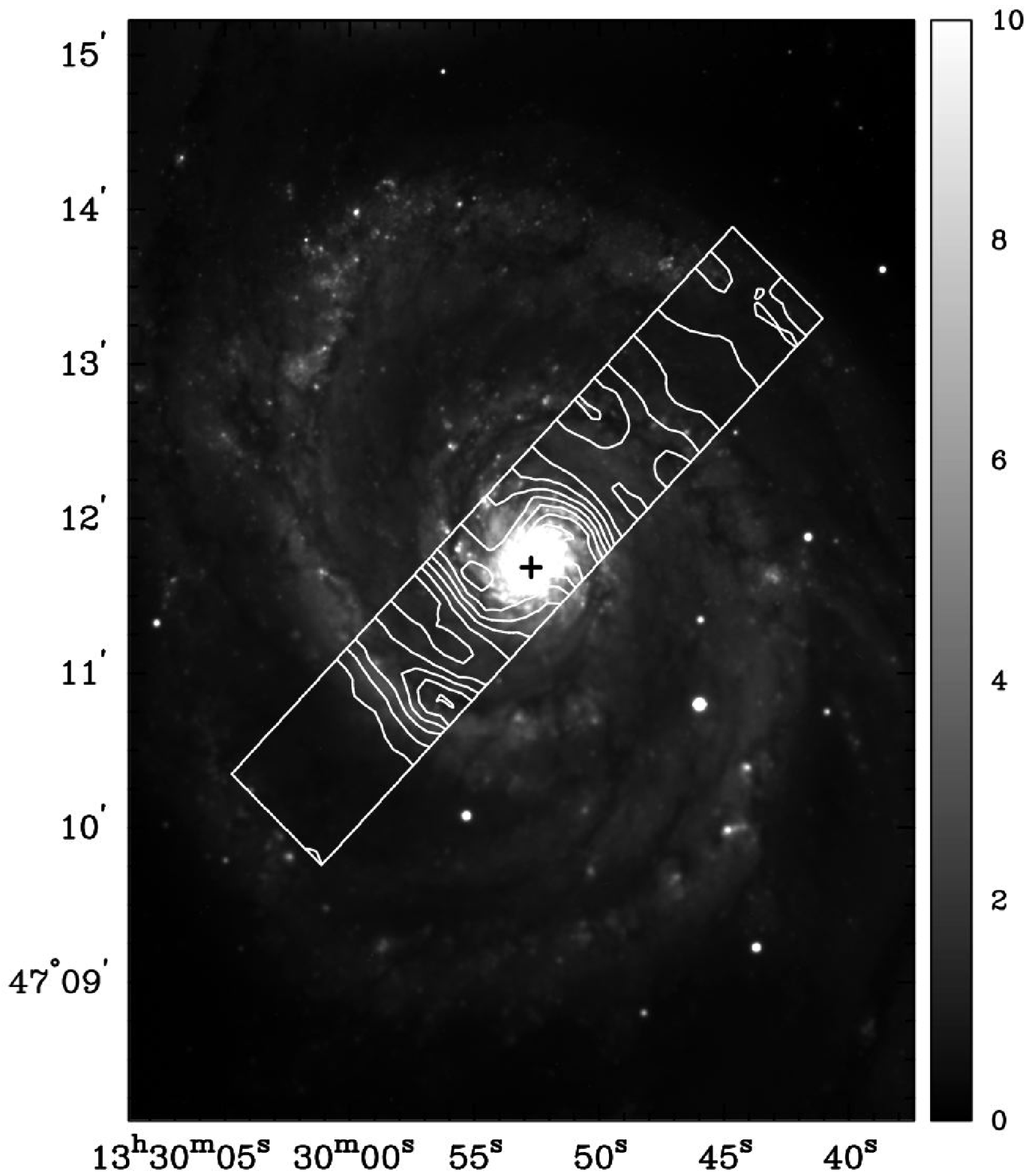}
\plottwo{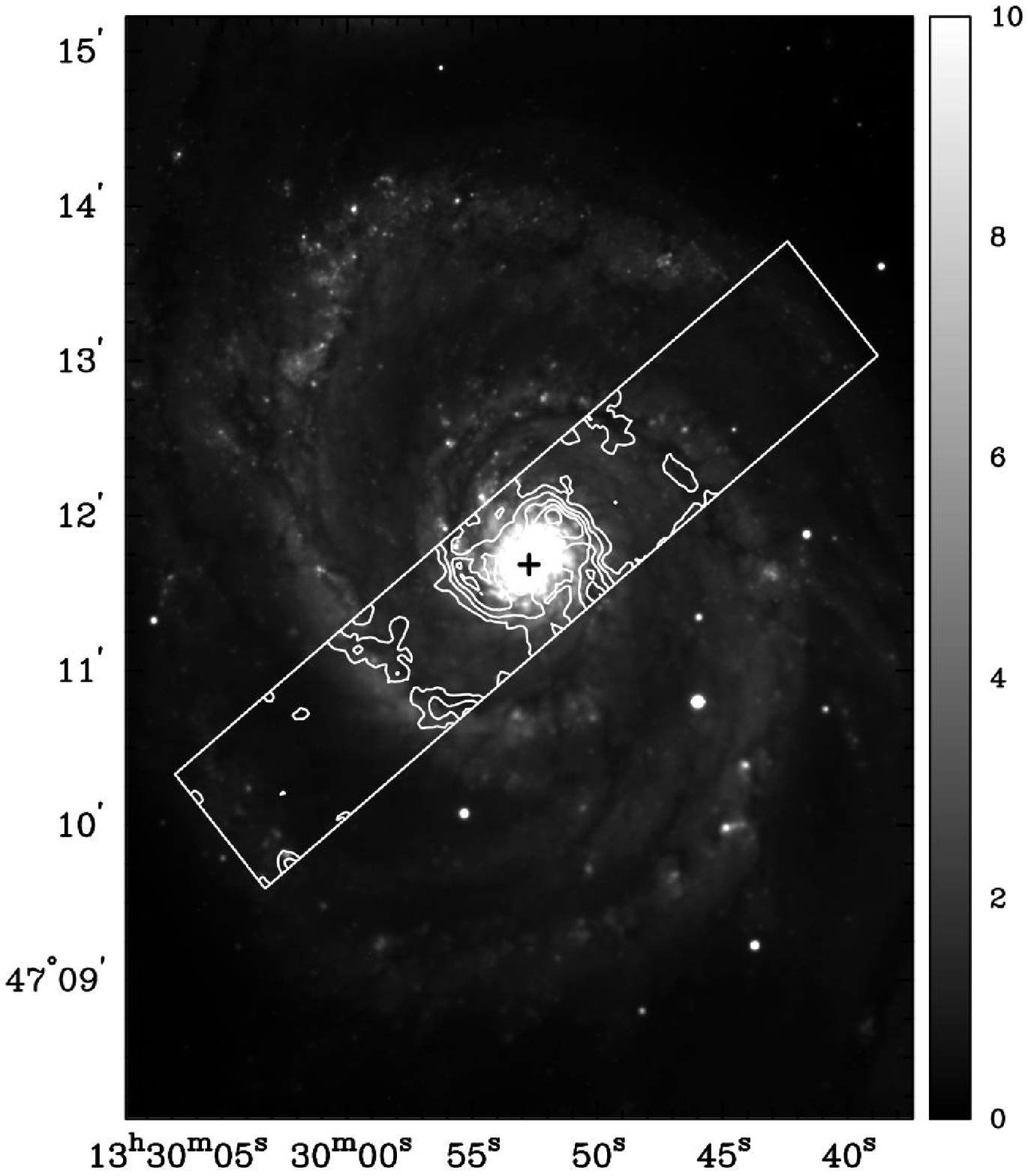}{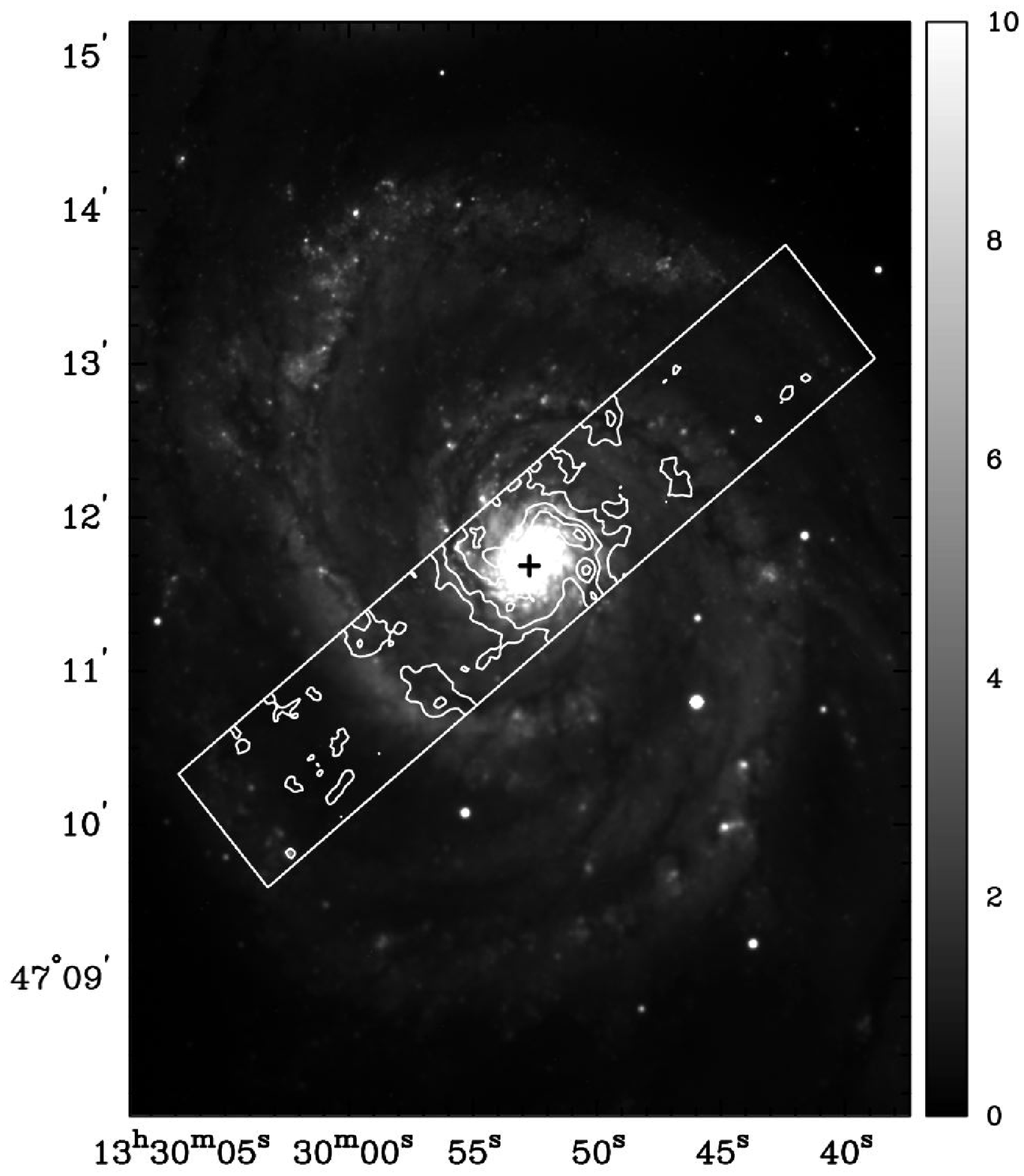}
\caption{Comparison of the $V$ band image to the H$_2$ S(0) ($top$ $left$),  
H$_2$ S(1) ($top$ $right$),  H$_2$ S(2) ($bottom$ $left$),  and 
H$_2$ S(3) ($bottom$ $right$) emission.  The $V$ band image is in units 
of counts s$^{-1}$.  Contour levels for H$_2$ S(0), H$_2$ S(1), 
H$_2$ S(2), and H$_2$ S(3) are the same as in Figure \ref{figure-3}.}
\label{figure-14}
\end{figure}

\clearpage

\clearpage

\clearpage

\clearpage

\begin{deluxetable}{cccccc}
\tabletypesize{\scriptsize}
\rotate
\tablecaption{H$_2$ Parameters 
\tablenum{1}
\label{tbl1}}
\tablewidth{0pt}
\tablehead{
\colhead{Transition} & \colhead{Wavelength (\micron)} & \colhead{Rotational State (J)} & \colhead{Energy (E/k)} & \colhead{A ($\mathrm{s^{-1}}$)} & \colhead{Statistical Weight (g)} }
\startdata
H$_2$(0-0)S(0) & 28.22 & 2 & 510 & 2.94$\times$${10^{-11}}$ & 5 \\
H$_2$(0-0)S(1) & 17.04 & 3 & 1015 & 4.76$\times$${10^{-10}}$ & 21 \\
H$_2$(0-0)S(2) & 12.28 & 4 & 1682 & 2.76$\times$${10^{-9}}$ & 9 \\
H$_2$(0-0)S(3) & 9.66 & 5 & 2504 & 9.84$\times$${10^{-9}}$ & 33 \\
H$_2$(0-0)S(4) & 8.03 & 6 & 3474 & 2.64$\times$${10^{-8}}$ & 13 \\
H$_2$(0-0)S(5) & 6.91 & 7 & 4586 & 5.88$\times$${10^{-8}}$ & 45 \\
\enddata
\tablecomments{The statistical weight (g) is (2$J$ +1)(2$I$+1) where $I$ equals 1 for odd J transitions (ortho transitions) and $I$ equals 0 for even J transitions (para transitions).} 
\end{deluxetable}

\end{document}